\documentclass{aastex62}

\defcitealias{K16}{Paper~I}

\turnoffediting

\received{July 12, 2018}
\revised{September 9, 2018}
\accepted{\today}

\submitjournal{ApJS}

\shorttitle{Grain heating by cosmic rays in dark cores}
\shortauthors{J. Kalv\=ans}

\begin{document}

\title{Temperature spectra of interstellar dust grains heated by cosmic-rays II:\\ dark cloud cores}

\correspondingauthor{Juris Kalv\=ans}
\email{juris.kalvans@venta.lv}

\author[0000-0002-2962-7064]{Juris Kalv\=ans}
\affil{Engineering Research Institute ``Ventspils International Radio Astronomy Center'' of Ventspils University of Applied Sciences,\\
In$\check{z}$enieru~101, Ventspils, LV-3601, Latvia}

\begin{abstract}
Cosmic-ray (CR) induced heating of whole interstellar grains is an important desorption mechanism for grain surface molecules in interstellar molecular clouds. This study aims to provide a detailed temperature spectra for such CR-induced heating. For this, olivine grains with radius of 0.05, 0.1 and 0.2 microns shielded by interstellar gas with isotropic column densities characteristic to dark cores were considered. The accumulation of an ice mantle of increasing thickness was taken into account. The CR energy spectra was obtained for these column densities for 32 cosmic-ray constituents. We calculated the frequencies with which a CR nucleus with a known energy hits a grain, depositing a certain amount of energy. As a result, we obtain the energy and temperature spectra for grains affected by CR hits. This allows to improve the existing approaches on CR-induced whole-grain heating in astrochemical modeling.
\end{abstract}


\section{Introduction} \label{intrd}

Cosmic-ray (CR) particles impacting interstellar dust grains are able to induce grain heating \citep{deJong73}. For molecules adsorbed on grain surfaces, such a heating may allow overcoming energy barriers for evaporation, diffusion, and chemical reactions. In particular, desorption of surface molecules induced by whole-grain heating (WGH) by CRs  is able to influence the composition of interstellar gas and ices \citep{Hasegawa93}. WGH is considered in many contemporary astrochemical models \citep[e.g.,][]{Cazaux16,Esplugues16,Sipila16,Acharyya17,Awad17,Bisbas17,Coutens17,Holdship17,Kalvans17,Kamp17,Vasyunin17,Semenov17,Aikawa18,Ceccarelli18,Furuya18,Majumdar18}.

The above mentioned studies primarily employ CR \edit1{spectra} based on data from 1970s \citep{Leger85}. Since then, new insights on CR spectra and elemental abundances have been gained, largely thanks to data gathered by spacecraft \citep[e.g.,][]{Webber98,Moskalenko02}. Research has continued specifically for WGH-induced thermal evaporation \citep{Willacy98,Shen04,Roberts07,Iqbal18} and other WGH-induced molecular processes \citep{Reboussin14,Kalvans14,Kalvans15,Mainitz16,Shingledecker18}.

The aim of this study is to transform recently obtained data on CR spectra and elemental abundances into WGH frequencies that can easily \edit1{be} implemented in models simulating the chemistry of interstellar clouds. Data for a special case (translucent clouds) were calculated by \citet[hereafter Paper~I]{K16}. In the present study, the WGH data application ranges from diffuse molecular gas to dense and dark prestellar cores, among other improvements and updates.

In the references above, WGH events have a frequency $f_T$, which describes, how often \edit1{a single grain reaches} a certain WGH temperature $T_{\rm CR}$. \edit1{Here, $T_{\rm CR}$ is the temperature of a grain after the energy, deposited by a CR particle in grain material, has smoothly dissipated as heat over the whole volume of the grain and before any cooling processes have taken effect. Following \citet{Hasegawa93}, i}n most \edit1{current papers (see references above)} $T_{\rm CR}$ is assumed to be 70~K (corresponding to frequency $f_{70}$), considering only the impacts of iron CR particles. In practice, grain heating is induced by impacts of various CR nuclei that deposit a range of energies into the grains. For example, olivine grains with radius $a=0.1$~$\mu$m can receive energies $E_{\rm grain}$ up to $\approx2$K~MeV, corresponding to grain temperatures up to $T_{\rm CR}\approx100$K. To fully describe WGH, one needs to obtain a temperature spectra that describes the frequency of any temperature reached by the grain.

The major tasks of this study are: choosing grain parameters for a range of hydrogen column densities $N_H$ (Section~\ref{cncpt}), characteristic for interstellar clouds; calculating CR spectra for these column densities (\ref{spctr}); calculating the CR energy loss functions for these grains (\ref{grcr}); calculating the $T_{\rm CR}$ spectra for CR-heated grains (\ref{tcr}); and summarizing the new WGH data for applications in astrochemical models (Section~\ref{rslt}).

\section{Methodology} \label{mthd}

\subsection{Concept for calculations} \label{cncpt}
%
\begin{table}
\centering
\caption{Parameters for calculating the CR-induced WGH temperature spectra. Data adopted from \citet{Kalvans18}.} \label{tab-core}
\begin{tabular}{lcccc}
\tablewidth{0pt}
\hline
\hline
No. & $A_V$, mag & $N_H$, cm$^{-2}$ & $b/a$\tablenotemark{a} & $T_d$, K \\
\hline
1 & 1.24 & 2.48E+21 & 0\tablenotemark{b} & 13.7 \\
2 & 2.93 & 5.86E+21 & 0.10 & 11.8 \\
3 & 5.15 & 1.03E+22 & 0.20 & 10.4 \\
4 & 11.05 & 2.21E+22 & 0.30 & 8.7 \\
5 & 32.45 & 6.49E+22 & 0.35 & 6.9 \\
6 & 64 & 1.28E+23 & 0.35 & 6.3 \\
7 & 128 & 2.56E+23 & 0.35 & 5.8 \\
8 & 256 & 5.12E+23 & 0.35 & 5.6 \\
\hline
\end{tabular}
\tablenotetext{a}{ Ice thickness $b$ \edit1{relative to} grain radius $a$.}
\end{table}
Calculating a WGH temperature spectrum for a grain in a dark cloud core is a complex task because such cores may have a range of column densities $N_H$ (proportional to interstellar extinction $A_V$) that attenuate the initial CR spectra. Additionally, \edit1{the interstellar radiation field (ISRF)} is attenuated too. \edit1{Grains at higher extinctions experience less irradiation and are able to adsorb more molecules on their surfaces,} allowing the accumulation of icy mantles. An icy mantle affects grain size and heat capacity, which change $f_T$ \edit1{for a given $T_{\rm CR}$ value. Increased grain cross-section allows more encounters with CR particles; however these additional CR hits touch only the less-dense ice layer and thus deposit relatively low amounts of energy. On the other hand, an increased grain heat capacity means that more energetic (and rarer) encounters are required to lift grain temperature to a particular $T_{\rm CR}$.}

The thickness of the icy mantle depends on $A_V$ \edit1{and, thus,} $N_H$ \citep[e.g.,][]{Ysard16}. This means that one must obtain a correlation between $N_H$ (which determines the CR spectra) and a corresponding ice mantle thickness $b$ (which affects grain properties) over a range of $N_H$ values that are relevant to starless or prestellar cores. Here, we define $N_H$ as the column density from the edge of a cloud core to its center. The same regards also $A_V$ (see below).

The correlation between $N_H$ and $b$ was obtained from \citet{Kalvans18}, a 1D model \edit1{``Alchemic-Venta''} that follows the chemistry of a \edit1{collapsing} prestellar core. \edit1{This model reflects the growth of ice, tuned with the help of observational data. Thus, the modeling results are suited for the present study, where major ice properties, such as thickness, are of importance. Appendix~\ref{app-alchm} describes details relevant to the formation of the icy mantle in this model.}

In order to create relevant and convenient geometrical grain models for describing their interaction with CRs (Section~\ref{grn}), the ice thickness was expressed as whole decimal parts of grain radius $a$ whenever possible. Then, the $A_V$ value with corresponding ice thickness was found from the modeling results for the parcel of gas in the very center of the core. For the $N_H/A_V$ ratio, a conventional value of $2.0\times10^{21}$~cm$^{-2}$ was adopted \citep{Valencic15,Zhu17}.


Table~\ref{tab-core} shows the ice mantle thickness and their corresponding $A_V$ and $N_H$ values. When $A_V\ga32$~mag, a near-complete freeze-out happens, and the mantle does not thicken anymore with increasing $A_V$. The $N_H$ specified in Table~\ref{tab-core} was used to calculate the CR spectra, while ice thickness is important in determining the energy deposited by CR particles in grains, and grain heat capacities.

\edit1{A diffuse cloud in its initial contraction stages is permeated by the ISRF and thus contains mainly bare grains. This corresponds to the first point in our calculations (Table~\ref{tab-core}). The exact value of $A_V$ was taken to be 1.24~mag, at a point, where ice thickness reaches 1.0 monolayers for the first time. This ensures that the calculated data are relevant for grains actually undergoing molecule accretion. Strictly speaking, grains with a monolayer of ice are not bare. However, thickness of one monolayer is only 0.0035~$a$ and this practically does not affect grain size and heat capacity -- the properties important in this study. Meanwhile, the chosen $A_V$ is relevant for molecular clouds,} \edit2{where molecule adsorption onto grains occurs.}

\subsection{Cosmic-ray spectra} \label{spctr}
%
\begin{table}
\centering
\caption{Adopted abundances of cosmic-ray constituents, relative to protium.} \label{tab-ab}
\begin{tabular}{lccc}
\hline
\hline
X & No. & [X]/[H] & Ref. \\
\hline
H & 1 & 1.000E+00 &  \\
D & 1 & 2.100E-02 & (1) \\
He & 2 & 8.140E-02 & (2) \\
$^3$He & 2 & 6.626E-03 & (3) \\
Li & 3 & 9.753E-05 & (2) \\
Be & 4 & 4.383E-05 & (2) \\
B & 5 & 2.157E-04 & (2) \\
C & 6 & 1.671E-03 & (2) \\
N & 7 & 2.444E-04 & (2) \\
O & 8 & 1.570E-03 & (2) \\
F & 9 & 5.123E-06 & (2) \\
Ne & 10 & 1.507E-04 & (2) \\
Na & 11 & 1.784E-05 & (2) \\
Mg & 12 & 2.264E-04 & (2) \\
Al & 13 & 3.302E-05 & (2) \\
Si & 14 & 1.898E-04 & (2) \\
P & 15 & 3.036E-06 & (2) \\
S & 16 & 2.087E-05 & (2) \\
Cl & 17 & 1.898E-06 & (2) \\
Ar & 18 & 4.554E-06 & (2) \\
K & 19 & 2.657E-06 & (2) \\
Ca & 20 & 1.195E-05 & (2) \\
Sc & 21 & 1.898E-06 & (2) \\
Ti & 22 & 1.101E-05 & (2) \\
V & 23 & 5.123E-06 & (2) \\
Cr & 24 & 1.347E-05 & (2) \\
Mn & 25 & 5.693E-06 & (2) \\
Fe & 26 & 1.152E-04 & (2) \\
Co & 27 & 6.519E-07 & (4) \\
Ni & 28 & 6.452E-06 & (2) \\
Cu & 29 & 6.091E-08 & (5,6) \\
Zn & 30 & 7.130E-08 & (6) \\
\hline
\end{tabular} \\
References: (1) \citet{Adriani16}; (2) \citet{Cummings16}; (3) \citet{Webber17}; (4) \citet{George09}; (5) \citet{Binns14}; (6) \citet{Murphy16}.
\end{table}
%
\begin{figure}[ht!]
\vspace{-2cm}
\plotone{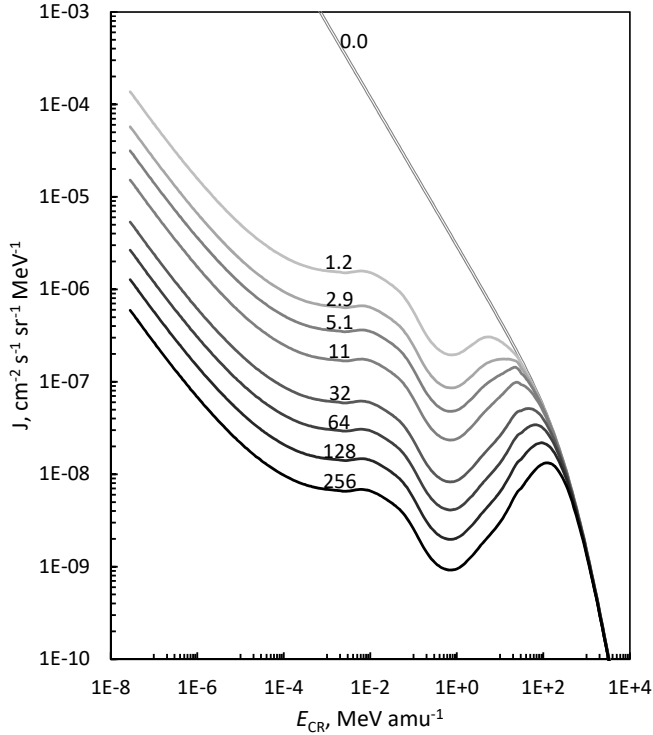}
\vspace{-9.5cm}
\caption{Example CR energy spectra used in this study: the spectra of potassium CR ions calculated at different values of $A_V$. Here $N_H=A_V\times2\times10^{21}$.}
\label{att-spctr}
\end{figure}
In order to comprehensively calculate the temperature spectra of grains subjected to WGH, 32 elemental CR constituents -- atomic nuclei -- were considered. To obtain the initial energy spectra of these particles, we start with the ``High'' proton spectrum of \citet[][their Equation~(1) and Table~1]{Ivlev15p}. This spectrum has abundant low-energy particles, which helps to explain high CR-induced ionization rates observed in the ISM \citep[e.g.,][]{McCall03}.

The initial spectra for nuclei other than protons were obtained by multiplying the proton flux density with [X]/[H], the abundance of CR element X, relative to protium. Table~\ref{tab-ab} shows the adopted abundances of the 32 CR constituents. These were primarily based on \textit{Voyager I} data. The abundance of CR deuterium nuclei was adopted from the PAMELA experiment \citep{Picozza07}, that of cobalt was taken from another space-borne instrument CRIS \citep{Stone98}, while the data for Cu and Zn were drawn from the airborne \textsc{SuperTIGER} experiment \citep{Binns14}. 

The spectra of CRs propagating towards the center of the cloud core is modified by atoms in gas and dust according to:
   \begin{equation}
   \label{cr1}
J_{k,N_H}=J_{k,0}\frac{L_k(E_{k,0})}{L_k(E_{k,N_H})},
   \end{equation}
where $J_{k,N_H}$ is the final spectrum of a CR particle $k$ at hydrogen atom column density $N_H$ (cm$^{-2}$), $L_k$ is the CR energy loss function, dependent on CR particle energy $E_k$ and column density traversed \citep{Padovani09}. $L_k$ was calculated with the help of the {\scshape srim} package \citep{Ziegler10}. Similarly to \citetalias{K16} and that of \citet{Chabot16}, the target medium here was a gas with an assumed elemental composition similar to that of the local ISM and with total column densities listed in Table~\ref{tab-core}. The relatively low column densities ($<10^{24}$~cm$^{-2}$) mean that energy losses via pion production and spallation are minor. Only for the densest case 8 of Table~\ref{tab-core}, spallation may reduce the intensity of the heaviest (iron group) elements by a factor of 2 \citep{Padovani18}. Figure~\ref{att-spctr} shows an example of the obtained energy spectra for potassium (K) CR particles.

\subsection{Dust grain model} \label{grn}

This study considers solid grains made of olivine nuclei, possibly coated with an icy mantle. Grains with three sizes of the nuclei were included -- 0.05, 0.1, and 0.2 $\mu$m. \edit1{As discussed in Section~\ref{cncpt}, ice thickness at various $A_V$ values was derived from \citet{Kalvans18}. This paper considers grains with radius $a=0.1$~$\mu$m.} The relative ice thickness $b/a$ was attributed \edit1{also} to 0.05 and 0.2~$\mu$m grains. Following \citetalias{K16}, the grain nuclei were assumed a chemical composition MgFeSiO$_4$ and density 3.32~g\,cm$^{-3}$. The ice mantle was assumed to consist of H$_2$O, CO, CO$_2$, and CH$_3$OH molecules with proportions 100:31:38:4 and density 1.0~g\,cm$^{-3}$.

The ambient dust temperature $T_d$ was adopted from \citet{Hocuk17}, who calculated $T_d$ as a function of $A_V$ for 0.1~$\mu$m grains. The equilibrium temperature for grains of similar sizes does not differ much \citep[e.g.,][]{Cuppen06}, so we adopted the same $T_d$ also for the 0.05 and 0.02~$\mu$m grains. With  respect to this study, $T_d$ can be approximated to 10~K in all cases. This would produce deviations only within a few per cent in the final WGH temperature spectra. The volumnic heat capacity of the olivine nuclei was described with Equation (18) of \citet{Cuppen06}, while that of (water) ice was adopted from Equation (1) of \citet{Leger85}. Table~\ref{tab-core} lists the thickness of the icy mantles and ambient temperatures for all types of grains.

\subsection{Delivery of CR energy to grains} \label{grcr}
%
\begin{figure}[ht!]
\vspace{-2cm}
\plotone{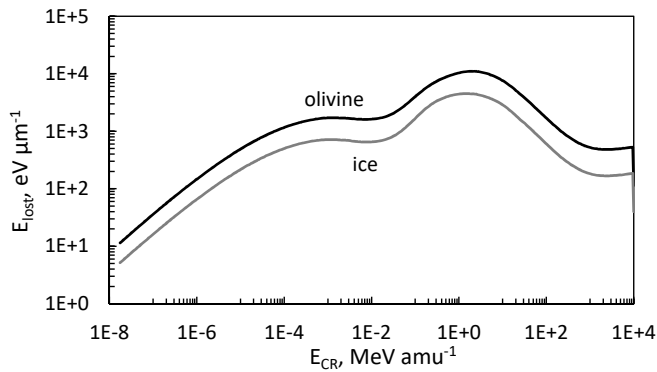}
\vspace{-14cm}
\caption{Example of the energy loss function (per atomic mass unit) for copper CR particles in grain material.}
\label{att-elost}
\end{figure}
%
\begin{figure*}
          \vspace{-3cm}
\gridline{\hspace{-0.5cm}
          \fig{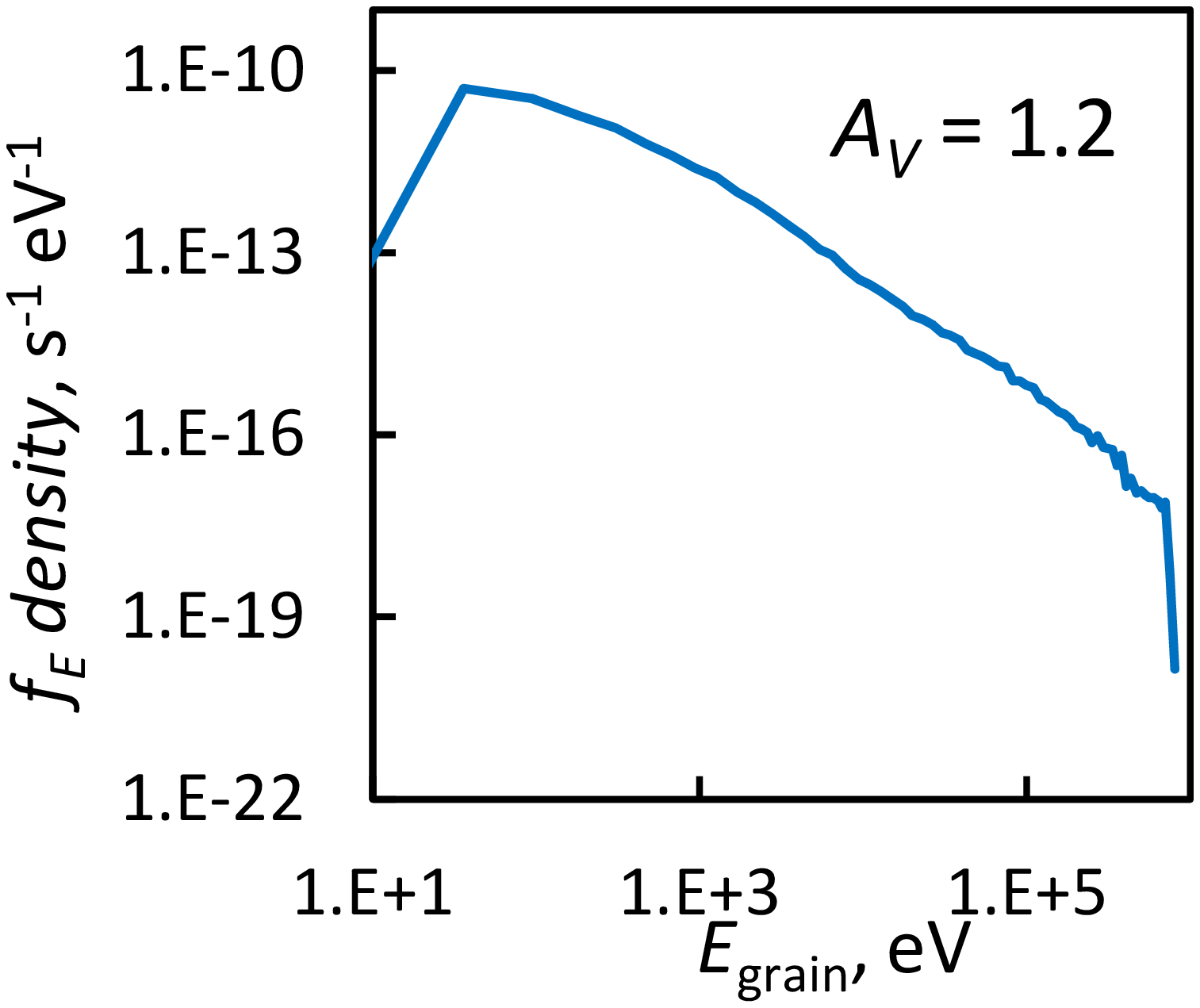}{0.4\textwidth}{}
          \hspace{-1.5cm}
					\fig{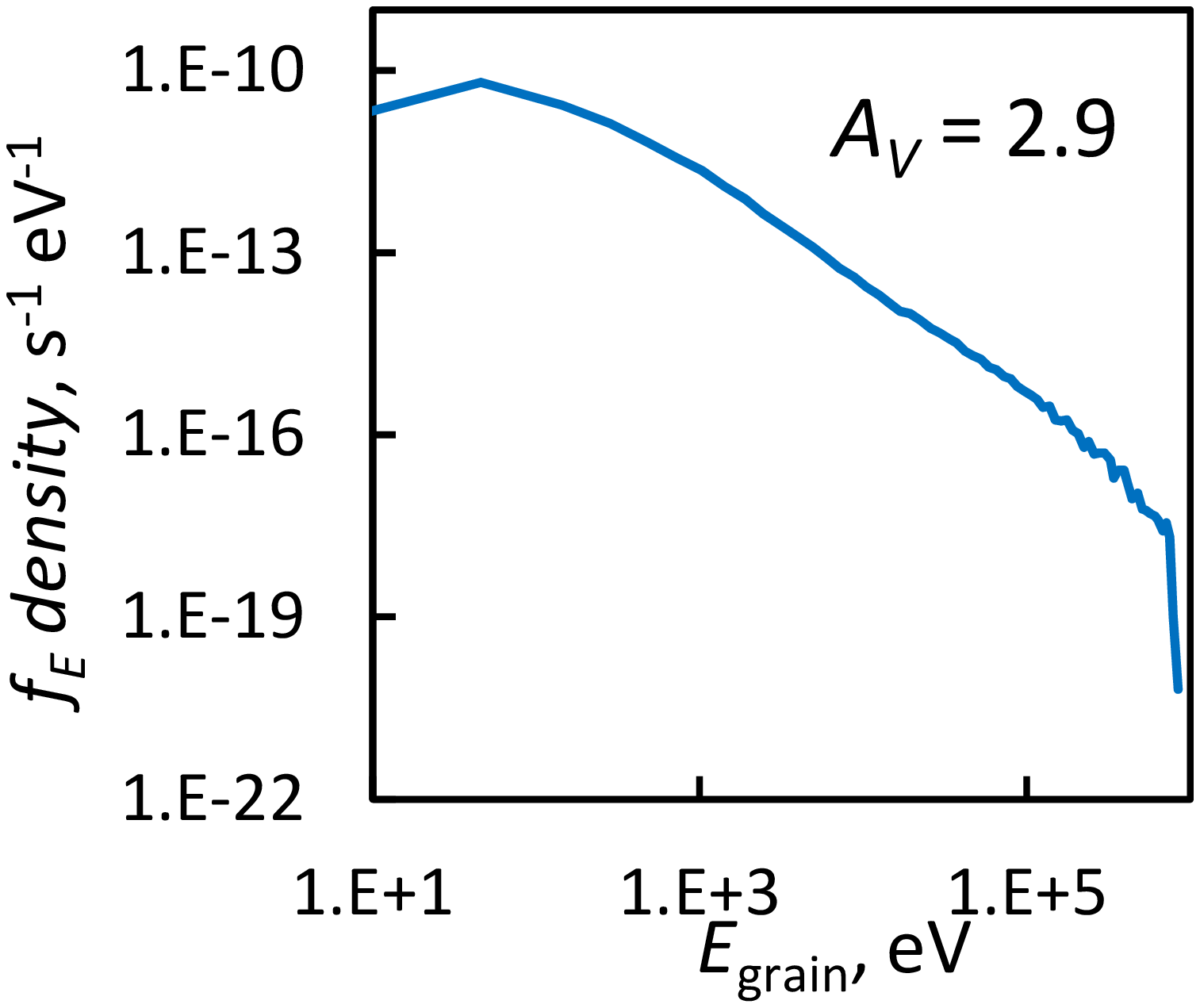}{0.4\textwidth}{}
          \hspace{-1.5cm}
          \fig{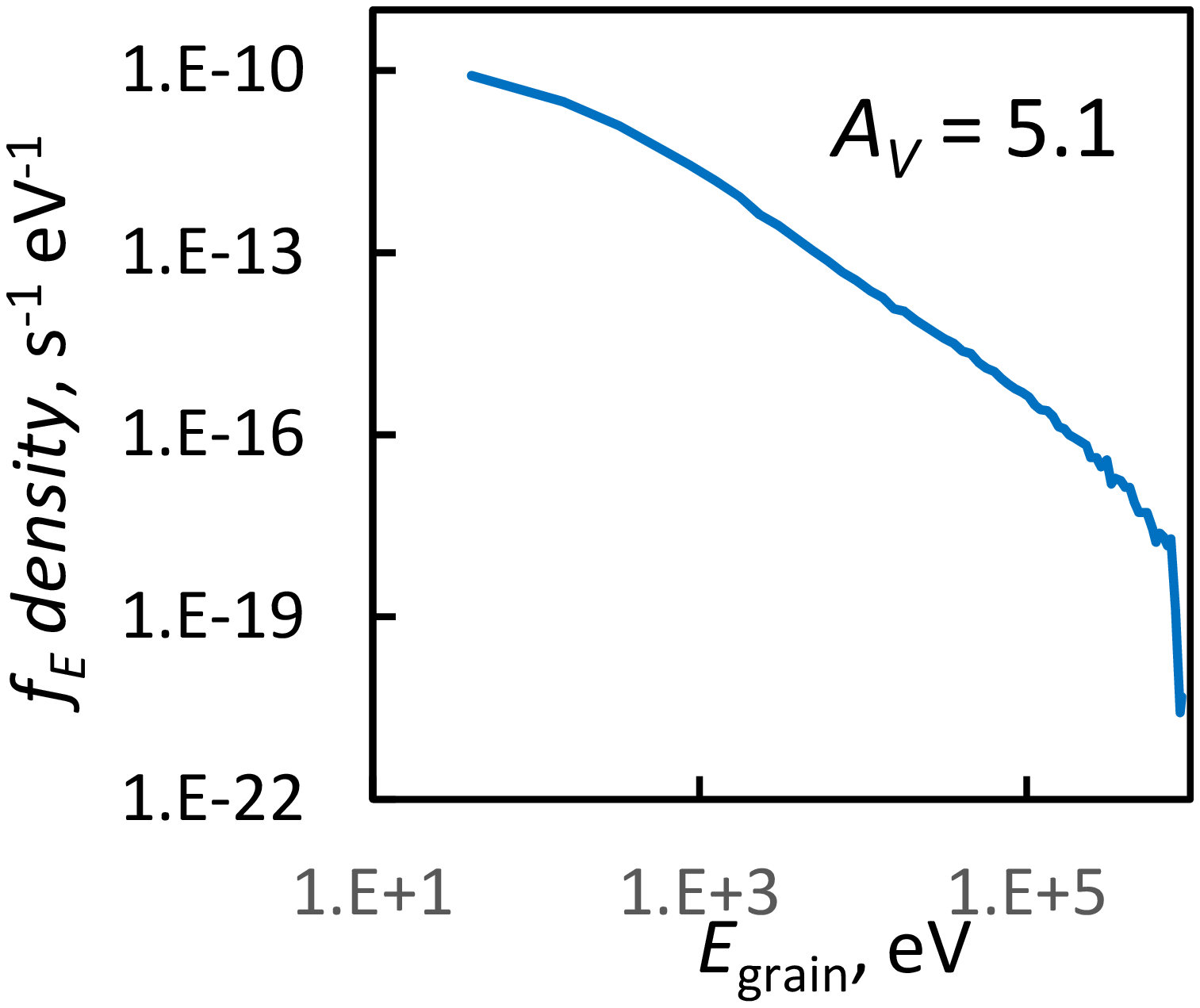}{0.4\textwidth}{}
          }
          \vspace{-6cm}
\gridline{\hspace{-0.5cm}
          \fig{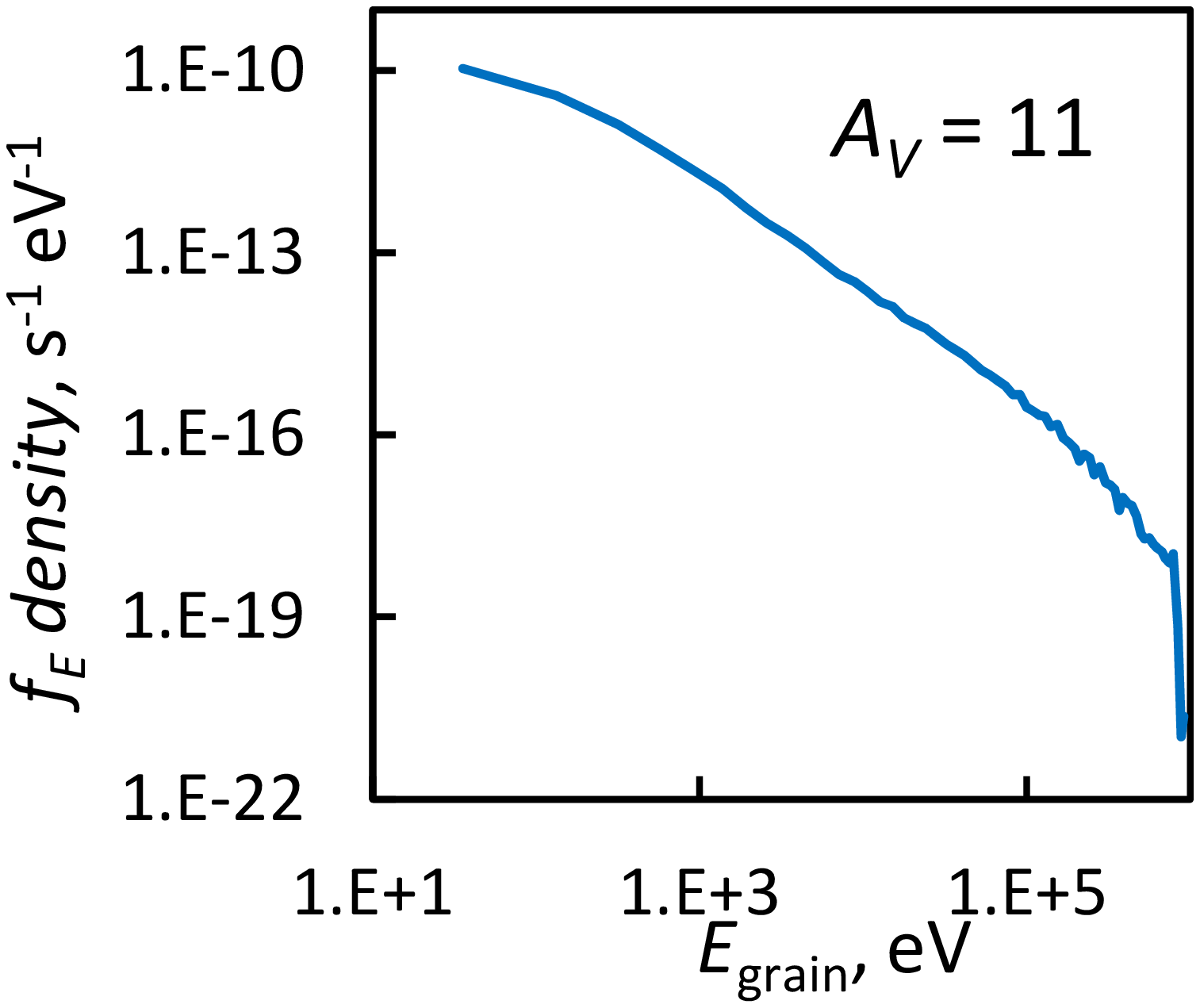}{0.4\textwidth}{}
          \hspace{-1.5cm}
          \fig{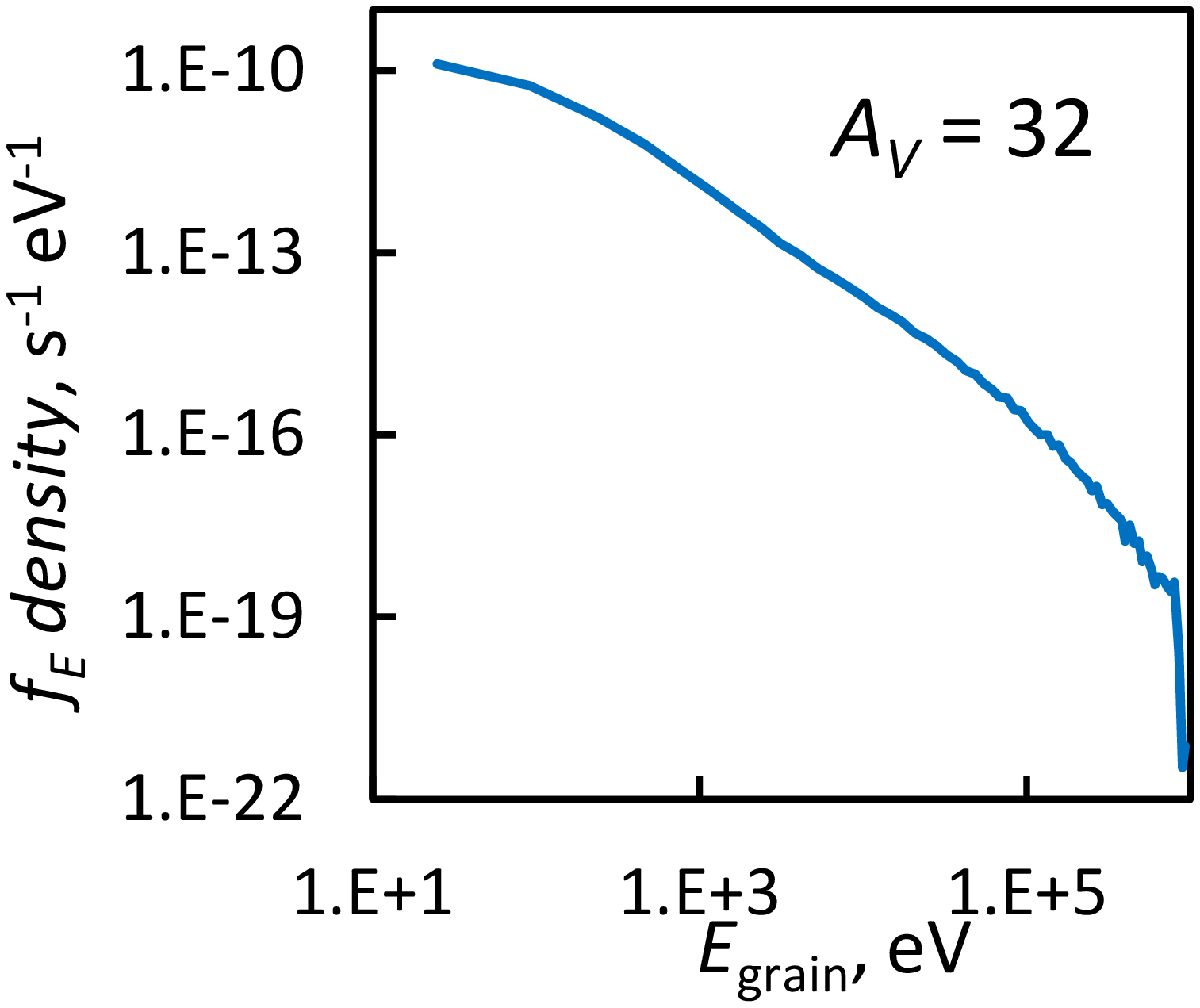}{0.4\textwidth}{}
          \hspace{-1.5cm}
          \fig{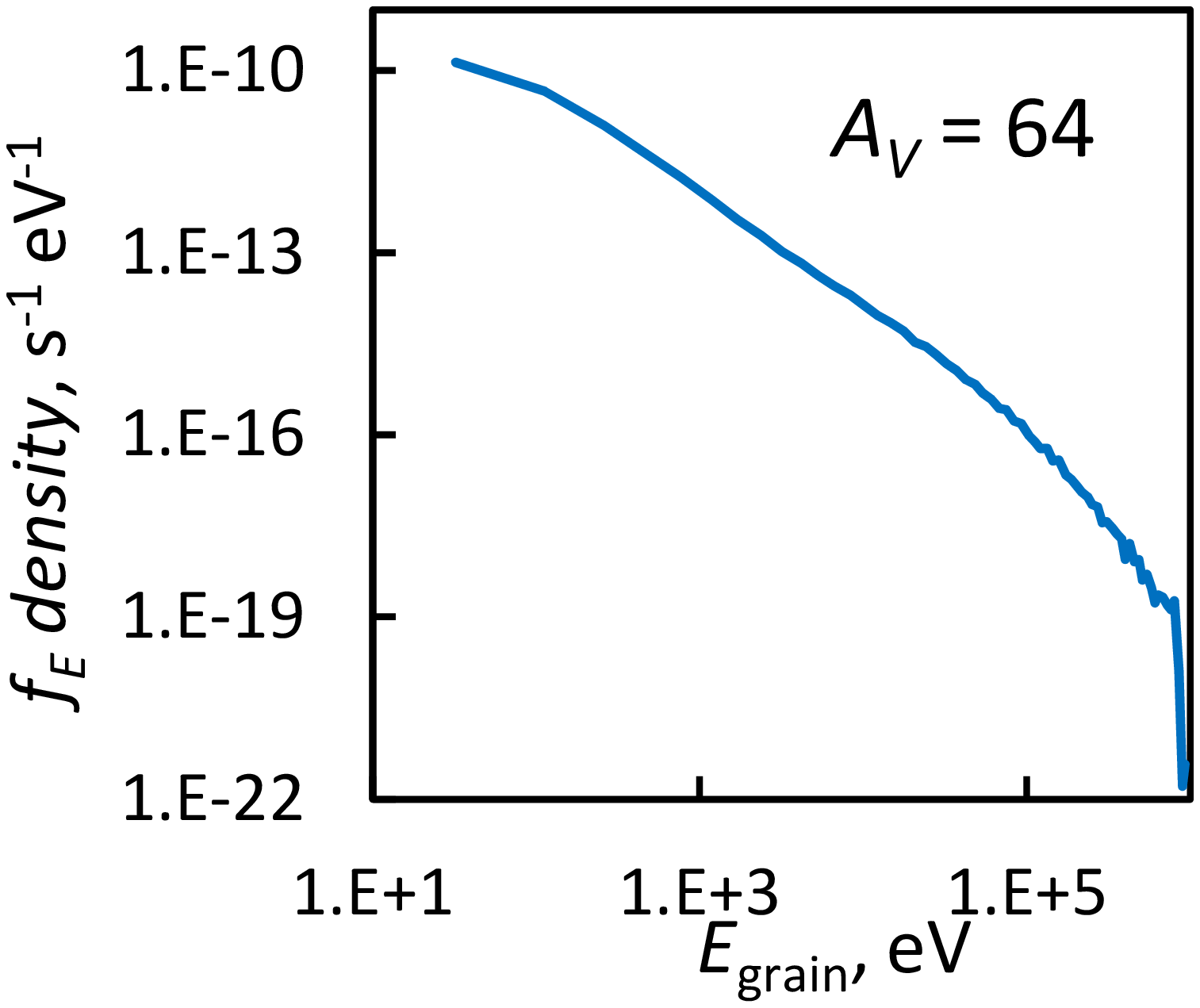}{0.4\textwidth}{}
          }
          \vspace{-6cm}
\gridline{\fig{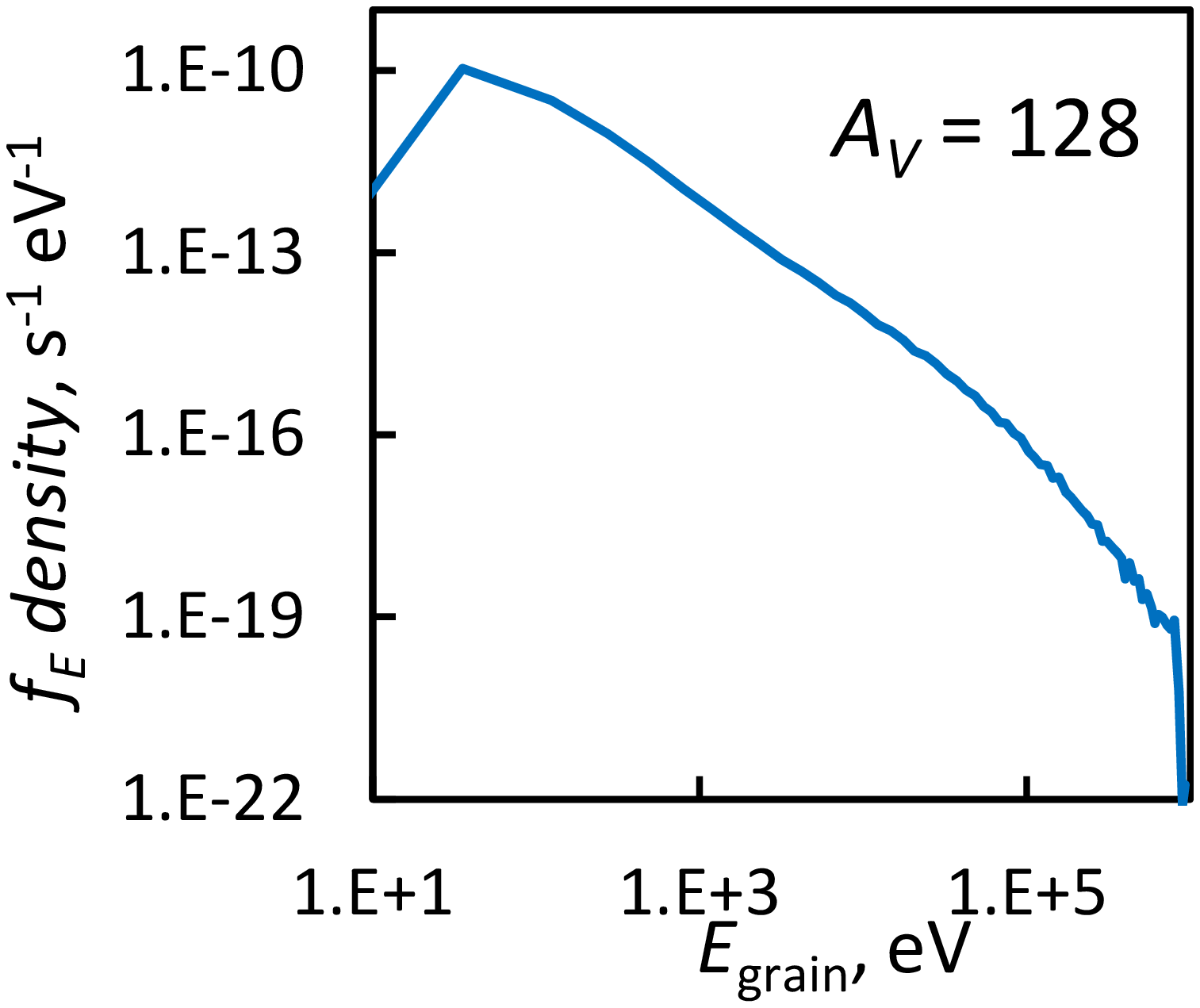}{0.4\textwidth}{}
          \fig{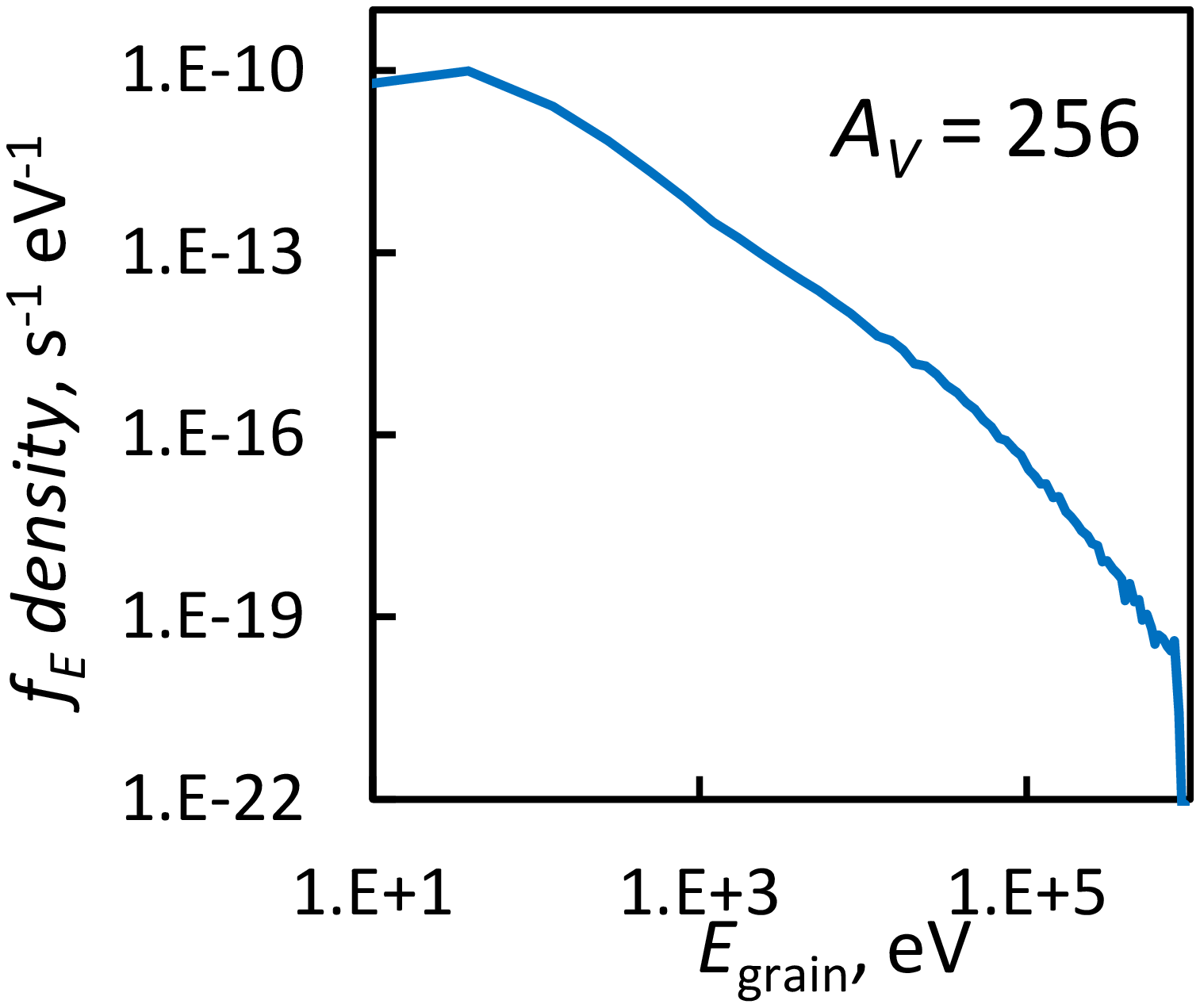}{0.4\textwidth}{}
          }
\vspace{-3cm}
\caption{Calculated spectra of energy received by 0.05~$\mu$m grains from CR impacts, shielded by different column densities of interstellar gas with $N_H=A_V\times2\times10^{21}$~H\,atoms\,cm$^{-2}$.}
\label{att-egr0.05}
\end{figure*}
%
\begin{figure*}
          \vspace{-3cm}
\gridline{\hspace{-0.5cm}
          \fig{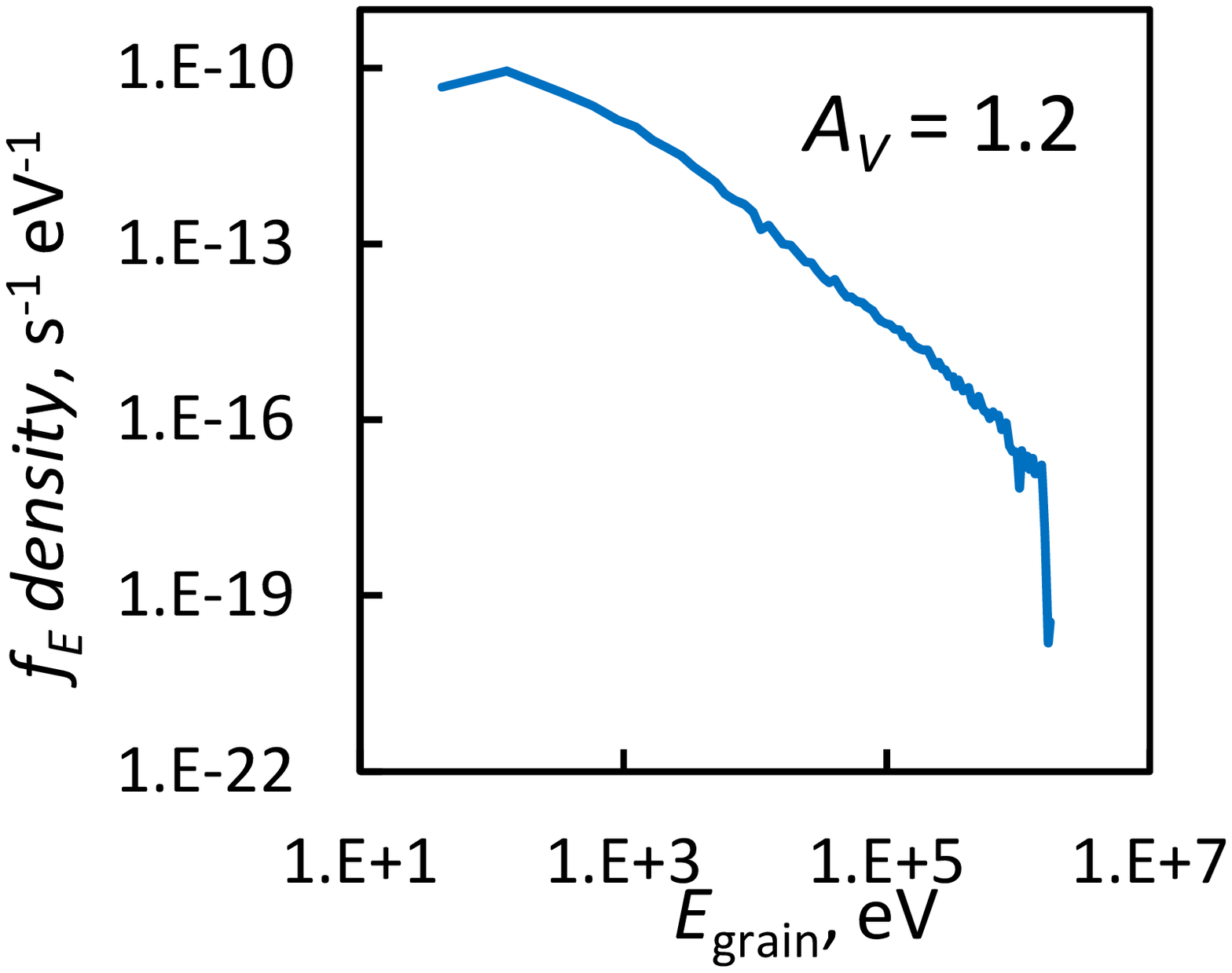}{0.4\textwidth}{}
          \hspace{-1.5cm}
					\fig{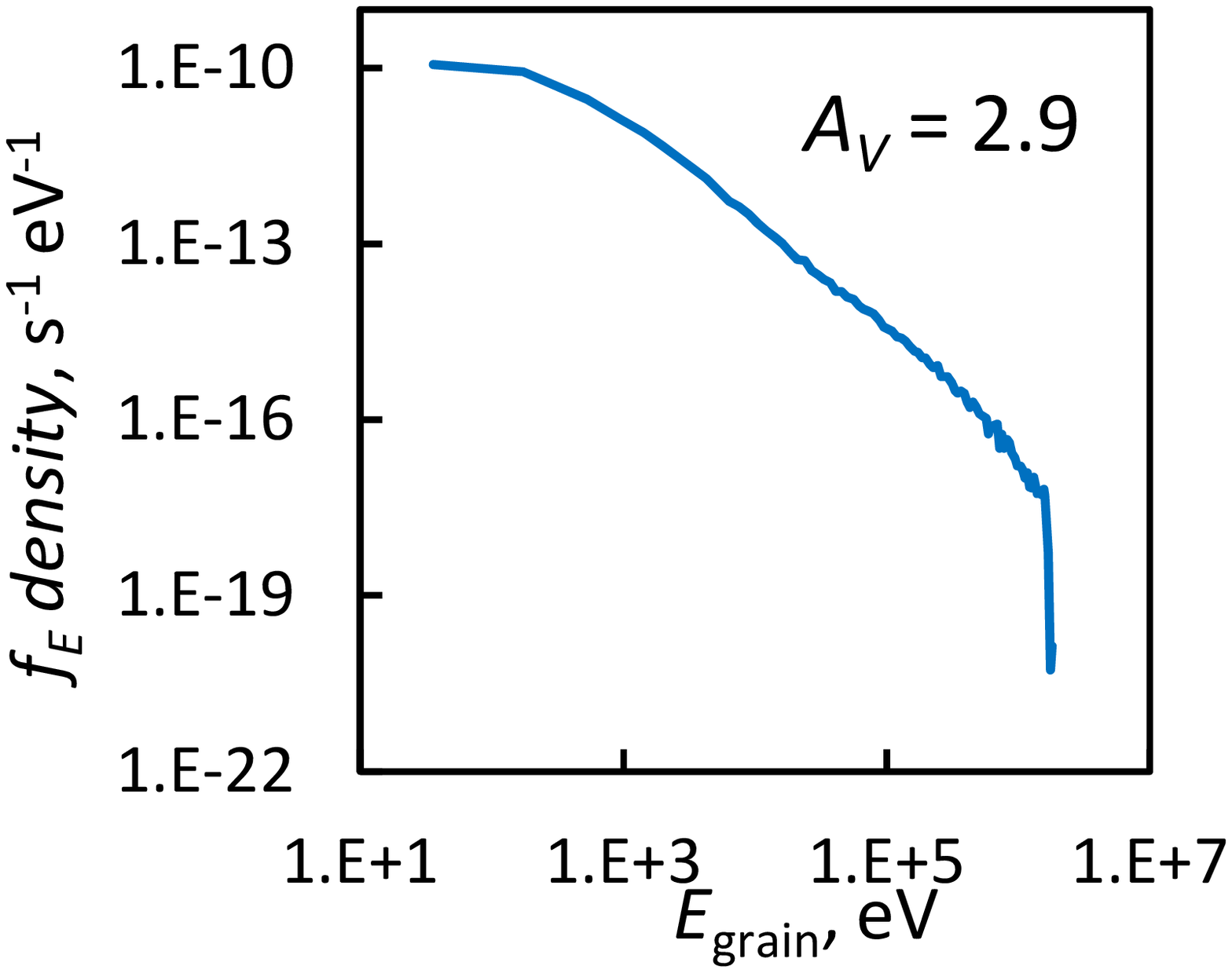}{0.4\textwidth}{}
          \hspace{-1.5cm}
          \fig{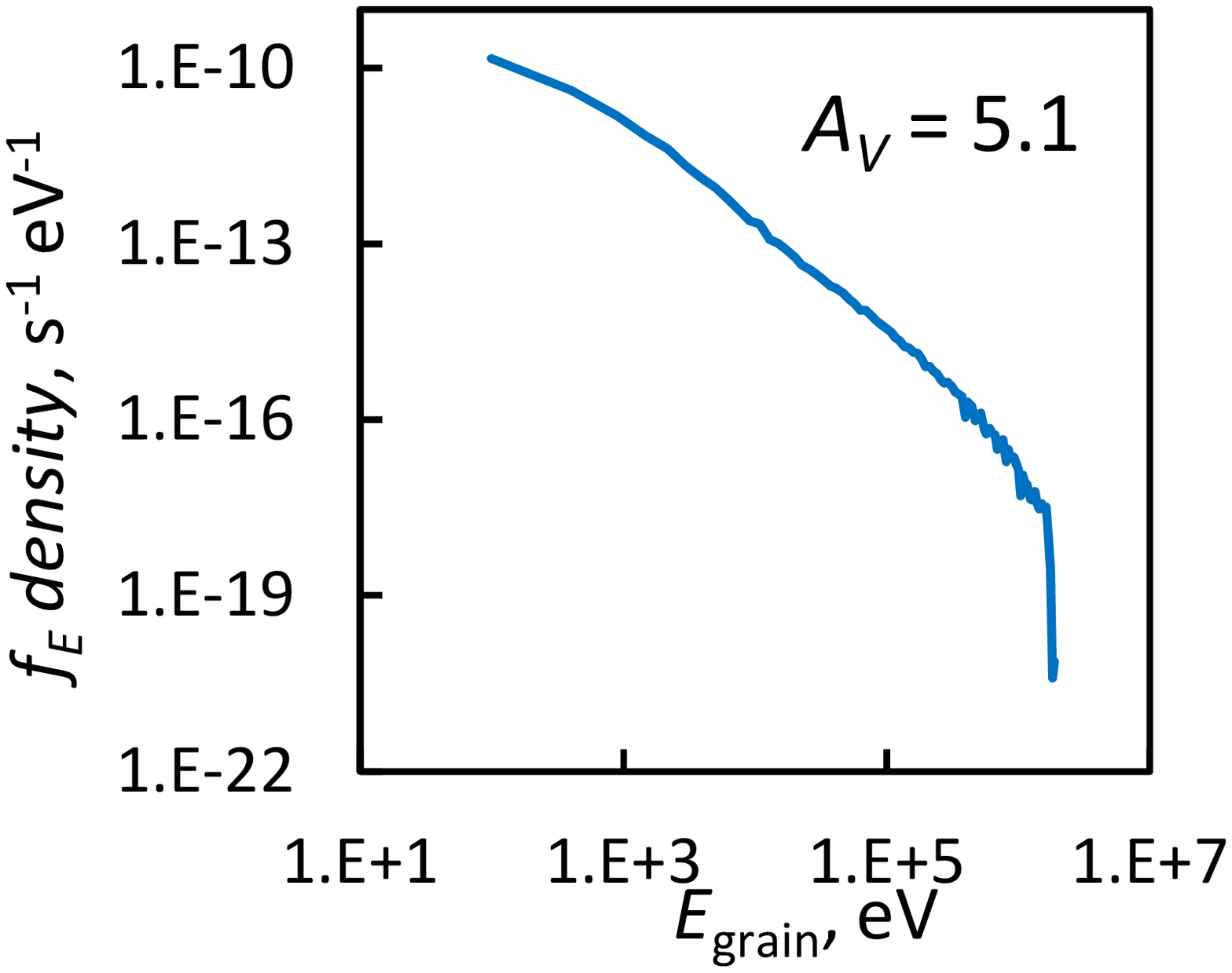}{0.4\textwidth}{}
          }
          \vspace{-6cm}
\gridline{\hspace{-0.5cm}
          \fig{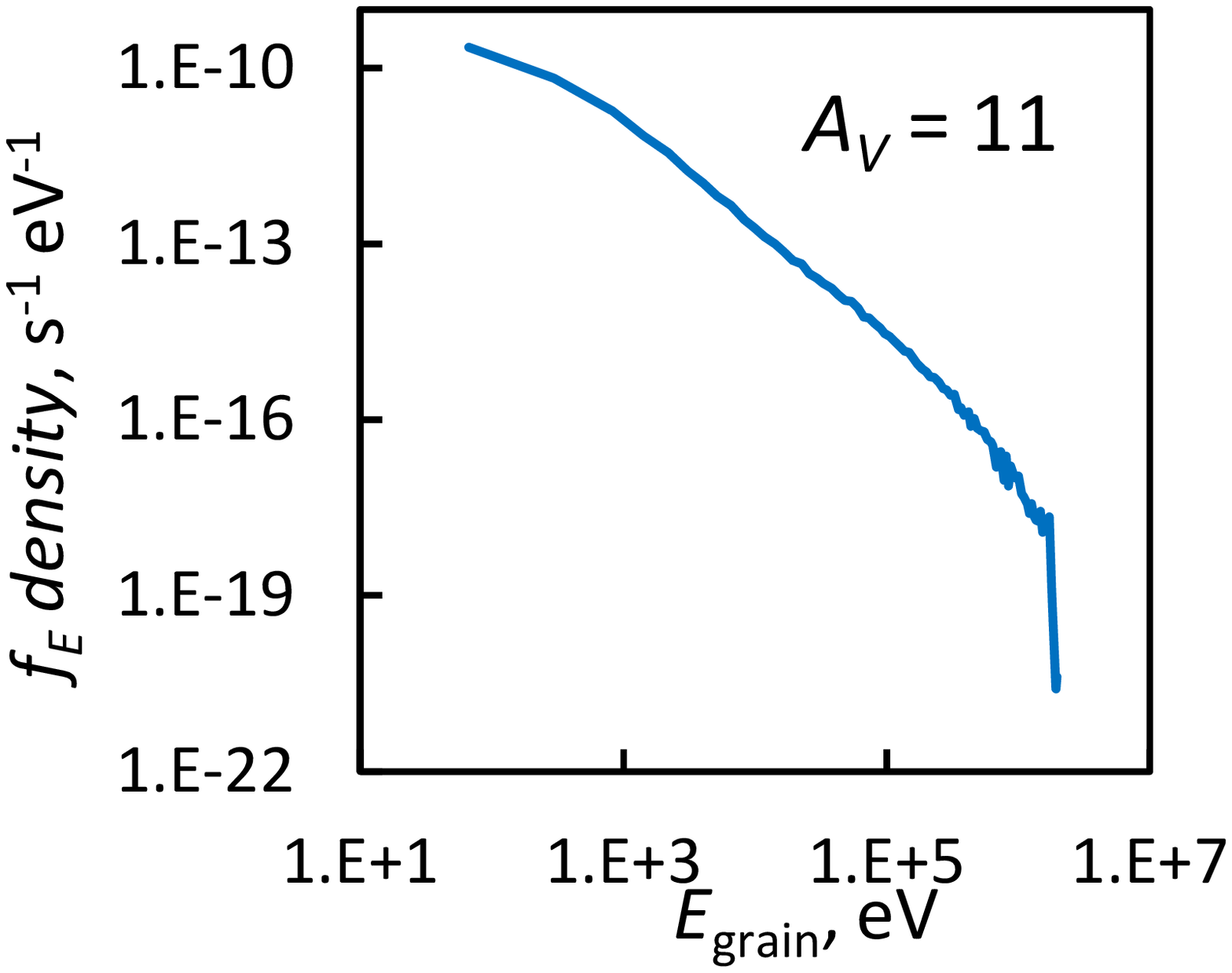}{0.4\textwidth}{}
          \hspace{-1.5cm}
          \fig{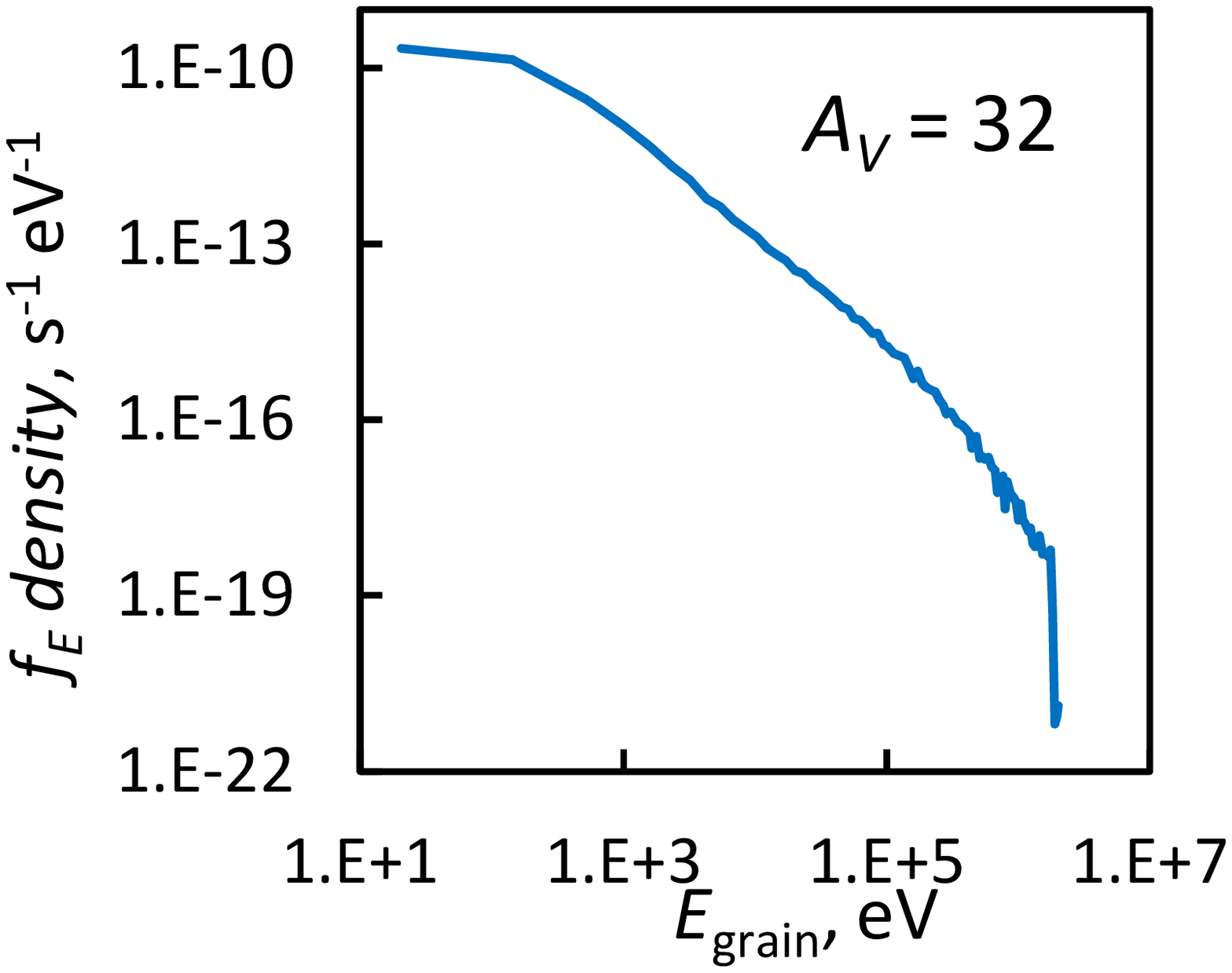}{0.4\textwidth}{}
          \hspace{-1.5cm}
          \fig{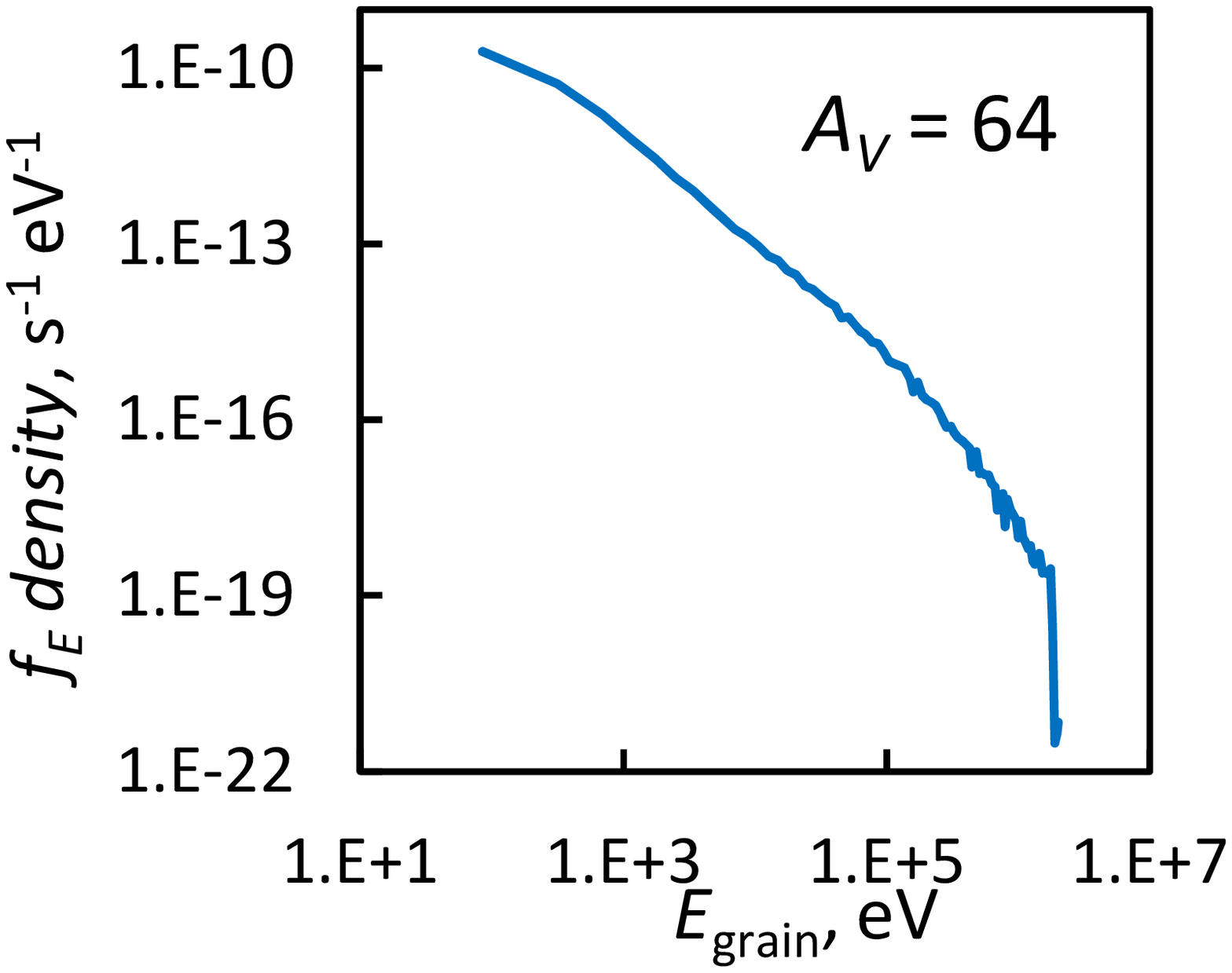}{0.4\textwidth}{}
          }
          \vspace{-6cm}
\gridline{\fig{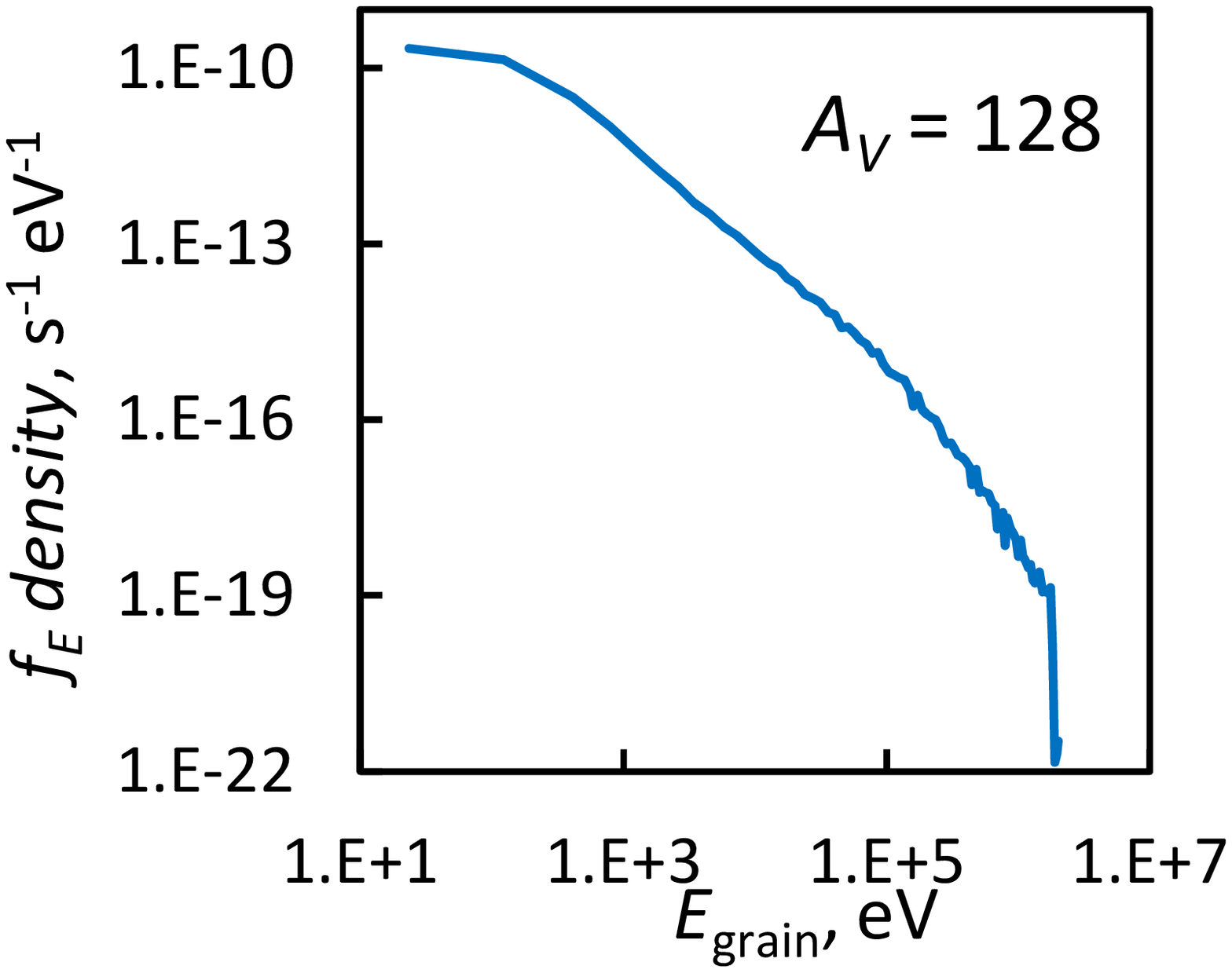}{0.4\textwidth}{}
          \fig{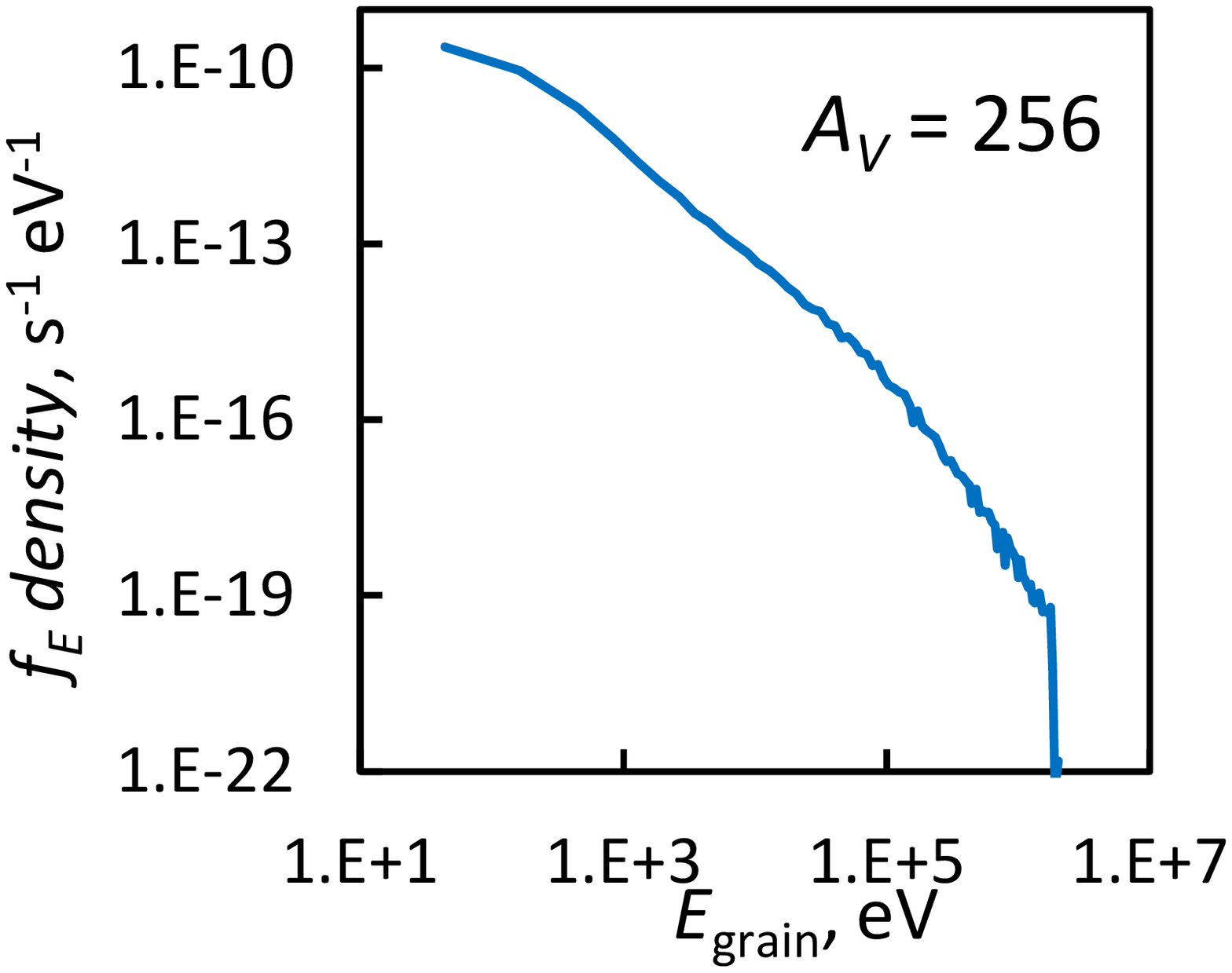}{0.4\textwidth}{}
          }
\vspace{-3cm}
\caption{Calculated spectra of energy received by 0.1~$\mu$m grains from CR impacts, shielded by different column densities of interstellar gas.}
\label{att-egr0.1}
\end{figure*}
%
\begin{figure*}
          \vspace{-3cm}
\gridline{\hspace{-0.5cm}
          \fig{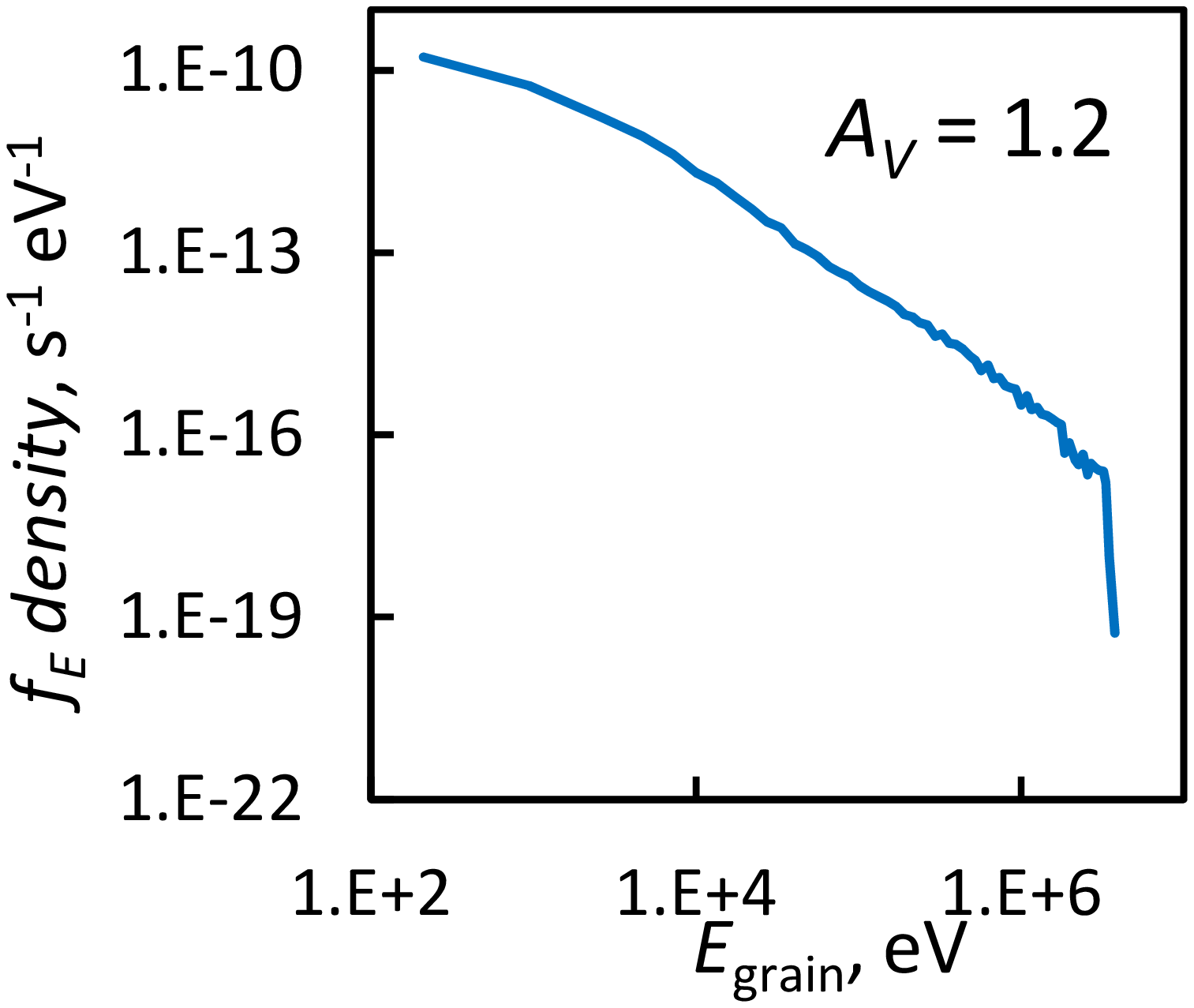}{0.4\textwidth}{}
          \hspace{-1.5cm}
					\fig{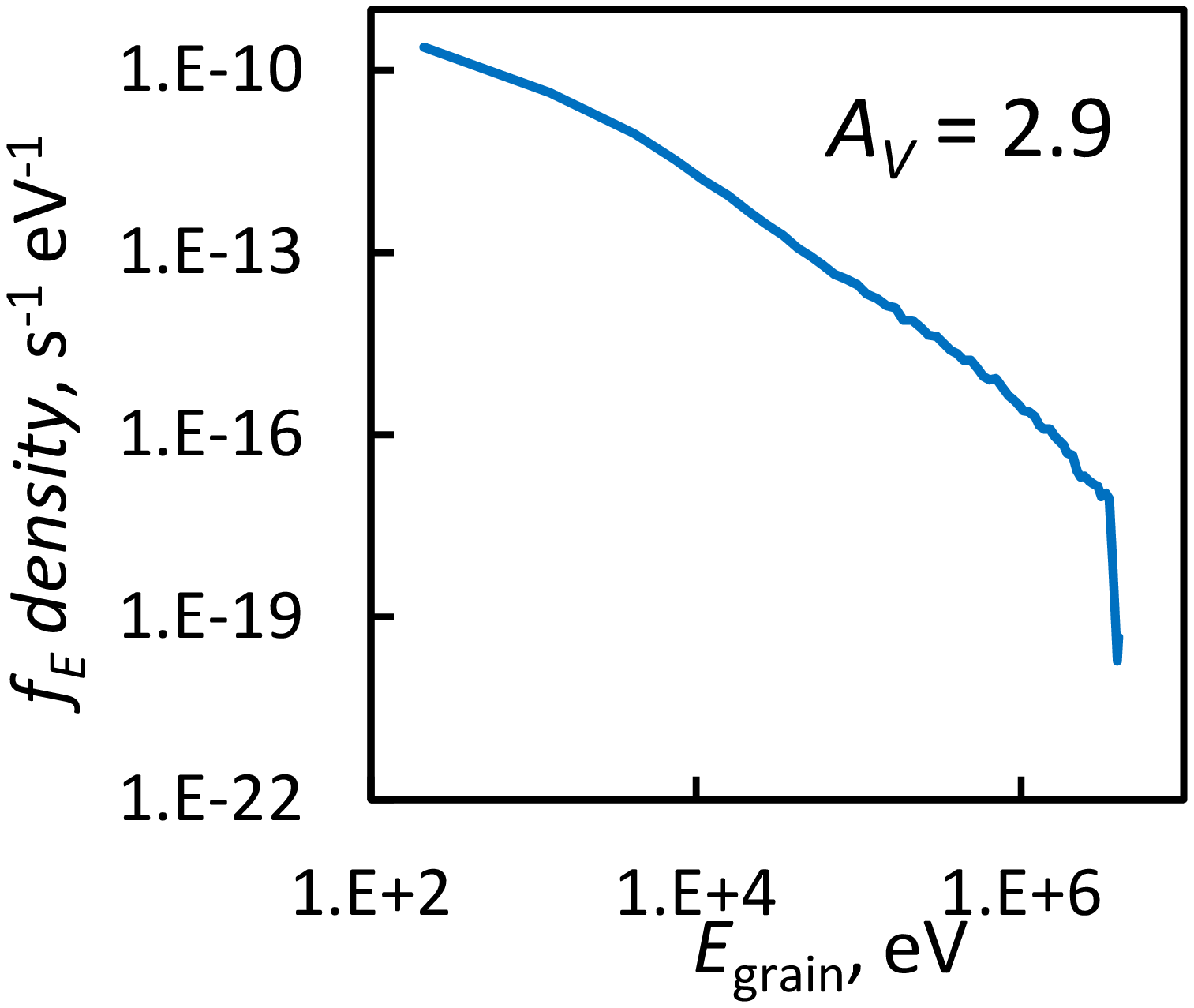}{0.4\textwidth}{}
          \hspace{-1.5cm}
          \fig{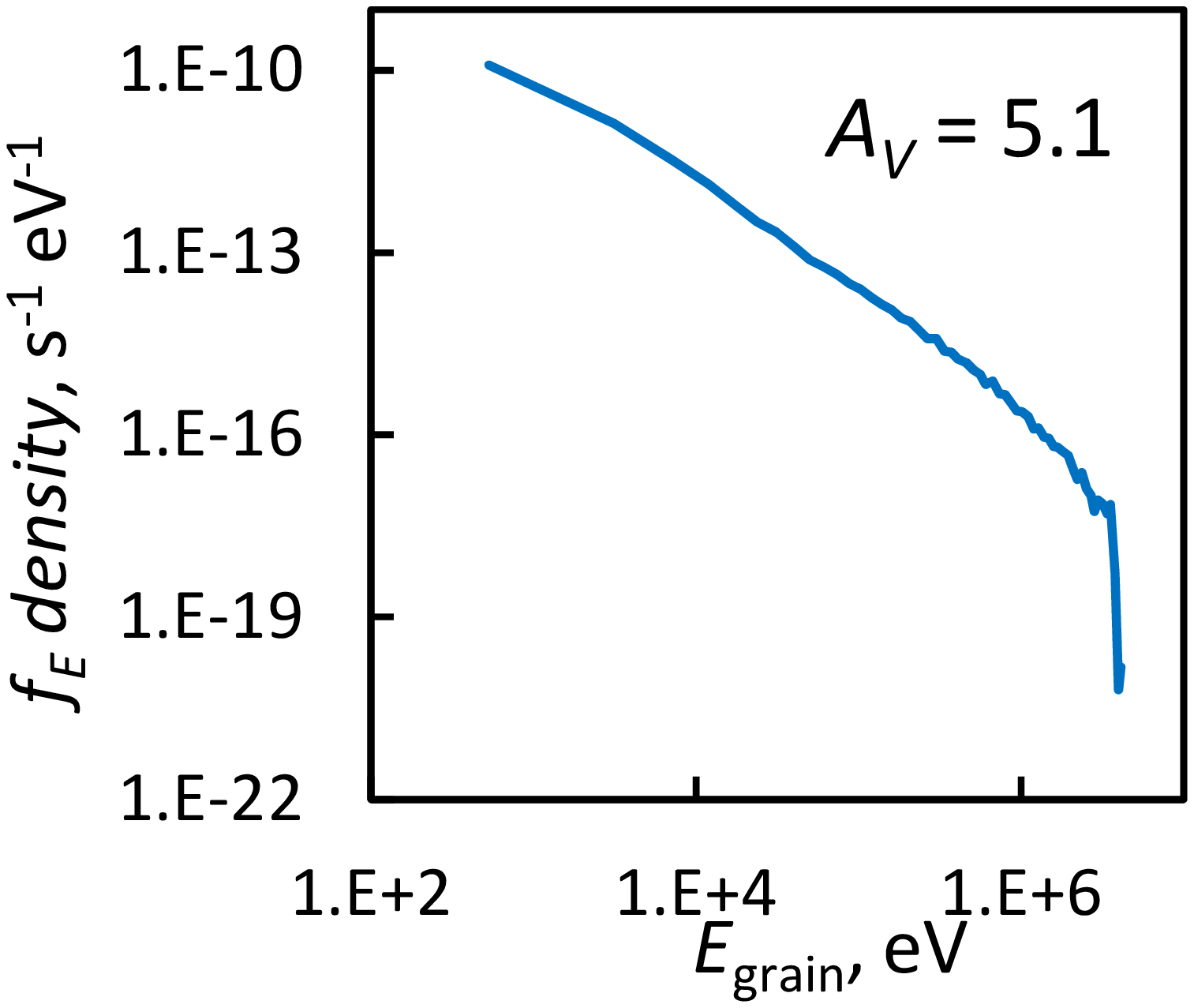}{0.4\textwidth}{}
          }
          \vspace{-6cm}
\gridline{\hspace{-0.5cm}
          \fig{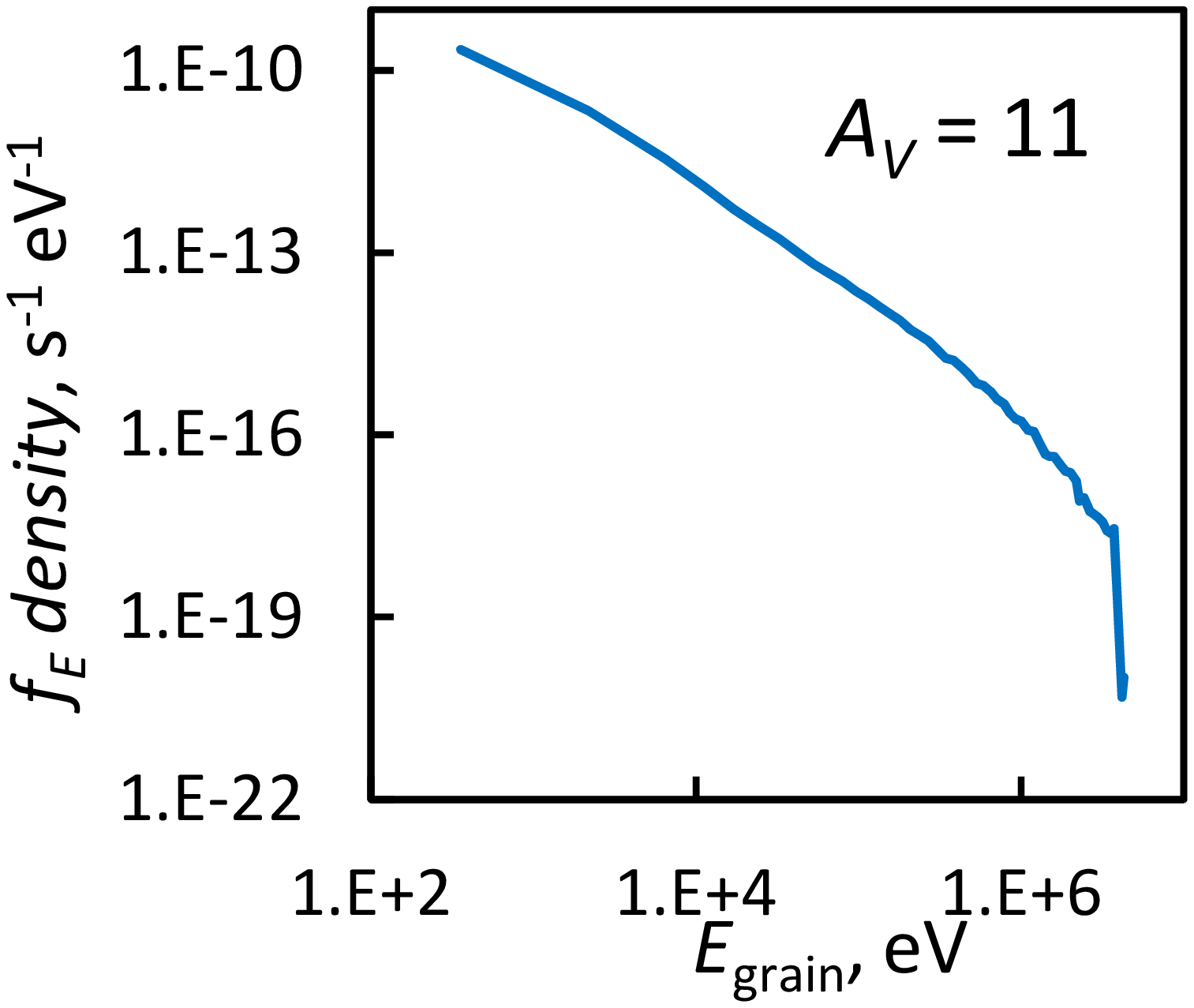}{0.4\textwidth}{}
          \hspace{-1.5cm}
          \fig{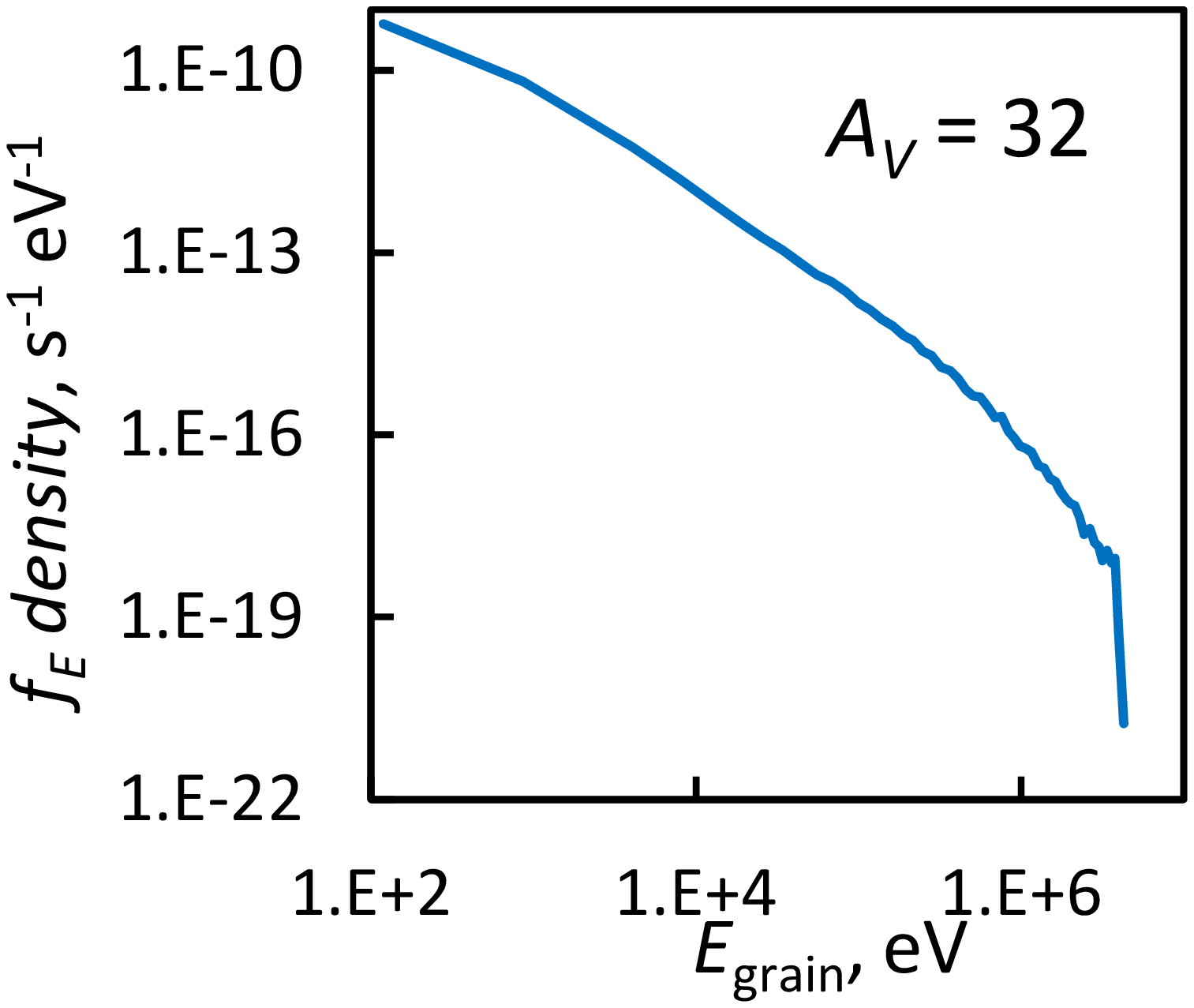}{0.4\textwidth}{}
          \hspace{-1.5cm}
          \fig{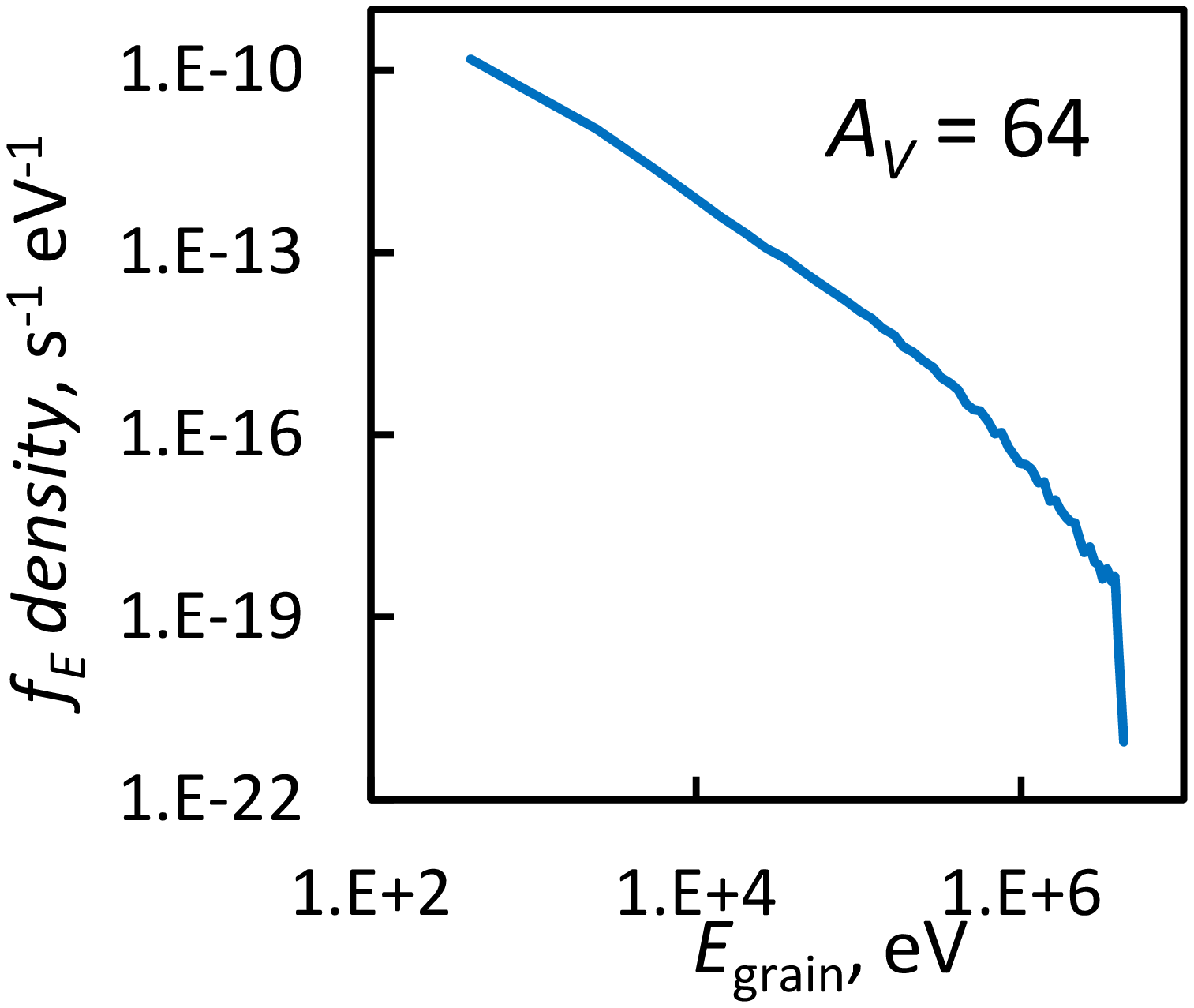}{0.4\textwidth}{}
          }
          \vspace{-6cm}
\gridline{\fig{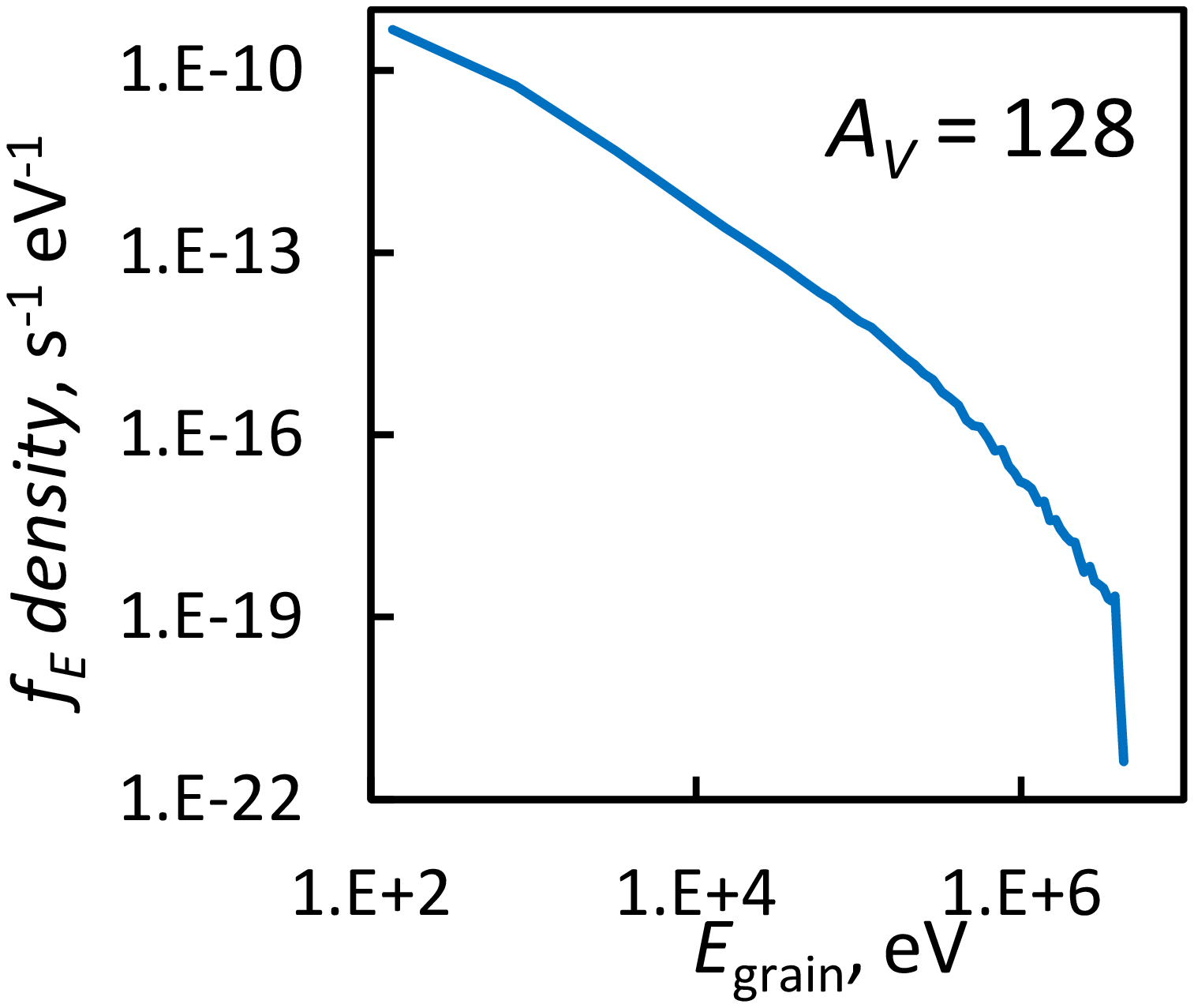}{0.4\textwidth}{}
          \fig{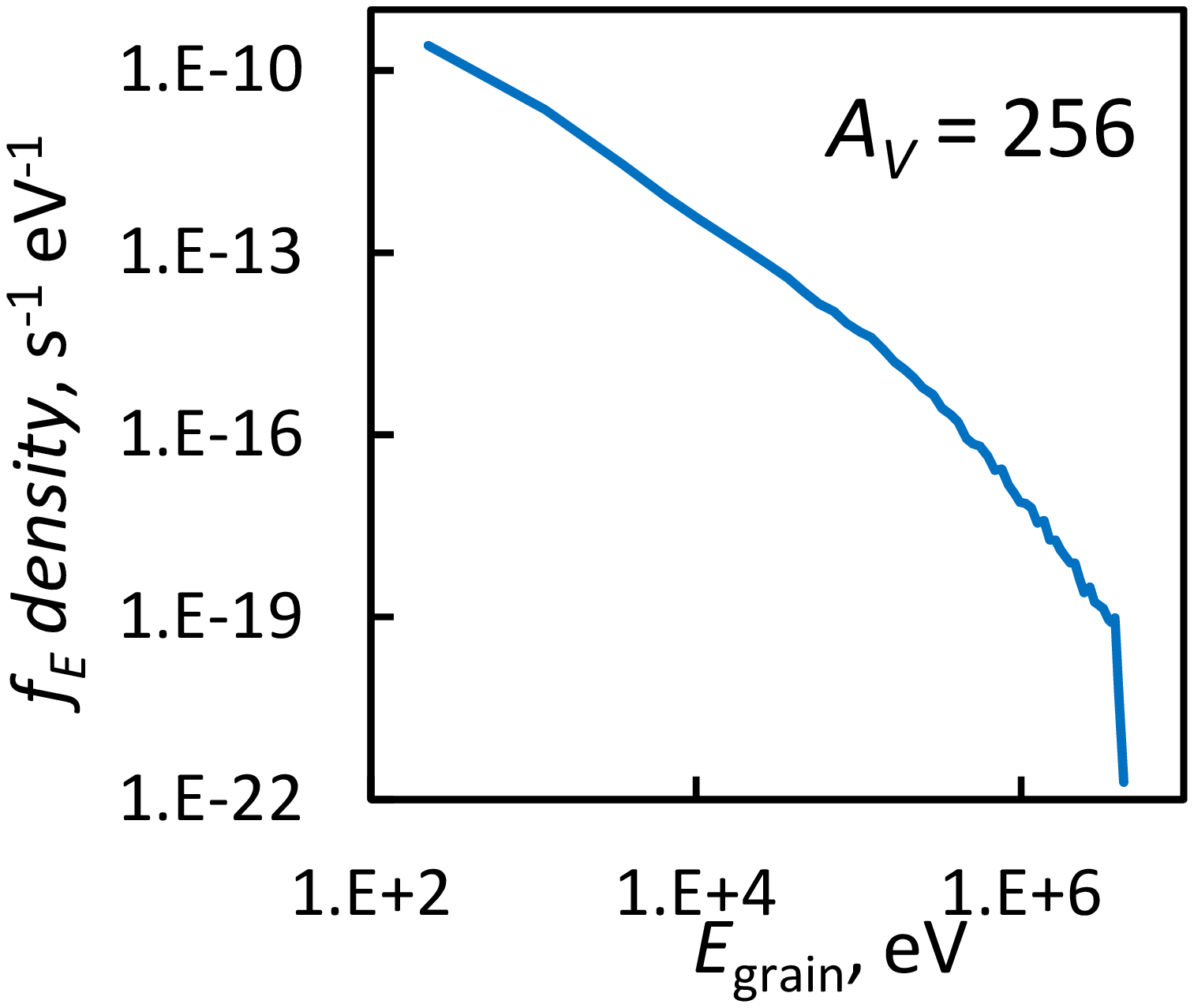}{0.4\textwidth}{}
          }
\vspace{-3cm}
\caption{Calculated spectra of energy received by 0.1~$\mu$m grains from CR impacts, shielded by different column densities of interstellar gas.}
\label{att-egr0.2}
\end{figure*}
\edit1{CRs hit a dust grain, interacting with material and transferring part (or all) of their energy. The amount of energy delivered depends on the nuclei and energy of the CR particles, as well as the CR track length across the grain, and grain density and elemental composition.} Figure~\ref{att-elost} shows the energy loss function for a CR particle (Cu nuclei) in grain \edit1{components}. The CR particles lose energy $E_{\rm lost}$ when passing through only olivine (for bare grains), olivine and ice (major heating of icy grains), or only ice (minor heating of icy grains).

\edit1{The track length and respective cross sections for CRs passing through the grains were calculated with the help of a simple, angular grain model. Previous studies \citep[primarily based on the work of][]{Leger85} often assumed that CRs pass through the center part of a spherical olivine grain, i.e., the longest possible track. The full cross-section of a grain was used, which practically means that the grain was taken to be cylindrical and the total energy received from CR hits is overestimated by $V_{\rm cylinder}/V_{\rm sphere}=1.5$ times. On the other hand, CR energy delivered to the icy part of the grain has been ignored, with a few exceptions \citep[e.g.,][]{Shen04}.}

We employed geometrical grain models consisting of cuboids, similarly to \citetalias{K16}, \edit1{so that the track length and respective cross section was calculated for only a few characteristic CR paths through the grain.} These models were \edit1{tuned} so that \edit1{CR maximum track lengths, volumes, and cross sections of olivine and ice parts} are identical to \edit1{those of} spherical grains. \edit1{This ensures that the maximum energy received by the grain in a single WGH event, the energy averaged over time, and the frequency of CR hits (WGH events) are similar to those of spherical olivine grains, coated with icy mantles of uniform thickness.}

\edit1{In total,} two \edit1{CR tracks were considered} for bare olivine grains and four for all ice-covered grain types. \edit1{The track lengths and their respective cross sections (hence, the corresponding frequency for such events $f_E$) were calculated for each CR nuclei listed in Table~\ref{tab-ab} with CR energies in the range $1...10^{10}$~eV\,amu$^{-1}$. The respective CR energies lost in the grain $E_{\rm lost}$ were calculated using the} \textsc{srim} \edit1{package.} Ions with energies up to $1...100$~keV (depending on grain size) may lose their entire energy and come to a full stop in the grain.

\edit1{$E_{\rm lost}$ is not fully delivered to the grain.} Following \citet{Leger85}, we assume that a portion of up to 0.4 of $E_{\rm lost}$ is carried away from the grain by fast electrons. The remaining energy \edit1{is $E_{\rm grain}$ and} primarily is deposited in a cylinder along the CR track. Appendix~\ref{app-ft} lists the spectra of energy received by the grain from CR impacts: the $E_{\rm grain}$ values with corresponding frequencies $f_E=f_T$. These spectra are depicted graphically in Figures \ref{att-egr0.05}, \ref{att-egr0.1} and \ref{att-egr0.2}. 

Interestingly, low-energy $E_{\rm grain}$ collisions (left-hand side of the plots) have higher frequencies $f_E$ for grains deeper into the cloud core. This is because the ice layer has appeared on the grains, increasing their cross-section. This effect has little practical significance \edit1{because the additional} CR particles that are passing through ice are able to deposit only small amounts of energy in the grains because of the short track and low \edit1{ice} density. \edit1{Consequently,} these collisions are able to raise grain temperature only by a few K.

\subsection{Obtaining the whole-grain heating temperature spectra} \label{tcr}

Following Section~\ref{grcr}, we now have a dataset characterizing \edit1{how often a specific CR particle hits an interstellar grain. It is also known how long is the track of this particle through olivine and ice, and thus, the energy it delivers to the grain.} These data were calculated for each of the 24 grain types and constitute $1.9\times10^4$ entries for each of the bare grains and $3.8\times10^4$ entries for the ice-covered grains.

\edit1{Within $10^{-11}$~s, $E_{\rm grain}$ is converted into heat. The heated cylinder in the grain} expands and transfers heat to the whole grain within $\approx 10^{-9}$~s \citep[e.g.,][]{Leger85,Mainitz16}. For this point, when the grain has an uniform temperature and cooling processes have not yet taken effect, we calculated the WGH temperature spectra.

The amount of energy is now converted to WGH temperature $T_{\rm CR}$ using the heat capacities outlined in Section~\ref{grn}. Each $T_{\rm CR}$ value has its corresponding occurrence frequency $f_T$, obtained by combining the frequencies for CR impacts that deliver similar amounts of energy to the grain. $T_{\rm CR}$ were calculated with an accuracy of 0.01~K. In order to obtain usable $T_{\rm CR}$ spectra, $f_T$ were combined in data clumps spanning 2~K for 0.05~$\mu$m grains and 1~K for 0.1 and 0.2~$\mu$m grains. For example, for 0.1~$\mu$m grains with a 0.01~$\mu$m icy mantle ($A_V=2.93$~mag) the frequency $f_{96}$ corresponding to WGH interval with $T_{\rm CR}$ in the range 96.01--97.00~K arises because of grain impacts by Fe, Co, Ni, Cu and Zn nuclei, all passing through the center part of the grain, with energies 30--300~MeV per CR particle. More than 90~\% of the value $f_{96}=3.17\times10^{-13}$~s$^{-1}$ is provided by the Fe ion impacts. The full calculated $T_{\rm CR}$ spectra for all types of grains can be found in Appendix~\ref{app-ft}.


The primary uncertainties in the calculated WGH heating frequencies are associated with the flux of CRs. This study considers CR energy spectrum with abundant low-energy nuclei. While such spectrum is plausible \citep{Indriolo09}, it may overestimate the intensity of low-energy nuclei by up to an order of magnitude \citep{Padovani09}. On the other hand, the adopted new abundances of heavy CR nuclei (Table~\ref{tab-ab}) are lower at least by a factor of few than those in other studies.

Another uncertainty is associated with the fact that cloud cores often are not spherical and isolated in ISM. In the calculations above, we assumed an uniform flux of CRs coming from all directions and hitting a grain at the center of a core. This is often not the case, especially for \edit1{cores} residing in giant molecular cloud complexes. The proximity of massive interstellar gas reserves likely reduces the amount of (low-energy) CRs reaching the cloud core in consideration. The situation is made more complex by the fact that the CR flux is not isotropic as the particles follow curved magnetic field lines, requiring them to traverse more space (and interstellar gas) than by judging from purely geometrical viewpoint. This further affects the actual CR flux in the cloud \citep{Padovani13h,Padovani18,Silsbee18}. Thus, the application of such more precise data requires also more detailed knowledge about the studied object.

\section{Results} \label{rslt}
%
\begin{table*}
\centering
\caption{Frequency $f_T$, s$^{-1}$, of CR-induced WGH events for 0.05~$\mu$m grains that lift the grain temperature above a minimum $T_{\rm CR}$ threshold. CRs are allowed to reach the grains with equal intensity from all directions (spherical geometry).} \label{tab-sum0.05}
\begin{tabular}{lcccccccc}
\tablewidth{0pt}
\hline
\hline
 & \multicolumn{8}{c}{$A_V=N_H/(2\times10^{21})$} \\
$T_{\rm CR}$, K & 1.24 & 2.93 & 5.15 & 11.05 & 32.45 & 64 & 128 & 256 \\
\hline
$>28$ & 2.94E-09 & 1.39E-09 & 8.48E-10 & 5.13E-10 & 3.15E-10 & 2.30E-10 & 1.61E-10 & 1.08E-10 \\
$>30$ & 2.29E-09 & 1.11E-09 & 6.82E-10 & 4.22E-10 & 2.58E-10 & 1.88E-10 & 1.31E-10 & 8.61E-11 \\
$>40$ & 7.96E-10 & 4.41E-10 & 2.90E-10 & 1.82E-10 & 1.02E-10 & 7.07E-11 & 4.65E-11 & 2.88E-11 \\
$>50$ & 3.77E-10 & 2.27E-10 & 1.46E-10 & 8.40E-11 & 4.19E-11 & 2.70E-11 & 1.63E-11 & 9.20E-12 \\
$>60$ & 2.08E-10 & 1.22E-10 & 7.40E-11 & 3.96E-11 & 1.79E-11 & 1.08E-11 & 5.97E-12 & 3.05E-12 \\
$>70$ & 1.18E-10 & 6.66E-11 & 3.90E-11 & 1.90E-11 & 7.56E-12 & 4.19E-12 & 2.15E-12 & 1.03E-12 \\
$>80$ & 6.53E-11 & 3.56E-11 & 2.01E-11 & 8.55E-12 & 3.06E-12 & 1.60E-12 & 7.90E-13 & 3.69E-13 \\
$>90$ & 3.58E-11 & 1.84E-11 & 9.93E-12 & 4.04E-12 & 1.29E-12 & 6.37E-13 & 3.08E-13 & 1.42E-13 \\
$>100$ & 1.90E-11 & 9.16E-12 & 4.60E-12 & 1.65E-12 & 4.89E-13 & 2.44E-13 & 1.16E-13 & 5.33E-14 \\
$>110$ & 9.50E-12 & 3.97E-12 & 1.91E-12 & 6.02E-13 & 1.64E-13 & 8.14E-14 & 3.88E-14 & 1.77E-14 \\
$>120$ & 3.56E-12 & 1.53E-12 & 6.45E-13 & 2.14E-13 & 5.67E-14 & 2.80E-14 & 1.33E-14 & 6.05E-15 \\
$>130$ & 1.56E-12 & 4.85E-13 & 1.49E-13 & 3.35E-15 & 3.31E-17 & 1.63E-17 & 7.62E-18 & 3.43E-18 \\
$>140$ & 2.32E-14 & 2.19E-16 &  &  &  &  &  &  \\
\hline
\end{tabular}
\end{table*}
%
\begin{table*}
\centering
\caption{Frequency $f_T$, s$^{-1}$, of CR-induced WGH events for 0.1~$\mu$m grains that lift the grain temperature above a minimum $T_{\rm CR}$ threshold. CRs are allowed to reach the grains with equal intensity from all directions.} \label{tab-sum0.1}
\begin{tabular}{lcccccccc}
\tablewidth{0pt}
\hline
\hline
 & \multicolumn{8}{c}{$A_V=N_H/(2\times10^{21})$} \\
$T_{\rm CR}$, K & 1.24 & 2.93 & 5.15 & 11.05 & 32.45 & 64 & 128 & 256 \\
\hline
$>27$ & 4.27E-09 & 1.94E-09 & 1.20E-09 & 7.14E-10 & 3.84E-10 & 2.61E-10 & 1.70E-10 & 1.05E-10 \\
$>30$ & 2.88E-09 & 1.41E-09 & 8.77E-10 & 5.05E-10 & 2.62E-10 & 1.73E-10 & 1.09E-10 & 6.43E-11 \\
$>40$ & 1.07E-09 & 5.67E-10 & 3.23E-10 & 1.62E-10 & 6.80E-11 & 4.03E-11 & 2.20E-11 & 1.11E-11 \\
$>50$ & 4.84E-10 & 2.33E-10 & 1.22E-10 & 5.18E-11 & 1.83E-11 & 9.78E-12 & 4.90E-12 & 2.31E-12 \\
$>60$ & 2.15E-10 & 9.37E-11 & 4.24E-11 & 1.54E-11 & 4.58E-12 & 2.31E-12 & 1.11E-12 & 5.12E-13 \\
$>70$ & 9.02E-11 & 3.49E-11 & 1.45E-11 & 4.21E-12 & 1.07E-12 & 5.32E-13 & 2.53E-13 & 1.16E-13 \\
$>80$ & 3.56E-11 & 1.15E-11 & 3.48E-12 & 8.95E-13 & 2.26E-13 & 1.12E-13 & 5.29E-14 & 2.41E-14 \\
$>90$ & 9.34E-12 & 2.62E-12 & 5.40E-13 & 3.13E-16 &  &  &  &  \\
$>100$ & 5.93E-13 &  &  &  &  &  &  &  \\
\hline
\end{tabular}
\end{table*}
%
\begin{table*}
\centering
\caption{Frequency $f_T$, s$^{-1}$, of CR-induced WGH events for 0.2~$\mu$m grains that lift the grain temperature above a minimum $T_{\rm CR}$ threshold. CRs are allowed to reach the grains with equal intensity from all directions.} \label{tab-sum0.2}
\begin{tabular}{lcccccccc}
\tablewidth{0pt}
\hline
\hline
 & \multicolumn{8}{c}{$A_V=N_H/(2\times10^{21})$} \\
$T_{\rm CR}$, K & 1.24 & 2.93 & 5.15 & 11.05 & 32.45 & 64 & 128 & 256 \\
\hline
$>27$ & 5.45E-09 & 2.55E-09 & 1.35E-09 & 6.21E-10 & 2.59E-10 & 1.51E-10 & 8.22E-11 & 4.15E-11 \\
$>30$ & 3.79E-09 & 1.72E-09 & 8.83E-10 & 3.81E-10 & 1.43E-10 & 8.01E-11 & 4.16E-11 & 2.01E-11 \\
$>40$ & 1.26E-09 & 4.79E-10 & 2.01E-10 & 6.57E-11 & 1.85E-11 & 9.37E-12 & 4.52E-12 & 2.08E-12 \\
$>50$ & 3.80E-10 & 1.19E-10 & 3.92E-11 & 7.62E-12 & 1.89E-12 & 9.35E-13 & 4.44E-13 & 2.03E-13 \\
$>60$ & 9.21E-11 & 1.90E-11 & 3.73E-12 & 1.32E-15 &  &  &  &  \\
$>70$ & 1.23E-11 & 6.77E-16 &  &  &  &  &  &  \\
\hline
\end{tabular}
\end{table*}
%
\begin{figure}[ht!]
\vspace{-1cm}
\plotone{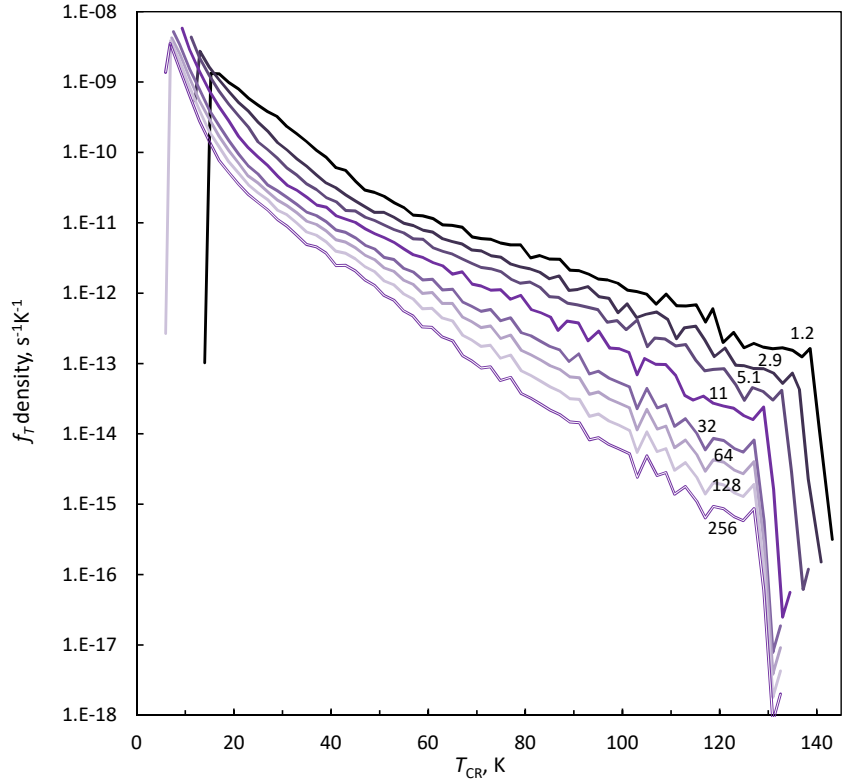}
\vspace{-10cm}
\caption{CR-induced WGH temperature spectra for 0.05~$\mu$m grains. The different column densities $N_H$ of interstellar gas are indicated with interstellar extinction $A_V$ values, where $N_H=A_V\times2\times10^{21}$~H\,atoms\,cm$^{-2}$.}
\label{att-tcr0.05}
\end{figure}
%
\begin{figure}[ht!]
\vspace{-1cm}
\plotone{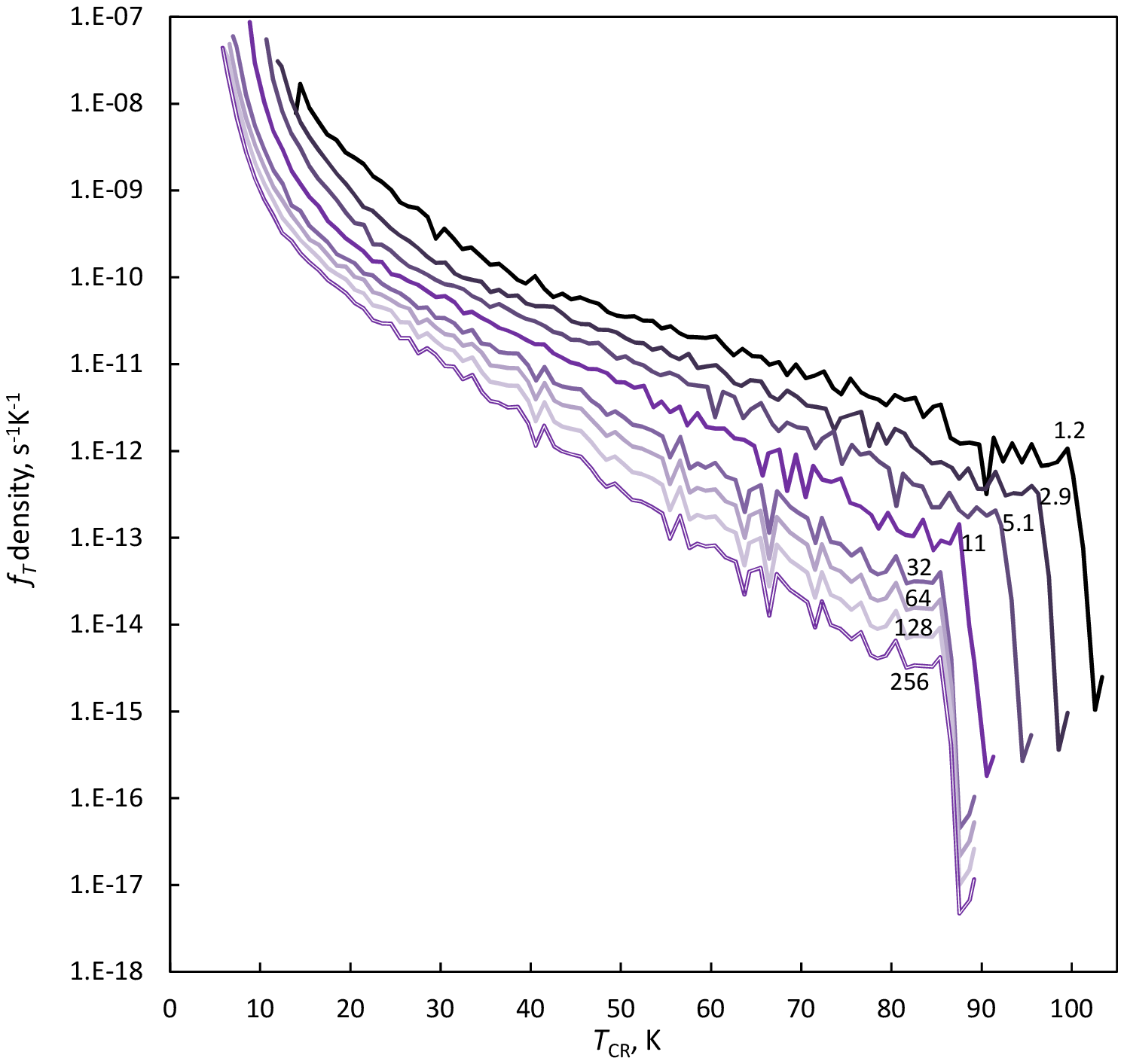}
\vspace{-10cm}
\caption{CR-induced WGH temperature spectra for 0.1~$\mu$m grains. The column densities of interstellar gas are indicated with $A_V$.}
\label{att-tcr0.1}
\end{figure}
%
\begin{figure}[ht!]
\vspace{-1cm}
\plotone{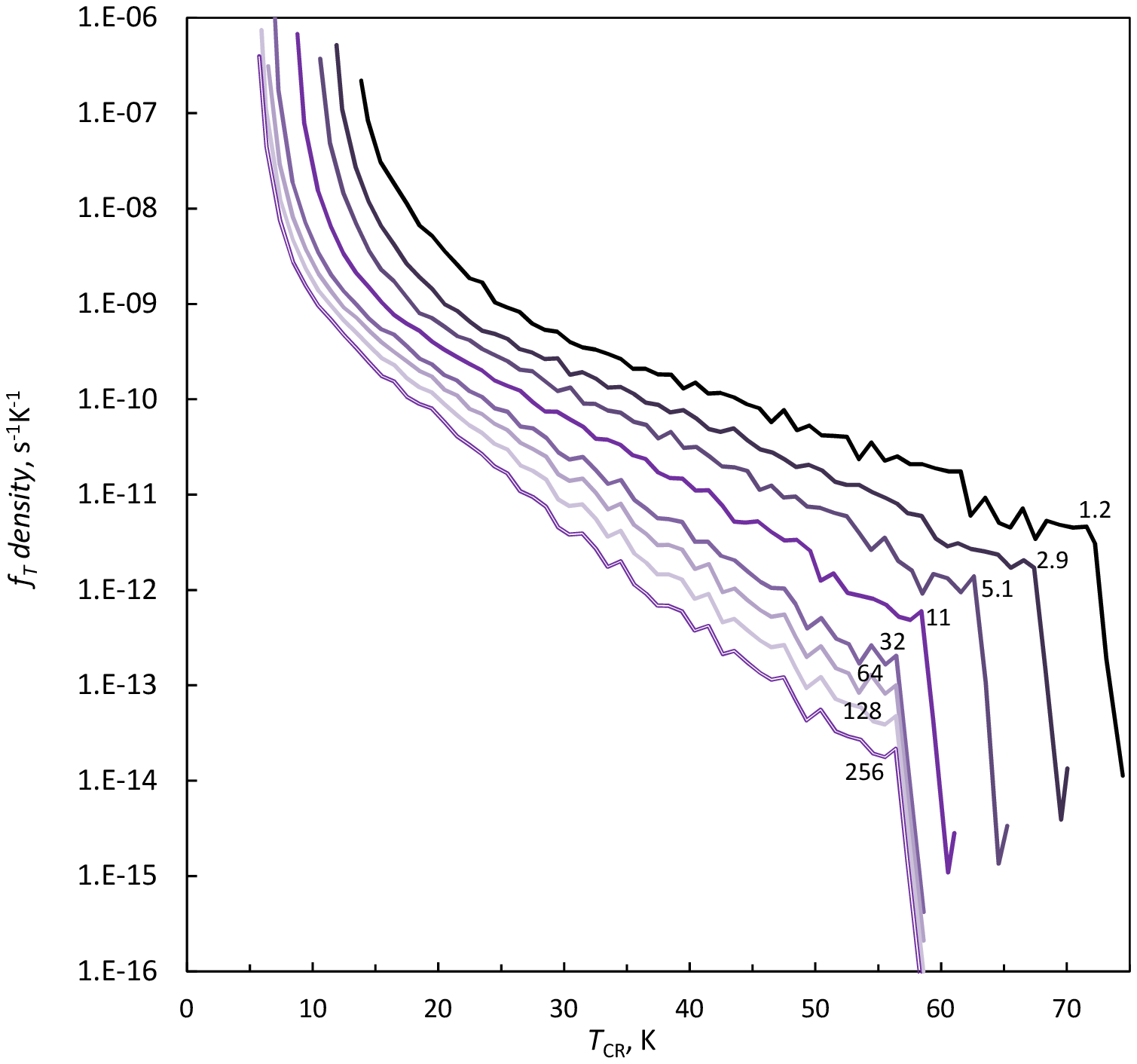}
\vspace{-10cm}
\caption{CR-induced WGH temperature spectra for 0.2~$\mu$m grains. The column densities of interstellar gas are indicated with $A_V$.}
\label{att-tcr0.2}
\end{figure}
%
\begin{figure}[ht!]
\vspace{-1cm}
\plotone{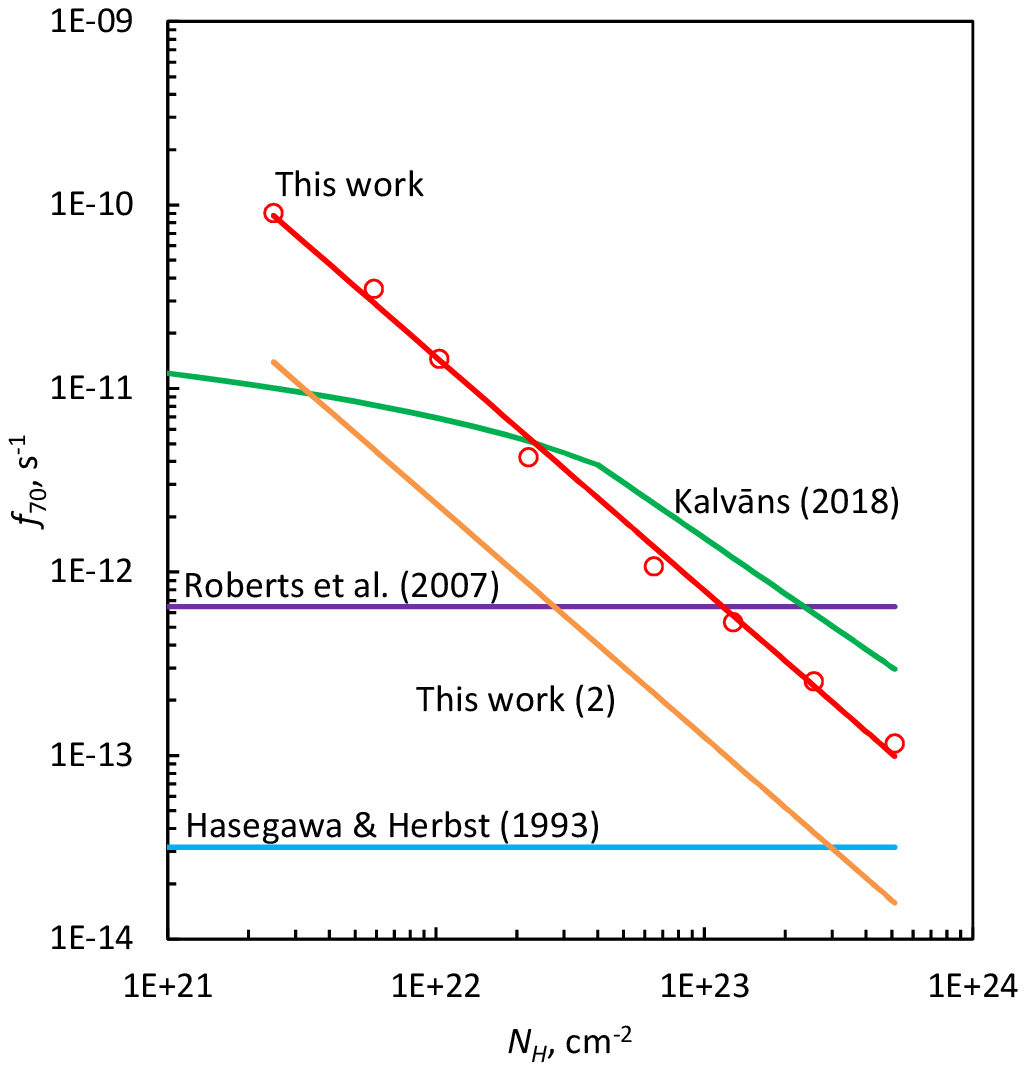}
\vspace{-12cm}
\caption{Comparison of WGH frequencies for 0.1~$\mu$m grains $f_{>70}$ for heating to temperatures $T_{\rm CR}$ of 70~K and above. The second curve from this work is for a reduced CR flux with plane-parallel geometry, irradiated from both sides \edit1{similarly to} \citet{Kalvans18}.}
\label{att-f70}
\end{figure}
Figures \ref{att-tcr0.05}, \ref{att-tcr0.1} and \ref{att-tcr0.2} show graphically the main result of the study -- the temperature spectra for CR-heated grains. The tables of Appendix~\ref{app-ft} list the corresponding \edit2{numerical} data. Abundant CR ions -- H, He, C, O, Mg, Si, Fe -- contribute most to grain heating to progressively higher temperatures. Super-heavy elements with atomic numbers higher than 28 can heat grains to the highest temperatures but are so rare that such heating occurs with intervals of tens of Myr and thus is practically unimportant.

For application of the calculated spectra in astrochemical models, Tables \ref{tab-sum0.05}, \ref{tab-sum0.1} and \ref{tab-sum0.2} display the WGH frequency for $T_{\rm CR}$ values above a threshold value at different column densities. The most common $T_{\rm CR}$ threshold used in astrochemistry is 70~K, \edit1{introduced by} \citet{Hasegawa93}. The dependence of the corresponding WGH frequency $f_{>70}$ on the hydrogen atom column density $N_H$ can be described with the analytical function:
   \begin{equation}
   \label{res1}
f_{70}=g \times 1.197\times10^{16} \times {N_H}^{-1.273},
   \end{equation}
where $g=4\pi$ is a factor characterizing geometrical factors in the vicinity of the cloud core in consideration. If the core is substantially shielded, the value of $g$ can be lower than $4\pi$. \edit1{Equation~(\ref{res1}) can be used to calculate the fraction of time or `duty cycle' $f_{>70}t_{\rm cool}$ an interstellar grain spends in the temperature $T_{\rm CR}$ \citep[Equation~(15) of][]{Hasegawa93}. Here, $t_{\rm cool}$ is the cooling time of a heated grain.}

Figure~\ref{att-f70} shows a comparison of different values of the corresponding WGH frequency $f_{>70}$ for 0.1~$\mu$m grains as a function of hydrogen atom column density $N_H$. Interesting is the difference with the $f_{>70}$ curve of \citet[][Appendix~B]{Kalvans18}, where similar methods were employed. This difference can be explained with a more comprehensive approach on CR spectra and evolution of the icy mantle in this study. This demonstrates the necessity of proper treatment for all relevant effects affecting grain heating.

\begin{table*}
\centering
\caption{\edit1{List of parameters used in the study.}} \label{tab-param}
\begin{tabular}{llllll}
\tablewidth{0pt}
\hline
\hline
Parameter & Description & Refers to & Value & Notes  & unit \\
\hline
$A_V$ & interstellar extinction & cloud core & see Table \ref{tab-core} &  & mag \\
$N_H$ & column density & cloud core & see Table \ref{tab-core} &  & cm$^{-2}$ \\
$N_H/A_V$ & NH per magnitude of extinction & cloud core & 2$\times 10^{21}$ & $\pm$20 \% & cm$^{-2}$\,mag$^{-1}$ \\
\hline
$a$ & radius & grains & 0.05; 0.1; 0.2 &  & $\mu$m \\
$b$ & ice thickness & grains & see Table \ref{tab-core} &  & $a$ \\
$T_d$ & ambient dust temperature & cold grains & see Table \ref{tab-core} & $\approx$10 & K \\
$T_{\rm CR}$ & temperature of WGH & grains hit by CR &  &  & K \\
$f_T$ & $T_{\rm CR}$ frequency & grains hit by CR &  &  & s$^{-1}$ \\
\hline
$J_k$ & spectra of ion k & CRs & see Eq. (\ref{cr1}) &  & cm$^{-2}$\,(s\,sr\,MeV\,)$^{-1}$ \\
$L_k$ & energy loss function of ion k & CRs in cloud &  & SRIM & eV\,cm$^2$ \\
\hline
$E_{\rm CR}$ & CR particle energy & CRs &  &  & eV\,amu$^{-1}$ \\
$E_{\rm lost}$ & CR energy lost in grains & CRs in grains &  & SRIM & eV \\
$E_{\rm grain}$ & energy received by grain from CR hit & grains hit by CR &  &  & eV \\
\hline
$g$ & cloud geometrical exposure factor & cloud core & $\leq4\pi$ &  &  \\
\hline
\end{tabular}
\end{table*}
\edit1{T}his study presents the temperature and energy spectra for medium-sized interstellar grains affected by CR impacts. \edit1{Table~\ref{tab-param} summarizes the various parameters employed in the calculations.} The heating frequencies to temperatures above 70~K (and other $T_{\rm CR}$ thresholds) have been pinpointed for uses in astrochemical modeling. The changes of CR spectra with increasing cloud core column densities were considered, as well as grain growth via accumulation of ice and relevant ambient dust temperature $T_d$, although the latter can reasonably be approximated to 10~K or other $T_d$ distribution, relevant for cold cores.

\edit1{While aimed primarily for the case of starless cloud cores, this study is relevant also for prestellar cores with line-of-sight interstellar extinctions from 2 to 512~mag. Given that the WGH frequency in this interval is stable and predictable, extrapolation to higher $A_V$ values is permitted. Additionally, while this research covers only cold cores, it can be attributed also to mildly heated cores, because ambient grain temperature of 20~K affects $f_T$ only by a few per cent, while WGH effects cease to be important for surface chemistry at $T_d\approx30$~K \citepalias[see][]{K16}. At 30~K ambient temperature, most surface processes are faster than processes facilitated by the short intervals of WGH heating. However, because we considered near total freeze-out conditions, the results are best applicable to cold, quiescent cores, where such freeze out of CO and other species occurs. The main property of clouds that must be considered when applying these data, is their exposure to CRs. This is affected by proximity to massive star forming regions, presence of gas clumps that may shield the core in consideration, and the mirroring and focusing of CRs by magnetic fields \citep[see, e.g.,][]{Padovani13g}.}

\acknowledgments
This publication has been funded by ERDF postdoctoral grant No. 1.1.1.2/VIAA/I/16/194 ``Chemical effects of cosmic-ray induced heating of interstellar dust grains'' being implemented in Ventspils University of Applied Sciences. I am grateful to Alexei Ivlev for necessary comments on the manuscript. I thank to Ventspils City Council for support. This research has made use of NASA’s Astrophysics Data System.

\vspace{5mm}
\software{SRIM \citep[http://www.srim.org]{Ziegler10}}

\bibliography{CRput3}

\begin{thebibliography}{}
\expandafter\ifx\csname natexlab\endcsname\relax\def\natexlab#1{#1}\fi
\providecommand{\url}[1]{\href{#1}{#1}}
\providecommand{\dodoi}[1]{doi:~\href{http://doi.org/#1}{\nolinkurl{#1}}}
\providecommand{\doeprint}[1]{\href{http://ascl.net/#1}{\nolinkurl{http://ascl.net/#1}}}
\providecommand{\doarXiv}[1]{\href{https://arxiv.org/abs/#1}{\nolinkurl{https://arxiv.org/abs/#1}}}

\bibitem[{{Acharyya} \& {Herbst}(2017)}]{Acharyya17}
{Acharyya}, K., \& {Herbst}, E. 2017, \apj, 850, 105,
  \dodoi{10.3847/1538-4357/aa937e}

\bibitem[{{Adriani} {et~al.}(2016){Adriani}, {Barbarino}, {Bazilevskaya},
  {Bellotti}, {Boezio}, {Bogomolov}, {Bongi}, {Bonvicini}, {Bottai}, {Bruno},
  {Cafagna}, {Campana}, {Carlson}, {Casolino}, {Castellini}, {De Donato}, {De
  Santis}, {De Simone}, {Di Felice}, {Formato}, {Galper}, {Karelin},
  {Koldashov}, {Koldobskiy}, {Krutkov}, {Kvashnin}, {Leonov}, {Malakhov},
  {Marcelli}, {Martucci}, {Mayorov}, {Menn}, {Merg{\`e}}, {Mikhailov},
  {Mocchiutti}, {Monaco}, {Mori}, {Munini}, {Osteria}, {Palma}, {Panico},
  {Papini}, {Pearce}, {Picozza}, {Ricci}, {Ricciarini}, {Sarkar}, {Scotti},
  {Simon}, {Sparvoli}, {Spillantini}, {Stozhkov}, {Vacchi}, {Vannuccini},
  {Vasilyev}, {Voronov}, {Yurkin}, {Zampa}, \& {Zampa}}]{Adriani16}
{Adriani}, O., {Barbarino}, G.~C., {Bazilevskaya}, G.~A., {et~al.} 2016, \apj,
  818, 68, \dodoi{10.3847/0004-637X/818/1/68}

\bibitem[{{Aikawa} {et~al.}(2018){Aikawa}, {Furuya}, {Hincelin}, \&
  {Herbst}}]{Aikawa18}
{Aikawa}, Y., {Furuya}, K., {Hincelin}, U., \& {Herbst}, E. 2018, \apj, 855,
  119, \dodoi{10.3847/1538-4357/aaad6c}

\bibitem[{{Awad} \& {Shalabeia}(2017)}]{Awad17}
{Awad}, Z., \& {Shalabeia}, O.~M. 2017, \apss, 362, 83,
  \dodoi{10.1007/s10509-017-3061-8}

\bibitem[{{Binns} {et~al.}(2014){Binns}, {Bose}, {Braun}, {Brandt}, {Daniels},
  {Dowkontt}, {Fitzsimmons}, {Hahne}, {Hams}, {Israel}, {Klemic}, {Labrador},
  {Link}, {Mewaldt}, {Mitchell}, {Moore}, {Murphy}, {Olevitch}, {Rauch},
  {Sakai}, {San Sebastian}, {Sasaki}, {Simburger}, {Stone}, {Waddington},
  {Ward}, \& {Wiedenbeck}}]{Binns14}
{Binns}, W.~R., {Bose}, R.~G., {Braun}, D.~L., {et~al.} 2014, \apj, 788, 18,
  \dodoi{10.1088/0004-637X/788/1/18}

\bibitem[{{Bisbas} {et~al.}(2017){Bisbas}, {van Dishoeck}, {Papadopoulos},
  {Sz{\H u}cs}, {Bialy}, \& {Zhang}}]{Bisbas17}
{Bisbas}, T.~G., {van Dishoeck}, E.~F., {Papadopoulos}, P.~P., {et~al.} 2017,
  \apj, 839, 90, \dodoi{10.3847/1538-4357/aa696d}

\bibitem[{{Cazaux} {et~al.}(2016){Cazaux}, {Minissale}, {Dulieu}, \&
  {Hocuk}}]{Cazaux16}
{Cazaux}, S., {Minissale}, M., {Dulieu}, F., \& {Hocuk}, S. 2016, \aap, 585,
  A55, \dodoi{10.1051/0004-6361/201527187}

\bibitem[{{Ceccarelli} {et~al.}(2018){Ceccarelli}, {Viti}, {Balucani}, \&
  {Taquet}}]{Ceccarelli18}
{Ceccarelli}, C., {Viti}, S., {Balucani}, N., \& {Taquet}, V. 2018, \mnras,
  476, 1371, \dodoi{10.1093/mnras/sty313}

\bibitem[{{Cecchi-Pestellini} \& {Aiello}(1992)}]{Cecchi92}
{Cecchi-Pestellini}, C., \& {Aiello}, S. 1992, MNRAS, 258, 125,
  \dodoi{10.1093/mnras/258.1.125}

\bibitem[{{Chabot}(2016)}]{Chabot16}
{Chabot}, M. 2016, \aap, 585, A15, \dodoi{10.1051/0004-6361/201425441}

\bibitem[{{Coutens} {et~al.}(2017){Coutens}, {Rawlings}, {Viti}, \&
  {Williams}}]{Coutens17}
{Coutens}, A., {Rawlings}, J.~M.~C., {Viti}, S., \& {Williams}, D.~A. 2017,
  \mnras, 467, 737, \dodoi{10.1093/mnras/stx119}

\bibitem[{{Cummings} {et~al.}(2016){Cummings}, {Stone}, {Heikkila}, {Lal},
  {Webber}, {J{\'o}hannesson}, {Moskalenko}, {Orlando}, \&
  {Porter}}]{Cummings16}
{Cummings}, A.~C., {Stone}, E.~C., {Heikkila}, B.~C., {et~al.} 2016, \apj, 831,
  18, \dodoi{10.3847/0004-637X/831/1/18}

\bibitem[{{Cuppen} {et~al.}(2006){Cuppen}, {Morata}, \& {Herbst}}]{Cuppen06}
{Cuppen}, H.~M., {Morata}, O., \& {Herbst}, E. 2006, \mnras, 367, 1757,
  \dodoi{10.1111/j.1365-2966.2006.10079.x}

\bibitem[{{de Jong} \& {Kamijo}(1973)}]{deJong73}
{de Jong}, T., \& {Kamijo}, F. 1973, \aap, 25, 363

\bibitem[{{Esplugues} {et~al.}(2016){Esplugues}, {Cazaux}, {Meijerink},
  {Spaans}, \& {Caselli}}]{Esplugues16}
{Esplugues}, G.~B., {Cazaux}, S., {Meijerink}, R., {Spaans}, M., \& {Caselli},
  P. 2016, \aap, 591, A52, \dodoi{10.1051/0004-6361/201528001}

\bibitem[{{Furuya} \& {Persson}(2018)}]{Furuya18}
{Furuya}, K., \& {Persson}, M.~V. 2018, \mnras, 476, 4994,
  \dodoi{10.1093/mnras/sty553}

\bibitem[{{Garrod} {et~al.}(2008){Garrod}, {Weaver}, \& {Herbst}}]{Garrod08}
{Garrod}, R.~T., {Weaver}, S.~L.~W., \& {Herbst}, E. 2008, \apj, 682, 283,
  \dodoi{10.1086/588035}

\bibitem[{{George} {et~al.}(2009){George}, {Lave}, {Wiedenbeck}, {Binns},
  {Cummings}, {Davis}, {de Nolfo}, {Hink}, {Israel}, {Leske}, {Mewaldt},
  {Scott}, {Stone}, {von Rosenvinge}, \& {Yanasak}}]{George09}
{George}, J.~S., {Lave}, K.~A., {Wiedenbeck}, M.~E., {et~al.} 2009, \apj, 698,
  1666, \dodoi{10.1088/0004-637X/698/2/1666}

\bibitem[{{Hasegawa} \& {Herbst}(1993)}]{Hasegawa93}
{Hasegawa}, T.~I., \& {Herbst}, E. 1993, \mnras, 261, 83

\bibitem[{{Hasegawa} {et~al.}(1992){Hasegawa}, {Herbst}, \&
  {Leung}}]{Hasegawa92}
{Hasegawa}, T.~I., {Herbst}, E., \& {Leung}, C.~M. 1992, \apjs, 82, 167,
  \dodoi{10.1086/191713}

\bibitem[{{Hocuk} {et~al.}(2017){Hocuk}, {Sz{\H u}cs}, {Caselli}, {Cazaux},
  {Spaans}, \& {Esplugues}}]{Hocuk17}
{Hocuk}, S., {Sz{\H u}cs}, L., {Caselli}, P., {et~al.} 2017, A\&A, 604, A58,
  \dodoi{10.1051/0004-6361/201629944}

\bibitem[{{Holdship} {et~al.}(2017){Holdship}, {Viti}, {Jim{\'e}nez-Serra},
  {Makrymallis}, \& {Priestley}}]{Holdship17}
{Holdship}, J., {Viti}, S., {Jim{\'e}nez-Serra}, I., {Makrymallis}, A., \&
  {Priestley}, F. 2017, \aj, 154, 38, \dodoi{10.3847/1538-3881/aa773f}

\bibitem[{{Indriolo} {et~al.}(2009){Indriolo}, {Fields}, \&
  {McCall}}]{Indriolo09}
{Indriolo}, N., {Fields}, B.~D., \& {McCall}, B.~J. 2009, \apj, 694, 257,
  \dodoi{10.1088/0004-637X/694/1/257}

\bibitem[{{Iqbal} \& {Wakelam}(2018)}]{Iqbal18}
{Iqbal}, W., \& {Wakelam}, V. 2018, \aap, 615, A20,
  \dodoi{10.1051/0004-6361/201732486}

\bibitem[{{Ivlev} {et~al.}(2015){Ivlev}, {Padovani}, {Galli}, \&
  {Caselli}}]{Ivlev15p}
{Ivlev}, A.~V., {Padovani}, M., {Galli}, D., \& {Caselli}, P. 2015, \apj, 812,
  135, \dodoi{10.1088/0004-637X/812/2/135}

\bibitem[{{Kalv{\= a}ns}(2014)}]{Kalvans14}
{Kalv{\= a}ns}, J. 2014, Baltic Astronomy, 23, 137

\bibitem[{{Kalv{\= a}ns}(2015)}]{Kalvans15}
---. 2015, A\&A, 573, A38

\bibitem[{{Kalv{\= a}ns}(2016)}]{K16}
---. 2016, \apjs, 224, 42 (Paper~I), \dodoi{10.3847/0067-0049/224/2/42}

\bibitem[{{Kalv{\= a}ns}(2018)}]{Kalvans18}
---. 2018, \mnras, 478, 2753, \dodoi{10.1093/mnras/sty1172}

\bibitem[{{Kalv{\= a}ns} {et~al.}(2017){Kalv{\= a}ns}, {Shmeld}, {Kalnin}, \&
  {Hocuk}}]{Kalvans17}
{Kalv{\= a}ns}, J., {Shmeld}, I., {Kalnin}, J.~R., \& {Hocuk}, S. 2017, MNRAS,
  \dodoi{10.1093/mnras/stx174}

\bibitem[{{Kamp} {et~al.}(2017){Kamp}, {Thi}, {Woitke}, {Rab}, {Bouma}, \&
  {M{\'e}nard}}]{Kamp17}
{Kamp}, I., {Thi}, W.-F., {Woitke}, P., {et~al.} 2017, \aap, 607, A41,
  \dodoi{10.1051/0004-6361/201730388}

\bibitem[{{Leger} {et~al.}(1985){Leger}, {Jura}, \& {Omont}}]{Leger85}
{Leger}, A., {Jura}, M., \& {Omont}, A. 1985, \aap, 144, 147

\bibitem[{{Mainitz} {et~al.}(2016){Mainitz}, {Anders}, \&
  {Urbassek}}]{Mainitz16}
{Mainitz}, M., {Anders}, C., \& {Urbassek}, H.~M. 2016, \aap, 592, A35,
  \dodoi{10.1051/0004-6361/201628525}

\bibitem[{{Majumdar} {et~al.}(2018){Majumdar}, {Loison}, {Ruaud}, {Gratier},
  {Wakelam}, \& {Coutens}}]{Majumdar18}
{Majumdar}, L., {Loison}, J.-C., {Ruaud}, M., {et~al.} 2018, \mnras, 473, L59,
  \dodoi{10.1093/mnrasl/slx157}

\bibitem[{{McCall} {et~al.}(2003){McCall}, {Huneycutt}, {Saykally}, {Geballe},
  {Djuric}, {Dunn}, {Semaniak}, {Novotny}, {Al-Khalili}, {Ehlerding},
  {Hellberg}, {Kalhori}, {Neau}, {Thomas}, {{\"O}sterdahl}, \&
  {Larsson}}]{McCall03}
{McCall}, B.~J., {Huneycutt}, A.~J., {Saykally}, R.~J., {et~al.} 2003, \nat,
  422, 500, \dodoi{10.1038/nature01498}

\bibitem[{{Moskalenko} {et~al.}(2002){Moskalenko}, {Strong}, {Ormes}, \&
  {Potgieter}}]{Moskalenko02}
{Moskalenko}, I.~V., {Strong}, A.~W., {Ormes}, J.~F., \& {Potgieter}, M.~S.
  2002, \apj, 565, 280, \dodoi{10.1086/324402}

\bibitem[{{Murphy} {et~al.}(2016){Murphy}, {Sasaki}, {Binns}, {Brandt}, {Hams},
  {Israel}, {Labrador}, {Link}, {Mewaldt}, {Mitchell}, {Rauch}, {Sakai},
  {Stone}, {Waddington}, {Walsh}, {Ward}, \& {Wiedenbeck}}]{Murphy16}
{Murphy}, R.~P., {Sasaki}, M., {Binns}, W.~R., {et~al.} 2016, \apj, 831, 148,
  \dodoi{10.3847/0004-637X/831/2/148}

\bibitem[{{Padovani} \& {Galli}(2013)}]{Padovani13g}
{Padovani}, M., \& {Galli}, D. 2013, in ASSP, Vol.~34, Cosmic Rays in
  Star-Forming Environments, ed. D.~F. {Torres} \& O.~{Reimer}, 61

\bibitem[{{Padovani} {et~al.}(2009){Padovani}, {Galli}, \&
  {Glassgold}}]{Padovani09}
{Padovani}, M., {Galli}, D., \& {Glassgold}, A.~E. 2009, \aap, 501, 619,
  \dodoi{10.1051/0004-6361/200911794}

\bibitem[{{Padovani} {et~al.}(2013){Padovani}, {Hennebelle}, \&
  {Galli}}]{Padovani13h}
{Padovani}, M., {Hennebelle}, P., \& {Galli}, D. 2013, \aap, 560, A114,
  \dodoi{10.1051/0004-6361/201322407}

\bibitem[{{Padovani} {et~al.}(2018){Padovani}, {Ivlev}, {Galli}, \&
  {Caselli}}]{Padovani18}
{Padovani}, M., {Ivlev}, A.~V., {Galli}, D., \& {Caselli}, P. 2018, \aap, 614,
  A111, \dodoi{10.1051/0004-6361/201732202}

\bibitem[{{Picozza} {et~al.}(2007){Picozza}, {Galper}, {Castellini}, {Adriani},
  {Altamura}, {Ambriola}, {Barbarino}, {Basili}, {Bazilevskaja}, {Bencardino},
  {Boezio}, {Bogomolov}, {Bonechi}, {Bongi}, {Bongiorno}, {Bonvicini},
  {Cafagna}, {Campana}, {Carlson}, {Casolino}, {de Marzo}, {de Pascale}, {de
  Rosa}, {Fedele}, {Hofverberg}, {Koldashov}, {Krutkov}, {Kvashnin}, {Lund},
  {Lundquist}, {Maksumov}, {Malvezzi}, {Marcelli}, {Menn}, {Mikhailov},
  {Minori}, {Misin}, {Mocchiutti}, {Morselli}, {Nikonov}, {Orsi}, {Osteria},
  {Papini}, {Pearce}, {Ricci}, {Ricciarini}, {Runtso}, {Russo}, {Simon},
  {Sparvoli}, {Spillantini}, {Stozhkov}, {Taddei}, {Vacchi}, {Vannuccini},
  {Voronov}, {Yurkin}, {Zampa}, {Zampa}, \& {Zverev}}]{Picozza07}
{Picozza}, P., {Galper}, A.~M., {Castellini}, G., {et~al.} 2007, Astroparticle
  Physics, 27, 296, \dodoi{10.1016/j.astropartphys.2006.12.002}

\bibitem[{{Reboussin} {et~al.}(2014){Reboussin}, {Wakelam}, {Guilloteau}, \&
  {Hersant}}]{Reboussin14}
{Reboussin}, L., {Wakelam}, V., {Guilloteau}, S., \& {Hersant}, F. 2014,
  \mnras, 440, 3557, \dodoi{10.1093/mnras/stu462}

\bibitem[{{Roberts} {et~al.}(2007){Roberts}, {Rawlings}, {Viti}, \&
  {Williams}}]{Roberts07}
{Roberts}, J.~F., {Rawlings}, J.~M.~C., {Viti}, S., \& {Williams}, D.~A. 2007,
  \mnras, 382, 733, \dodoi{10.1111/j.1365-2966.2007.12402.x}

\bibitem[{{Semenov}(2017)}]{Semenov17}
{Semenov}, D.~A. 2017, {ALCHEMIC: Advanced time-dependent chemical kinetics},
  Astrophysics Source Code Library.
\newblock \doeprint{1708.008}

\bibitem[{{Shen} {et~al.}(2004){Shen}, {Greenberg}, {Schutte}, \& {van
  Dishoeck}}]{Shen04}
{Shen}, C.~J., {Greenberg}, J.~M., {Schutte}, W.~A., \& {van Dishoeck}, E.~F.
  2004, \aap, 415, 203, \dodoi{10.1051/0004-6361:20031669}

\bibitem[{{Shingledecker} \& {Herbst}(2018)}]{Shingledecker18}
{Shingledecker}, C.~N., \& {Herbst}, E. 2018, PCCP, 20, 5359,
  \dodoi{10.1039/C7CP05901A}

\bibitem[{{Silsbee} {et~al.}(2018){Silsbee}, {Ivlev}, {Padovani}, \&
  {Caselli}}]{Silsbee18}
{Silsbee}, K., {Ivlev}, A.~V., {Padovani}, M., \& {Caselli}, P. 2018, \apj,
  863, 188, \dodoi{10.3847/1538-4357/aad3cf}

\bibitem[{{Sipil{\"a}} {et~al.}(2016){Sipil{\"a}}, {Caselli}, \&
  {Taquet}}]{Sipila16}
{Sipil{\"a}}, O., {Caselli}, P., \& {Taquet}, V. 2016, \aap, 591, A9,
  \dodoi{10.1051/0004-6361/201628272}

\bibitem[{{Stone} {et~al.}(1998){Stone}, {Cohen}, {Cook}, {Cummings}, {Gauld},
  {Kecman}, {Leske}, {Mewaldt}, {Thayer}, {Dougherty}, {Grumm}, {Milliken},
  {Radocinski}, {Wiedenbeck}, {Christian}, {Shuman}, {Trexel}, {von
  Rosenvinge}, {Binns}, {Crary}, {Dowkontt}, {Epstein}, {Hink}, {Klarmann},
  {Lijowski}, \& {Olevitch}}]{Stone98}
{Stone}, E.~C., {Cohen}, C.~M.~S., {Cook}, W.~R., {et~al.} 1998, \ssr, 86, 285,
  \dodoi{10.1023/A:1005075813033}

\bibitem[{{Thi} {et~al.}(2010){Thi}, {Woitke}, \& {Kamp}}]{Thi10}
{Thi}, W.-F., {Woitke}, P., \& {Kamp}, I. 2010, \mnras, 407, 232,
  \dodoi{10.1111/j.1365-2966.2009.16162.x}

\bibitem[{{Valencic} \& {Smith}(2015)}]{Valencic15}
{Valencic}, L.~A., \& {Smith}, R.~K. 2015, ApJ, 809, 66,
  \dodoi{10.1088/0004-637X/809/1/66}

\bibitem[{{Vasyunin} {et~al.}(2017){Vasyunin}, {Caselli}, {Dulieu}, \&
  {Jim{\'e}nez-Serra}}]{Vasyunin17}
{Vasyunin}, A.~I., {Caselli}, P., {Dulieu}, F., \& {Jim{\'e}nez-Serra}, I.
  2017, \apj, 842, 33, \dodoi{10.3847/1538-4357/aa72ec}

\bibitem[{{Webber}(1998)}]{Webber98}
{Webber}, W.~R. 1998, \apj, 506, 329, \dodoi{10.1086/306222}

\bibitem[{{Webber}(2017)}]{Webber17}
---. 2017, ArXiv e-prints.
\newblock \doarXiv{1711.11584}

\bibitem[{{Willacy} \& {Millar}(1998)}]{Willacy98}
{Willacy}, K., \& {Millar}, T.~J. 1998, \mnras, 298, 562,
  \dodoi{10.1046/j.1365-8711.1998.01648.x}

\bibitem[{{Ysard} {et~al.}(2016){Ysard}, {K{\"o}hler}, {Jones}, {Dartois},
  {Godard}, \& {Gavilan}}]{Ysard16}
{Ysard}, N., {K{\"o}hler}, M., {Jones}, A., {et~al.} 2016, \aap, 588, A44,
  \dodoi{10.1051/0004-6361/201527487}

\bibitem[{{Zhu} {et~al.}(2017){Zhu}, {Tian}, {Li}, \& {Zhang}}]{Zhu17}
{Zhu}, H., {Tian}, W., {Li}, A., \& {Zhang}, M. 2017, MNRAS, 471, 3494,
  \dodoi{10.1093/mnras/stx1580}

\bibitem[{{Ziegler} {et~al.}(2010){Ziegler}, {Ziegler}, \&
  {Biersack}}]{Ziegler10}
{Ziegler}, J.~F., {Ziegler}, M.~D., \& {Biersack}, J.~P. 2010, Nuclear
  Instruments and Methods in Physics Research B, 268, 1818,
  \dodoi{10.1016/j.nimb.2010.02.091}

\end{thebibliography}
\bibliographystyle{aasjournal}

\appendix
\section{Accumulation of the icy mantle in the astrochemical model ``Alchemic-Venta''}
\label{app-alchm}

\begin{table*}
\centering
\caption{\edit1{List of parameters for the model ``Alchemic-Venta'' of \citet{Kalvans18}, used for determining $A_V$ for a given ice thickness.}} \label{tab-model}
\begin{tabular}{lllll}
\tablewidth{0pt}
\hline
\hline
 & \multicolumn{2}{c}{Value at simulation} &  &  \\
Parameter & start & end & Unit & Main reference \\
\hline
Integration time & 0 & 1.39 & Myr &  \\
H number density & 2800 & $1.5\times10^6$ & cm$^{-3}$ &  \\
Sticking probability & 1\tablenotemark{a} & \nodata &  &  \\
$A_V$ to core center & 1.05 & 31.8 & mag &  \\
Gas temperature & 23 & 7.0 & K & \citet{Kalvans17} \\
Dust temperature & 13 & 6.3 & K & \citet{Hocuk17} \\
CR ionization rate $\zeta$ & $2.9\times10^{-16}$ & $8.3\times10^{-17}$ & s$^{-1}$ & \citet{Ivlev15p} \\
CR-induced photon flux & $1.1\times10^5$ & $3.1\times10^4$ & cm$^{-2}$s$^{-1}$ & \citet{Cecchi92} \\
ISRF photon flux & 10$^8$ & \nodata & cm$^{-2}$s$^{-1}$ &  \\
Grain size & 0.1 & 0.135 & $\mu$m &  \\
Adsorption site density &  &  &  &  \\
on grain surface & $1.5\times10^{-15}$ & \nodata & cm$^{-2}$ & \citet{Hasegawa92} \\
WGH $f_{>70}$\tablenotemark{b} & $1.1\times10^{-11}$ & $2.4\times10^{-12}$ & s$^{-1}$ & \citetalias{K16} \\
\hline
\end{tabular}
\tablenotetext{a}{ Sticking probabilities for light species H and H$_2$ calculated according to \citet{Thi10}.}
\tablenotetext{b}{ See text.}
\end{table*}
\edit1{The latest iteration of the} \edit2{rate-equation} \edit1{model ``Alchemic-Venta'' follows the collapse of a diffuse Bonnor-Ebert sphere to a \edit1{dense} core with a mass of 4~M$_\odot$. The core is exposed to the ISRF from two opposite sides.} \edit2{In the model,} \edit1{surface chemistry is treated with the modified rate-equation method. Adsorbing species first become part of the grain surface. When mantle thickness exceeds 1~monolayer, surface species are transferred to bulk-ice layers below. The icy mantle can grow to a thickness of up to 100~monolayers; molecules in the surface layer and bulk ice layers both are chemically active.}

\edit1{The ranges of parameters relevant for the accretion of molecules are listed in Table~\ref{tab-model}. Desorption energies for surface species were adopted from \citet[OSU chemical network]{Garrod08}. Importantly, the subject of the present study -- WGH -- itself delays ice accumulation on grains. CR-induced desorption with new CR spectra was already included in the model of \citet{Kalvans18}, based on the (limited) data of \citetalias{K16}. Therefore, the present study is a second iteration pinpointing the frequency of WGH events for grains. Other desorption mechanisms included in the model are evaporation, photodesorption by the ISRF- and CR-induced photons, reactive desorption, and encounter desorption for the light species H and H$_2$.}

\edit1{The model depicts ice composition (proportions of major ice species) rather accurately up to $A_V=11$~mag, when the calculated composition (mainly, the abundance of CO$_2$ ice) starts to deviate from observational data. The deviations indicate about 10~\% reduction in the final ice thickness at freeze-out conditions, which may result in a small margin of error of a few per cent for the WGH $f_T$ values calculated in the present study.}

\section{Calculated WGH temperature $T_{\rm CR}$ and grain received energy $E_{\rm grain}$ spectra with frequency $f_T$ for interstellar dark, cold cloud core conditions}
\label{app-ft}

%
\startlongtable
\begin{deluxetable}{lccccccc}
\tabletypesize{\scriptsize}
\tablecaption{
Bare grains with $a=0.05$~$\mu$m and $T_d=13.71$~K, shielded by interstellar gas with column density $N_H=2.48\times10^{21}$~H\,atoms\,cm$^{-2}$.
}
\label{tab0.05-1}
\tablehead{
\colhead{No.} & \colhead{$T_{\rm CR}$, K} & \colhead{interval, K} & \colhead{$\Delta T_{\rm CR}$, K} & \colhead{$E_{\rm grain}$, eV} & \colhead{interval, eV} & \colhead{$\Delta E_{\rm grain}$, eV} & \colhead{$f_T$, s$^{-1}$}
}
\startdata
1 & 13.98 & 13.76--14.00 & 0.25 & 5.669E+00 & 0.995--6.027 & 5.033E+00 & 2.49E-14 \\
2 & 15.23 & 14.01--16.00 & 2.00 & 3.617E+01 & . . .--5.813E+01 & 5.210E+01 & 2.65E-09 \\
3 & 16.96 & 16.01--18.00 & 2.00 & 9.240E+01 & . . .--1.339E+02 & 7.575E+01 & 2.61E-09 \\
4 & 18.93 & 18.01--20.00 & 2.00 & 1.804E+02 & . . .--2.396E+02 & 1.057E+02 & 1.96E-09 \\
5 & 20.96 & 20.01--22.00 & 2.00 & 3.050E+02 & . . .--3.821E+02 & 1.426E+02 & 1.59E-09 \\
6 & 22.99 & 22.01--24.00 & 2.00 & 4.708E+02 & . . .--5.694E+02 & 1.872E+02 & 1.18E-09 \\
7 & 24.92 & 24.01--26.00 & 2.00 & 6.755E+02 & . . .--8.098E+02 & 2.404E+02 & 9.41E-10 \\
8 & 26.87 & 26.01--28.00 & 2.00 & 9.357E+02 & . . .--1.112E+03 & 3.027E+02 & 7.61E-10 \\
\enddata
\end{deluxetable}

%
\startlongtable
\begin{deluxetable}{lccccccc}
\tabletypesize{\scriptsize}
\tablecaption{
Grains with $a=0.05$~$\mu$m, 0.1$a$ icy mantle and $T_d=11.81$~K, shielded by interstellar gas with column density $N_H=5.86\times10^{21}$~H\,atoms\,cm$^{-2}$.
}
\label{tab0.05-2}
\tablehead{
\colhead{No.} & \colhead{$T_{\rm CR}$, K} & \colhead{interval, K} & \colhead{$\Delta T_{\rm CR}$, K} & \colhead{$E_{\rm grain}$, eV} & \colhead{interval, eV} & \colhead{$\Delta E_{\rm grain}$, eV} & \colhead{$f_T$, s$^{-1}$}
}
\startdata
1 & 11.99 & 11.84--12.00 & 0.17 & 6.490E+00 & 1.017--6.370 & 5.353E+00 & 8.65E-11 \\
2 & 12.99 & 12.01--14.00 & 2.00 & 4.612E+01 & . . .--9.158E+01 & 8.521E+01 & 5.45E-09 \\
3 & 14.93 & 14.01--16.00 & 2.00 & 1.445E+02 & . . .--2.119E+02 & 1.203E+02 & 3.24E-09 \\
4 & 16.94 & 16.01--18.00 & 2.00 & 2.852E+02 & . . .--3.751E+02 & 1.633E+02 & 2.15E-09 \\
5 & 18.97 & 18.01--20.00 & 2.00 & 4.755E+02 & . . .--5.901E+02 & 2.150E+02 & 1.45E-09 \\
6 & 20.95 & 20.01--22.00 & 2.00 & 7.175E+02 & . . .--8.663E+02 & 2.761E+02 & 1.02E-09 \\
7 & 22.97 & 22.01--24.00 & 2.00 & 1.031E+03 & . . .--1.214E+03 & 3.474E+02 & 7.73E-10 \\
8 & 24.99 & 24.01--26.00 & 2.00 & 1.422E+03 & . . .--1.643E+03 & 4.296E+02 & 5.36E-10 \\
\enddata
\end{deluxetable}

%
\startlongtable
\begin{deluxetable}{lccccccc}
\tabletypesize{\scriptsize}
\tablecaption{
Grains with $a=0.05$~$\mu$m, 0.2$a$ icy mantle and $T_d=10.42$~K, shielded by interstellar gas with column density $N_H=1.03\times10^{22}$~H\,atoms\,cm$^{-2}$.
}
\label{tab0.05-3}
\tablehead{
\colhead{No.} & \colhead{$T_{\rm CR}$, K} & \colhead{interval, K} & \colhead{$\Delta T_{\rm CR}$, K} & \colhead{$E_{\rm grain}$, eV} & \colhead{interval, eV} & \colhead{$\Delta E_{\rm grain}$, eV} & \colhead{$f_T$, s$^{-1}$}
}
\startdata
1 & 11.22 & 10.44--12.00 & 1.57 & 4.008E+01 & 0.900--83.207 & 8.231E+01 & 6.79E-09 \\
2 & 12.90 & 12.01--14.00 & 2.00 & 1.455E+02 & . . .--2.300E+02 & 1.468E+02 & 4.44E-09 \\
3 & 14.93 & 14.01--16.00 & 2.00 & 3.194E+02 & . . .--4.321E+02 & 2.022E+02 & 2.48E-09 \\
4 & 16.95 & 16.01--18.00 & 2.00 & 5.543E+02 & . . .--7.006E+02 & 2.685E+02 & 1.48E-09 \\
5 & 18.92 & 18.01--20.00 & 2.00 & 8.540E+02 & . . .--1.047E+03 & 3.463E+02 & 9.66E-10 \\
6 & 20.93 & 20.01--22.00 & 2.00 & 1.243E+03 & . . .--1.483E+03 & 4.365E+02 & 6.58E-10 \\
7 & 22.99 & 22.01--24.00 & 2.00 & 1.745E+03 & . . .--2.023E+03 & 5.398E+02 & 4.52E-10 \\
8 & 24.98 & 24.01--26.00 & 2.00 & 2.336E+03 & . . .--2.680E+03 & 6.568E+02 & 2.80E-10 \\
\enddata
\end{deluxetable}

%
\startlongtable
\begin{deluxetable}{lccccccc}
\tabletypesize{\scriptsize}
\tablecaption{
Grains with $a=0.05$~$\mu$m, 0.3$a$ icy mantle and $T_d=8.67$~K, shielded by interstellar gas with column density $N_H=2.21\times10^{22}$~H\,atoms\,cm$^{-2}$.
}
\label{tab0.05-4}
\tablehead{
\colhead{No.} & \colhead{$T_{\rm CR}$, K} & \colhead{interval, K} & \colhead{$\Delta T_{\rm CR}$, K} & \colhead{$E_{\rm grain}$, eV} & \colhead{interval, eV} & \colhead{$\Delta E_{\rm grain}$, eV} & \colhead{$f_T$, s$^{-1}$}
}
\startdata
1 & 9.36 & 8.69--10.00 & 1.32 & 3.550E+01 & 0.928--71.753 & 7.082E+01 & 7.62E-09 \\
2 & 10.87 & 10.01--12.00 & 2.00 & 1.338E+02 & . . .--2.254E+02 & 1.536E+02 & 5.77E-09 \\
3 & 12.89 & 12.01--14.00 & 2.00 & 3.182E+02 & . . .--4.449E+02 & 2.195E+02 & 2.82E-09 \\
4 & 14.92 & 14.01--16.00 & 2.00 & 5.762E+02 & . . .--7.439E+02 & 2.990E+02 & 1.50E-09 \\
5 & 16.91 & 16.01--18.00 & 2.00 & 9.161E+02 & . . .--1.137E+03 & 3.927E+02 & 8.91E-10 \\
6 & 18.96 & 18.01--20.00 & 2.00 & 1.370E+03 & . . .--1.638E+03 & 5.015E+02 & 5.59E-10 \\
7 & 20.95 & 20.01--22.00 & 2.00 & 1.927E+03 & . . .--2.264E+03 & 6.261E+02 & 3.37E-10 \\
8 & 22.87 & 22.01--24.00 & 2.00 & 2.587E+03 & . . .--3.031E+03 & 7.671E+02 & 2.36E-10 \\
\enddata
\end{deluxetable}

%
\startlongtable
\begin{deluxetable}{lccccccc}
\tabletypesize{\scriptsize}
\tablecaption{
Grains with $a=0.05$~$\mu$m, 0.35$a$ icy mantle and $T_d=6.91$~K, shielded by interstellar gas with column density $N_H=6.49\times10^{22}$~H\,atoms\,cm$^{-2}$.
}
\label{tab0.05-5}
\tablehead{
\colhead{No.} & \colhead{$T_{\rm CR}$, K} & \colhead{interval, K} & \colhead{$\Delta T_{\rm CR}$, K} & \colhead{$E_{\rm grain}$, eV} & \colhead{interval, eV} & \colhead{$\Delta E_{\rm grain}$, eV} & \colhead{$f_T$, s$^{-1}$}
}
\startdata
1 & 7.55 & 6.94--8.00 & 1.07 & 2.479E+01 & 1.040--43.906 & 4.287E+01 & 5.50E-09 \\
2 & 8.86 & 8.01--10.00 & 2.00 & 9.143E+01 & . . .--1.640E+02 & 1.201E+02 & 6.77E-09 \\
3 & 10.88 & 10.01--12.00 & 2.00 & 2.391E+02 & . . .--3.469E+02 & 1.829E+02 & 3.03E-09 \\
4 & 12.91 & 12.01--14.00 & 2.00 & 4.595E+02 & . . .--6.072E+02 & 2.604E+02 & 1.58E-09 \\
5 & 14.96 & 14.01--16.00 & 2.00 & 7.693E+02 & . . .--9.606E+02 & 3.533E+02 & 7.87E-10 \\
6 & 16.97 & 16.01--18.00 & 2.00 & 1.176E+03 & . . .--1.423E+03 & 4.625E+02 & 4.73E-10 \\
7 & 18.97 & 18.01--20.00 & 2.00 & 1.699E+03 & . . .--2.012E+03 & 5.887E+02 & 2.91E-10 \\
8 & 20.97 & 20.01--22.00 & 2.00 & 2.358E+03 & . . .--2.744E+03 & 7.325E+02 & 1.88E-10 \\
\enddata
\end{deluxetable}

%
\startlongtable
\begin{deluxetable}{lccccccc}
\tabletypesize{\scriptsize}
\tablecaption{
Grains with $a=0.05$~$\mu$m, 0.35$a$ icy mantle and $T_d=6.25$~K, shielded by interstellar gas with column density $N_H=1.28\times10^{23}$~H\,atoms\,cm$^{-2}$.
}
\label{tab0.05-6}
\tablehead{
\colhead{No.} & \colhead{$T_{\rm CR}$, K} & \colhead{interval, K} & \colhead{$\Delta T_{\rm CR}$, K} & \colhead{$E_{\rm grain}$, eV} & \colhead{interval, eV} & \colhead{$\Delta E_{\rm grain}$, eV} & \colhead{$f_T$, s$^{-1}$}
}
\startdata
1 & 7.20 & 6.28--8.00 & 1.73 & 3.211E+01 & 0.846--54.222 & 5.338E+01 & 7.27E-09 \\
2 & 8.87 & 8.01--10.00 & 2.00 & 1.127E+02 & . . .--1.640E+02 & 1.098E+02 & 4.96E-09 \\
3 & 10.88 & 10.01--12.00 & 2.00 & 2.601E+02 & . . .--3.469E+02 & 1.829E+02 & 2.26E-09 \\
4 & 12.87 & 12.01--14.00 & 2.00 & 4.739E+02 & . . .--6.072E+02 & 2.604E+02 & 1.10E-09 \\
5 & 14.87 & 14.01--16.00 & 2.00 & 7.750E+02 & . . .--9.606E+02 & 3.533E+02 & 6.06E-10 \\
6 & 16.93 & 16.01--18.00 & 2.00 & 1.187E+03 & . . .--1.423E+03 & 4.625E+02 & 3.39E-10 \\
7 & 18.92 & 18.01--20.00 & 2.00 & 1.705E+03 & . . .--2.012E+03 & 5.887E+02 & 2.11E-10 \\
8 & 20.96 & 20.01--22.00 & 2.00 & 2.375E+03 & . . .--2.744E+03 & 7.325E+02 & 1.41E-10 \\
\enddata
\end{deluxetable}

%
\startlongtable
\begin{deluxetable}{lccccccc}
\tabletypesize{\scriptsize}
\tablecaption{
Grains with $a=0.05$~$\mu$m, 0.35$a$ icy mantle and $T_d=5.84$~K, shielded by interstellar gas with column density $N_H=2.56\times10^{23}$~H\,atoms\,cm$^{-2}$.
}
\label{tab0.05-7}
\tablehead{
\colhead{No.} & \colhead{$T_{\rm CR}$, K} & \colhead{interval, K} & \colhead{$\Delta T_{\rm CR}$, K} & \colhead{$E_{\rm grain}$, eV} & \colhead{interval, eV} & \colhead{$\Delta E_{\rm grain}$, eV} & \colhead{$f_T$, s$^{-1}$}
}
\startdata
1 & 5.95 & 5.88--6.00 & 0.13 & 2.865E+00 & 0.982--3.885 & 2.902E+00 & 3.11E-14 \\
2 & 6.98 & 6.01--8.00 & 2.00 & 3.519E+01 & . . .--7.525E+01 & 7.137E+01 & 7.72E-09 \\
3 & 8.88 & 8.01--10.00 & 2.00 & 1.239E+02 & . . .--1.954E+02 & 1.201E+02 & 3.91E-09 \\
4 & 10.91 & 10.01--12.00 & 2.00 & 2.732E+02 & . . .--3.782E+02 & 1.829E+02 & 1.69E-09 \\
5 & 12.93 & 12.01--14.00 & 2.00 & 4.929E+02 & . . .--6.386E+02 & 2.603E+02 & 7.90E-10 \\
6 & 14.90 & 14.01--16.00 & 2.00 & 7.907E+02 & . . .--9.919E+02 & 3.533E+02 & 3.95E-10 \\
7 & 16.89 & 16.01--18.00 & 2.00 & 1.189E+03 & . . .--1.454E+03 & 4.625E+02 & 2.43E-10 \\
8 & 18.89 & 18.01--20.00 & 2.00 & 1.708E+03 & . . .--2.043E+03 & 5.886E+02 & 1.48E-10 \\
\enddata
\end{deluxetable}

%
\startlongtable
\begin{deluxetable}{lccccccc}
\tabletypesize{\scriptsize}
\tablecaption{
Grains with $a=0.05$~$\mu$m, 0.35$a$ icy mantle and $T_d=5.60$~K, shielded by interstellar gas with column density $N_H=5.12\times10^{23}$~H\,atoms\,cm$^{-2}$.
}
\label{tab0.05-8}
\tablehead{
\colhead{No.} & \colhead{$T_{\rm CR}$, K} & \colhead{interval, K} & \colhead{$\Delta T_{\rm CR}$, K} & \colhead{$E_{\rm grain}$, eV} & \colhead{interval, eV} & \colhead{$\Delta E_{\rm grain}$, eV} & \colhead{$f_T$, s$^{-1}$}
}
\startdata
1 & 5.93 & 5.64--6.00 & 0.37 & 7.780E+00 & 0.902--9.497 & 8.595E+00 & 4.86E-10 \\
2 & 6.90 & 6.01--8.00 & 2.00 & 3.804E+01 & . . .--8.088E+01 & 7.139E+01 & 7.03E-09 \\
3 & 8.84 & 8.01--10.00 & 2.00 & 1.273E+02 & . . .--2.010E+02 & 1.201E+02 & 3.03E-09 \\
4 & 10.86 & 10.01--12.00 & 2.00 & 2.744E+02 & . . .--3.839E+02 & 1.829E+02 & 1.29E-09 \\
5 & 12.91 & 12.01--14.00 & 2.00 & 4.959E+02 & . . .--6.443E+02 & 2.604E+02 & 5.55E-10 \\
6 & 14.93 & 14.01--16.00 & 2.00 & 8.014E+02 & . . .--9.976E+02 & 3.534E+02 & 2.88E-10 \\
7 & 16.96 & 16.01--18.00 & 2.00 & 1.210E+03 & . . .--1.460E+03 & 4.625E+02 & 1.52E-10 \\
8 & 18.90 & 18.01--20.00 & 2.00 & 1.716E+03 & . . .--2.049E+03 & 5.887E+02 & 1.04E-10 \\
\enddata
\end{deluxetable}

%
\startlongtable
\begin{deluxetable}{lccccccc}
\tabletypesize{\scriptsize}
\tablecaption{
Bare grains with $a=0.1$~$\mu$m and $T_d=13.71$~K, shielded by interstellar gas with column density $N_H=2.48\times10^{21}$~H\,atoms\,cm$^{-2}$.
}
\label{tab0.1-1}
\tablehead{
\colhead{No.} & \colhead{$T_{\rm CR}$, K} & \colhead{interval, K} & \colhead{$\Delta T_{\rm CR}$, K} & \colhead{$E_{\rm grain}$, eV} & \colhead{interval, eV} & \colhead{$\Delta E_{\rm grain}$, eV} & \colhead{$f_T$, s$^{-1}$}
}
\startdata
1 & 13.96 & 13.72--14.00 & 0.29 & 4.182E+01 & 1.584--48.219 & 4.663E+01 & 2.20E-09 \\
2 & 14.45 & 14.01--15.00 & 1.00 & 1.303E+02 & . . .--2.359E+02 & 1.876E+02 & 1.67E-08 \\
3 & 15.47 & 15.01--16.00 & 1.00 & 3.394E+02 & . . .--4.650E+02 & 2.292E+02 & 8.84E-09 \\
4 & 16.45 & 16.01--17.00 & 1.00 & 5.872E+02 & . . .--7.414E+02 & 2.764E+02 & 6.17E-09 \\
5 & 17.44 & 17.01--18.00 & 1.00 & 8.827E+02 & . . .--1.071E+03 & 3.297E+02 & 4.37E-09 \\
6 & 18.46 & 18.01--19.00 & 1.00 & 1.244E+03 & . . .--1.460E+03 & 3.894E+02 & 3.80E-09 \\
7 & 19.47 & 19.01--20.00 & 1.00 & 1.671E+03 & . . .--1.916E+03 & 4.560E+02 & 2.72E-09 \\
8 & 20.46 & 20.01--21.00 & 1.00 & 2.153E+03 & . . .--2.446E+03 & 5.297E+02 & 2.35E-09 \\
\enddata
\end{deluxetable}

%
\startlongtable
\begin{deluxetable}{lccccccc}
\tabletypesize{\scriptsize}
\tablecaption{
Grains with $a=0.1$~$\mu$m, 0.1$a$ icy mantle and $T_d=11.81$~K, shielded by interstellar gas with column density $N_H=5.86\times10^{21}$~H\,atoms\,cm$^{-2}$.
}
\label{tab0.1-2}
\tablehead{
\colhead{No.} & \colhead{$T_{\rm CR}$, K} & \colhead{interval, K} & \colhead{$\Delta T_{\rm CR}$, K} & \colhead{$E_{\rm grain}$, eV} & \colhead{interval, eV} & \colhead{$\Delta E_{\rm grain}$, eV} & \colhead{$f_T$, s$^{-1}$}
}
\startdata
1 & 11.94 & 11.82--12.00 & 0.19 & 3.585E+01 & 2.706--50.957 & 4.825E+01 & 5.55E-09 \\
2 & 12.40 & 12.01--13.00 & 1.00 & 1.728E+02 & . . .--3.605E+02 & 3.095E+02 & 2.63E-08 \\
3 & 13.45 & 13.01--14.00 & 1.00 & 5.276E+02 & . . .--7.326E+02 & 3.721E+02 & 1.08E-08 \\
4 & 14.48 & 14.01--15.00 & 1.00 & 9.418E+02 & . . .--1.175E+03 & 4.422E+02 & 6.01E-09 \\
5 & 15.47 & 15.01--16.00 & 1.00 & 1.417E+03 & . . .--1.695E+03 & 5.200E+02 & 4.09E-09 \\
6 & 16.48 & 16.01--17.00 & 1.00 & 1.983E+03 & . . .--2.301E+03 & 6.059E+02 & 2.90E-09 \\
7 & 17.49 & 17.01--18.00 & 1.00 & 2.645E+03 & . . .--3.001E+03 & 7.004E+02 & 2.11E-09 \\
8 & 18.50 & 18.01--19.00 & 1.00 & 3.406E+03 & . . .--3.805E+03 & 8.037E+02 & 1.53E-09 \\
\enddata
\end{deluxetable}

%
\startlongtable
\begin{deluxetable}{lccccccc}
\tabletypesize{\scriptsize}
\tablecaption{
Grains with $a=0.1$~$\mu$m, 0.2$a$ icy mantle and $T_d=10.42$~K, shielded by interstellar gas with column density $N_H=1.03\times10^{22}$~H\,atoms\,cm$^{-2}$.
}
\label{tab0.1-3}
\tablehead{
\colhead{No.} & \colhead{$T_{\rm CR}$, K} & \colhead{interval, K} & \colhead{$\Delta T_{\rm CR}$, K} & \colhead{$E_{\rm grain}$, eV} & \colhead{interval, eV} & \colhead{$\Delta E_{\rm grain}$, eV} & \colhead{$f_T$, s$^{-1}$}
}
\startdata
1 & 10.68 & 10.43--11.00 & 0.58 & 9.941E+01 & 3.596--219.366 & 2.158E+02 & 3.13E-08 \\
2 & 11.42 & 11.01--12.00 & 1.00 & 4.037E+02 & . . .--6.657E+02 & 4.463E+02 & 1.89E-08 \\
3 & 12.45 & 12.01--13.00 & 1.00 & 9.038E+02 & . . .--1.202E+03 & 5.368E+02 & 8.11E-09 \\
4 & 13.44 & 13.01--14.00 & 1.00 & 1.481E+03 & . . .--1.840E+03 & 6.373E+02 & 4.52E-09 \\
5 & 14.45 & 14.01--15.00 & 1.00 & 2.171E+03 & . . .--2.588E+03 & 7.480E+02 & 3.06E-09 \\
6 & 15.46 & 15.01--16.00 & 1.00 & 2.985E+03 & . . .--3.457E+03 & 8.694E+02 & 1.88E-09 \\
7 & 16.44 & 16.01--17.00 & 1.00 & 3.898E+03 & . . .--4.459E+03 & 1.002E+03 & 1.35E-09 \\
8 & 17.46 & 17.01--18.00 & 1.00 & 4.978E+03 & . . .--5.605E+03 & 1.146E+03 & 1.04E-09 \\
\enddata
\end{deluxetable}

%
\startlongtable
\begin{deluxetable}{lccccccc}
\tabletypesize{\scriptsize}
\tablecaption{
Grains with $a=0.1$~$\mu$m, 0.3$a$ icy mantle and $T_d=8.67$~K, shielded by interstellar gas with column density $N_H=2.21\times10^{22}$~H\,atoms\,cm$^{-2}$.
}
\label{tab0.1-4}
\tablehead{
\colhead{No.} & \colhead{$T_{\rm CR}$, K} & \colhead{interval, K} & \colhead{$\Delta T_{\rm CR}$, K} & \colhead{$E_{\rm grain}$, eV} & \colhead{interval, eV} & \colhead{$\Delta E_{\rm grain}$, eV} & \colhead{$f_T$, s$^{-1}$}
}
\startdata
1 & 8.84 & 8.68--9.00 & 0.33 & 6.691E+01 & 3.706--125.107 & 1.214E+02 & 2.76E-08 \\
2 & 9.39 & 9.01--10.00 & 1.00 & 2.954E+02 & . . .--5.740E+02 & 4.489E+02 & 2.94E-08 \\
3 & 10.44 & 10.01--11.00 & 1.00 & 8.111E+02 & . . .--1.129E+03 & 5.550E+02 & 1.04E-08 \\
4 & 11.47 & 11.01--12.00 & 1.00 & 1.438E+03 & . . .--1.803E+03 & 6.738E+02 & 4.79E-09 \\
5 & 12.48 & 12.01--13.00 & 1.00 & 2.186E+03 & . . .--2.608E+03 & 8.055E+02 & 2.93E-09 \\
6 & 13.49 & 13.01--14.00 & 1.00 & 3.070E+03 & . . .--3.559E+03 & 9.506E+02 & 1.65E-09 \\
7 & 14.46 & 14.01--15.00 & 1.00 & 4.061E+03 & . . .--4.668E+03 & 1.109E+03 & 1.18E-09 \\
8 & 15.44 & 15.01--16.00 & 1.00 & 5.229E+03 & . . .--5.951E+03 & 1.282E+03 & 8.36E-10 \\
\enddata
\end{deluxetable}

%
\startlongtable
\begin{deluxetable}{lccccccc}
\tabletypesize{\scriptsize}
\tablecaption{
Grains with $a=0.1$~$\mu$m, 0.35$a$ icy mantle and $T_d=6.91$~K, shielded by interstellar gas with column density $N_H=6.49\times10^{22}$~H\,atoms\,cm$^{-2}$.
}
\label{tab0.1-5}
\tablehead{
\colhead{No.} & \colhead{$T_{\rm CR}$, K} & \colhead{interval, K} & \colhead{$\Delta T_{\rm CR}$, K} & \colhead{$E_{\rm grain}$, eV} & \colhead{interval, eV} & \colhead{$\Delta E_{\rm grain}$, eV} & \colhead{$f_T$, s$^{-1}$}
}
\startdata
1 & 6.98 & 6.92--7.00 & 0.09 & 2.054E+01 & 2.766--23.779 & 2.101E+01 & 4.56E-09 \\
2 & 7.38 & 7.01--8.00 & 1.00 & 1.420E+02 & . . .--3.512E+02 & 3.275E+02 & 4.49E-08 \\
3 & 8.42 & 8.01--9.00 & 1.00 & 5.250E+02 & . . .--7.761E+02 & 4.249E+02 & 1.25E-08 \\
4 & 9.44 & 9.01--10.00 & 1.00 & 1.004E+03 & . . .--1.312E+03 & 5.361E+02 & 5.54E-09 \\
5 & 10.46 & 10.01--11.00 & 1.00 & 1.607E+03 & . . .--1.974E+03 & 6.615E+02 & 2.95E-09 \\
6 & 11.47 & 11.01--12.00 & 1.00 & 2.342E+03 & . . .--2.775E+03 & 8.014E+02 & 1.68E-09 \\
7 & 12.49 & 12.01--13.00 & 1.00 & 3.235E+03 & . . .--3.731E+03 & 9.563E+02 & 1.18E-09 \\
8 & 13.49 & 13.01--14.00 & 1.00 & 4.270E+03 & . . .--4.858E+03 & 1.127E+03 & 6.68E-10 \\
\enddata
\end{deluxetable}

%
\startlongtable
\begin{deluxetable}{lccccccc}
\tabletypesize{\scriptsize}
\tablecaption{
Grains with $a=0.1$~$\mu$m, 0.35$a$ icy mantle and $T_d=6.25$~K, shielded by interstellar gas with column density $N_H=1.28\times10^{23}$~H\,atoms\,cm$^{-2}$.
}
\label{tab0.1-6}
\tablehead{
\colhead{No.} & \colhead{$T_{\rm CR}$, K} & \colhead{interval, K} & \colhead{$\Delta T_{\rm CR}$, K} & \colhead{$E_{\rm grain}$, eV} & \colhead{interval, eV} & \colhead{$\Delta E_{\rm grain}$, eV} & \colhead{$f_T$, s$^{-1}$}
}
\startdata
1 & 6.60 & 6.26--7.00 & 0.75 & 8.479E+01 & 2.249--188.738 & 1.865E+02 & 3.58E-08 \\
2 & 7.41 & 7.01--8.00 & 1.00 & 3.178E+02 & . . .--5.163E+02 & 3.276E+02 & 1.73E-08 \\
3 & 8.44 & 8.01--9.00 & 1.00 & 6.965E+02 & . . .--9.413E+02 & 4.250E+02 & 6.77E-09 \\
4 & 9.46 & 9.01--10.00 & 1.00 & 1.180E+03 & . . .--1.477E+03 & 5.362E+02 & 3.22E-09 \\
5 & 10.47 & 10.01--11.00 & 1.00 & 1.779E+03 & . . .--2.139E+03 & 6.616E+02 & 1.83E-09 \\
6 & 11.48 & 11.01--12.00 & 1.00 & 2.519E+03 & . . .--2.941E+03 & 8.016E+02 & 1.08E-09 \\
7 & 12.47 & 12.01--13.00 & 1.00 & 3.383E+03 & . . .--3.897E+03 & 9.565E+02 & 7.62E-10 \\
8 & 13.51 & 13.01--14.00 & 1.00 & 4.464E+03 & . . .--5.024E+03 & 1.127E+03 & 5.12E-10 \\
\enddata
\end{deluxetable}

%
\startlongtable
\begin{deluxetable}{lccccccc}
\tabletypesize{\scriptsize}
\tablecaption{
Grains with $a=0.1$~$\mu$m, 0.35$a$ icy mantle and $T_d=5.84$~K, shielded by interstellar gas with column density $N_H=2.56\times10^{23}$~H\,atoms\,cm$^{-2}$.
}
\label{tab0.1-7}
\tablehead{
\colhead{No.} & \colhead{$T_{\rm CR}$, K} & \colhead{interval, K} & \colhead{$\Delta T_{\rm CR}$, K} & \colhead{$E_{\rm grain}$, eV} & \colhead{interval, eV} & \colhead{$\Delta E_{\rm grain}$, eV} & \colhead{$f_T$, s$^{-1}$}
}
\startdata
1 & 5.96 & 5.85--6.00 & 0.16 & 2.347E+01 & 1.955--31.079 & 2.912E+01 & 6.30E-09 \\
2 & 6.39 & 6.01--7.00 & 1.00 & 1.213E+02 & . . .--2.746E+02 & 2.435E+02 & 3.42E-08 \\
3 & 7.43 & 7.01--8.00 & 1.00 & 4.103E+02 & . . .--6.020E+02 & 3.274E+02 & 1.04E-08 \\
4 & 8.48 & 8.01--9.00 & 1.00 & 7.985E+02 & . . .--1.027E+03 & 4.248E+02 & 4.17E-09 \\
5 & 9.49 & 9.01--10.00 & 1.00 & 1.284E+03 & . . .--1.563E+03 & 5.360E+02 & 1.97E-09 \\
6 & 10.47 & 10.01--11.00 & 1.00 & 1.865E+03 & . . .--2.224E+03 & 6.614E+02 & 1.17E-09 \\
7 & 11.49 & 11.01--12.00 & 1.00 & 2.613E+03 & . . .--3.026E+03 & 8.014E+02 & 7.51E-10 \\
8 & 12.47 & 12.01--13.00 & 1.00 & 3.467E+03 & . . .--3.982E+03 & 9.563E+02 & 4.78E-10 \\
\enddata
\end{deluxetable}

%
\startlongtable
\begin{deluxetable}{lccccccc}
\tabletypesize{\scriptsize}
\tablecaption{
Grains with $a=0.1$~$\mu$m, 0.35$a$ icy mantle and $T_d=5.60$~K, shielded by interstellar gas with column density $N_H=5.12\times10^{23}$~H\,atoms\,cm$^{-2}$.
}
\label{tab0.1-8}
\tablehead{
\colhead{No.} & \colhead{$T_{\rm CR}$, K} & \colhead{interval, K} & \colhead{$\Delta T_{\rm CR}$, K} & \colhead{$E_{\rm grain}$, eV} & \colhead{interval, eV} & \colhead{$\Delta E_{\rm grain}$, eV} & \colhead{$f_T$, s$^{-1}$}
}
\startdata
1 & 5.83 & 5.61--6.00 & 0.40 & 4.398E+01 & 1.793--75.973 & 7.418E+01 & 1.70E-08 \\
2 & 6.38 & 6.01--7.00 & 1.00 & 1.648E+02 & . . .--3.195E+02 & 2.436E+02 & 2.22E-08 \\
3 & 7.43 & 7.01--8.00 & 1.00 & 4.536E+02 & . . .--6.471E+02 & 3.275E+02 & 6.79E-09 \\
4 & 8.45 & 8.01--9.00 & 1.00 & 8.322E+02 & . . .--1.072E+03 & 4.249E+02 & 2.71E-09 \\
5 & 9.45 & 9.01--10.00 & 1.00 & 1.306E+03 & . . .--1.608E+03 & 5.361E+02 & 1.35E-09 \\
6 & 10.43 & 10.01--11.00 & 1.00 & 1.883E+03 & . . .--2.270E+03 & 6.615E+02 & 7.83E-10 \\
7 & 11.46 & 11.01--12.00 & 1.00 & 2.633E+03 & . . .--3.071E+03 & 8.015E+02 & 5.13E-10 \\
8 & 12.44 & 12.01--13.00 & 1.00 & 3.486E+03 & . . .--4.028E+03 & 9.564E+02 & 3.24E-10 \\
\enddata
\end{deluxetable}

%
\startlongtable
\begin{deluxetable}{lccccccc}
\tabletypesize{\scriptsize}
\tablecaption{
Bare grains with $a=0.2$~$\mu$m and $T_d=13.71$~K, shielded by interstellar gas with column density $N_H=2.48\times10^{21}$~H\,atoms\,cm$^{-2}$.
}
\label{tab0.2-1}
\tablehead{
\colhead{No.} & \colhead{$T_{\rm CR}$, K} & \colhead{interval, K} & \colhead{$\Delta T_{\rm CR}$, K} & \colhead{$E_{\rm grain}$, eV} & \colhead{interval, eV} & \colhead{$\Delta E_{\rm grain}$, eV} & \colhead{$f_T$, s$^{-1}$}
}
\startdata
1 & 13.87 & 13.72--14.00 & 0.29 & 2.107E+02 & 12.676--385.755 & 3.731E+02 & 6.20E-08 \\
2 & 14.38 & 14.01--15.00 & 1.00 & 9.419E+02 & . . .--1.887E+03 & 1.501E+03 & 8.27E-08 \\
3 & 15.43 & 15.01--16.00 & 1.00 & 2.643E+03 & . . .--3.720E+03 & 1.833E+03 & 3.01E-08 \\
4 & 16.44 & 16.01--17.00 & 1.00 & 4.675E+03 & . . .--5.931E+03 & 2.211E+03 & 1.85E-08 \\
5 & 17.50 & 17.01--18.00 & 1.00 & 7.216E+03 & . . .--8.569E+03 & 2.637E+03 & 1.12E-08 \\
6 & 18.50 & 18.01--19.00 & 1.00 & 1.008E+04 & . . .--1.168E+04 & 3.115E+03 & 6.60E-09 \\
7 & 19.48 & 19.01--20.00 & 1.00 & 1.340E+04 & . . .--1.533E+04 & 3.648E+03 & 5.15E-09 \\
8 & 20.51 & 20.01--21.00 & 1.00 & 1.746E+04 & . . .--1.957E+04 & 4.238E+03 & 3.54E-09 \\
\enddata
\end{deluxetable}

%
\startlongtable
\begin{deluxetable}{lccccccc}
\tabletypesize{\scriptsize}
\tablecaption{
Grains with $a=0.2$~$\mu$m, 0.1$a$ icy mantle and $T_d=11.81$~K, shielded by interstellar gas with column density $N_H=5.86\times10^{21}$~H\,atoms\,cm$^{-2}$.
}
\label{tab0.2-2}
\tablehead{
\colhead{No.} & \colhead{$T_{\rm CR}$, K} & \colhead{interval, K} & \colhead{$\Delta T_{\rm CR}$, K} & \colhead{$E_{\rm grain}$, eV} & \colhead{interval, eV} & \colhead{$\Delta E_{\rm grain}$, eV} & \colhead{$f_T$, s$^{-1}$}
}
\startdata
1 & 11.90 & 11.82--12.00 & 0.19 & 2.124E+02 & 21.652--407.656 & 3.860E+02 & 9.32E-08 \\
2 & 12.35 & 12.01--13.00 & 1.00 & 1.252E+03 & . . .--2.884E+03 & 2.476E+03 & 1.08E-07 \\
3 & 13.43 & 13.01--14.00 & 1.00 & 4.152E+03 & . . .--5.861E+03 & 2.977E+03 & 2.70E-08 \\
4 & 14.45 & 14.01--15.00 & 1.00 & 7.453E+03 & . . .--9.398E+03 & 3.538E+03 & 1.18E-08 \\
5 & 15.45 & 15.01--16.00 & 1.00 & 1.126E+04 & . . .--1.356E+04 & 4.160E+03 & 6.52E-09 \\
6 & 16.47 & 16.01--17.00 & 1.00 & 1.581E+04 & . . .--1.841E+04 & 4.848E+03 & 4.17E-09 \\
7 & 17.48 & 17.01--18.00 & 1.00 & 2.107E+04 & . . .--2.401E+04 & 5.603E+03 & 2.63E-09 \\
8 & 18.49 & 18.01--19.00 & 1.00 & 2.713E+04 & . . .--3.044E+04 & 6.430E+03 & 1.90E-09 \\
\enddata
\end{deluxetable}

%
\startlongtable
\begin{deluxetable}{lccccccc}
\tabletypesize{\scriptsize}
\tablecaption{
Grains with $a=0.2$~$\mu$m, 0.2$a$ icy mantle and $T_d=10.42$~K, shielded by interstellar gas with column density $N_H=1.03\times10^{22}$~H\,atoms\,cm$^{-2}$.
}
\label{tab0.2-3}
\tablehead{
\colhead{No.} & \colhead{$T_{\rm CR}$, K} & \colhead{interval, K} & \colhead{$\Delta T_{\rm CR}$, K} & \colhead{$E_{\rm grain}$, eV} & \colhead{interval, eV} & \colhead{$\Delta E_{\rm grain}$, eV} & \colhead{$f_T$, s$^{-1}$}
}
\startdata
1 & 10.60 & 10.43--11.00 & 0.58 & 5.329E+02 & 28.771--1754.929 & 1.726E+03 & 2.12E-07 \\
2 & 11.38 & 11.01--12.00 & 1.00 & 3.103E+03 & . . .--5.325E+03 & 3.570E+03 & 4.82E-08 \\
3 & 12.43 & 12.01--13.00 & 1.00 & 7.166E+03 & . . .--9.620E+03 & 4.294E+03 & 1.43E-08 \\
4 & 13.47 & 13.01--14.00 & 1.00 & 1.199E+04 & . . .--1.472E+04 & 5.098E+03 & 6.85E-09 \\
5 & 14.49 & 14.01--15.00 & 1.00 & 1.763E+04 & . . .--2.070E+04 & 5.984E+03 & 3.55E-09 \\
6 & 15.45 & 15.01--16.00 & 1.00 & 2.380E+04 & . . .--2.766E+04 & 6.955E+03 & 2.27E-09 \\
7 & 16.47 & 16.01--17.00 & 1.00 & 3.141E+04 & . . .--3.567E+04 & 8.015E+03 & 1.72E-09 \\
8 & 17.52 & 17.01--18.00 & 1.00 & 4.037E+04 & . . .--4.484E+04 & 9.166E+03 & 1.15E-09 \\
\enddata
\end{deluxetable}

%
\startlongtable
\begin{deluxetable}{lccccccc}
\tabletypesize{\scriptsize}
\tablecaption{
Grains with $a=0.2$~$\mu$m, 0.3$a$ icy mantle and $T_d=8.67$~K, shielded by interstellar gas with column density $N_H=2.21\times10^{22}$~H\,atoms\,cm$^{-2}$.
}
\label{tab0.2-4}
\tablehead{
\colhead{No.} & \colhead{$T_{\rm CR}$, K} & \colhead{interval, K} & \colhead{$\Delta T_{\rm CR}$, K} & \colhead{$E_{\rm grain}$, eV} & \colhead{interval, eV} & \colhead{$\Delta E_{\rm grain}$, eV} & \colhead{$f_T$, s$^{-1}$}
}
\startdata
1 & 8.79 & 8.68--9.00 & 0.33 & 3.574E+02 & 29.645--1000.853 & 9.712E+02 & 2.14E-07 \\
2 & 9.34 & 9.01--10.00 & 1.00 & 2.173E+03 & . . .--4.592E+03 & 3.591E+03 & 7.84E-08 \\
3 & 10.43 & 10.01--11.00 & 1.00 & 6.440E+03 & . . .--9.032E+03 & 4.440E+03 & 1.53E-08 \\
4 & 11.44 & 11.01--12.00 & 1.00 & 1.138E+04 & . . .--1.442E+04 & 5.390E+03 & 6.41E-09 \\
5 & 12.45 & 12.01--13.00 & 1.00 & 1.725E+04 & . . .--2.087E+04 & 6.444E+03 & 3.35E-09 \\
6 & 13.44 & 13.01--14.00 & 1.00 & 2.414E+04 & . . .--2.847E+04 & 7.605E+03 & 2.12E-09 \\
7 & 14.47 & 14.01--15.00 & 1.00 & 3.259E+04 & . . .--3.735E+04 & 8.876E+03 & 1.49E-09 \\
8 & 15.49 & 15.01--16.00 & 1.00 & 4.235E+04 & . . .--4.761E+04 & 1.026E+04 & 1.03E-09 \\
\enddata
\end{deluxetable}

%
\startlongtable
\begin{deluxetable}{lccccccc}
\tabletypesize{\scriptsize}
\tablecaption{
Grains with $a=0.2$~$\mu$m, 0.35$a$ icy mantle and $T_d=6.91$~K, shielded by interstellar gas with column density $N_H=6.49\times10^{22}$~H\,atoms\,cm$^{-2}$.
}
\label{tab0.2-5}
\tablehead{
\colhead{No.} & \colhead{$T_{\rm CR}$, K} & \colhead{interval, K} & \colhead{$\Delta T_{\rm CR}$, K} & \colhead{$E_{\rm grain}$, eV} & \colhead{interval, eV} & \colhead{$\Delta E_{\rm grain}$, eV} & \colhead{$f_T$, s$^{-1}$}
}
\startdata
1 & 6.96 & 6.92--7.00 & 0.09 & 1.193E+02 & 22.129--190.228 & 1.681E+02 & 9.81E-08 \\
2 & 7.27 & 7.01--8.00 & 1.00 & 8.566E+02 & . . .--2.810E+03 & 2.620E+03 & 1.73E-07 \\
3 & 8.39 & 8.01--9.00 & 1.00 & 4.092E+03 & . . .--6.209E+03 & 3.399E+03 & 1.87E-08 \\
4 & 9.42 & 9.01--10.00 & 1.00 & 7.943E+03 & . . .--1.050E+04 & 4.289E+03 & 7.03E-09 \\
5 & 10.44 & 10.01--11.00 & 1.00 & 1.280E+04 & . . .--1.579E+04 & 5.292E+03 & 3.43E-09 \\
6 & 11.47 & 11.01--12.00 & 1.00 & 1.872E+04 & . . .--2.220E+04 & 6.411E+03 & 2.00E-09 \\
7 & 12.47 & 12.01--13.00 & 1.00 & 2.577E+04 & . . .--2.985E+04 & 7.651E+03 & 1.35E-09 \\
8 & 13.49 & 13.01--14.00 & 1.00 & 3.418E+04 & . . .--3.886E+04 & 9.012E+03 & 9.76E-10 \\
\enddata
\end{deluxetable}

%
\startlongtable
\begin{deluxetable}{lccccccc}
\tabletypesize{\scriptsize}
\tablecaption{
Grains with $a=0.2$~$\mu$m, 0.35$a$ icy mantle and $T_d=6.25$~K, shielded by interstellar gas with column density $N_H=1.28\times10^{23}$~H\,atoms\,cm$^{-2}$.
}
\label{tab0.2-6}
\tablehead{
\colhead{No.} & \colhead{$T_{\rm CR}$, K} & \colhead{interval, K} & \colhead{$\Delta T_{\rm CR}$, K} & \colhead{$E_{\rm grain}$, eV} & \colhead{interval, eV} & \colhead{$\Delta E_{\rm grain}$, eV} & \colhead{$f_T$, s$^{-1}$}
}
\startdata
1 & 6.46 & 6.26--7.00 & 0.75 & 4.108E+02 & 17.991--1509.903 & 1.492E+03 & 2.29E-07 \\
2 & 7.38 & 7.01--8.00 & 1.00 & 2.463E+03 & . . .--4.130E+03 & 2.621E+03 & 2.90E-08 \\
3 & 8.44 & 8.01--9.00 & 1.00 & 5.559E+03 & . . .--7.530E+03 & 3.400E+03 & 8.05E-09 \\
4 & 9.46 & 9.01--10.00 & 1.00 & 9.439E+03 & . . .--1.182E+04 & 4.290E+03 & 3.71E-09 \\
5 & 10.48 & 10.01--11.00 & 1.00 & 1.428E+04 & . . .--1.711E+04 & 5.293E+03 & 2.04E-09 \\
6 & 11.49 & 11.01--12.00 & 1.00 & 2.017E+04 & . . .--2.353E+04 & 6.413E+03 & 1.33E-09 \\
7 & 12.48 & 12.01--13.00 & 1.00 & 2.713E+04 & . . .--3.118E+04 & 7.652E+03 & 9.10E-10 \\
8 & 13.47 & 13.01--14.00 & 1.00 & 3.535E+04 & . . .--4.019E+04 & 9.014E+03 & 7.10E-10 \\
\enddata
\end{deluxetable}

%
\startlongtable
\begin{deluxetable}{lccccccc}
\tabletypesize{\scriptsize}
\tablecaption{
Grains with $a=0.2$~$\mu$m, 0.35$a$ icy mantle and $T_d=5.84$~K, shielded by interstellar gas with column density $N_H=2.56\times10^{23}$~H\,atoms\,cm$^{-2}$.
}
\label{tab0.2-7}
\tablehead{
\colhead{No.} & \colhead{$T_{\rm CR}$, K} & \colhead{interval, K} & \colhead{$\Delta T_{\rm CR}$, K} & \colhead{$E_{\rm grain}$, eV} & \colhead{interval, eV} & \colhead{$\Delta E_{\rm grain}$, eV} & \colhead{$f_T$, s$^{-1}$}
}
\startdata
1 & 5.92 & 5.85--6.00 & 0.16 & 1.361E+02 & 15.637--248.629 & 2.330E+02 & 1.09E-07 \\
2 & 6.29 & 6.01--7.00 & 1.00 & 7.756E+02 & . . .--2.197E+03 & 1.948E+03 & 1.11E-07 \\
3 & 7.40 & 7.01--8.00 & 1.00 & 3.213E+03 & . . .--4.816E+03 & 2.620E+03 & 1.26E-08 \\
4 & 8.43 & 8.01--9.00 & 1.00 & 6.226E+03 & . . .--8.215E+03 & 3.399E+03 & 4.66E-09 \\
5 & 9.46 & 9.01--10.00 & 1.00 & 1.015E+04 & . . .--1.250E+04 & 4.288E+03 & 2.32E-09 \\
6 & 10.48 & 10.01--11.00 & 1.00 & 1.497E+04 & . . .--1.779E+04 & 5.292E+03 & 1.36E-09 \\
7 & 11.48 & 11.01--12.00 & 1.00 & 2.080E+04 & . . .--2.421E+04 & 6.411E+03 & 9.40E-10 \\
8 & 12.48 & 12.01--13.00 & 1.00 & 2.786E+04 & . . .--3.186E+04 & 7.650E+03 & 6.65E-10 \\
\enddata
\end{deluxetable}

%
\startlongtable
\begin{deluxetable}{lccccccc}
\tabletypesize{\scriptsize}
\tablecaption{
Grains with $a=0.2$~$\mu$m, 0.35$a$ icy mantle and $T_d=5.60$~K, shielded by interstellar gas with column density $N_H=5.12\times10^{23}$~H\,atoms\,cm$^{-2}$.
}
\label{tab0.2-8}
\tablehead{
\colhead{No.} & \colhead{$T_{\rm CR}$, K} & \colhead{interval, K} & \colhead{$\Delta T_{\rm CR}$, K} & \colhead{$E_{\rm grain}$, eV} & \colhead{interval, eV} & \colhead{$\Delta E_{\rm grain}$, eV} & \colhead{$f_T$, s$^{-1}$}
}
\startdata
1 & 5.75 & 5.61--6.00 & 0.40 & 2.257E+02 & 14.348--607.786 & 5.934E+02 & 1.53E-07 \\
2 & 6.32 & 6.01--7.00 & 1.00 & 1.189E+03 & . . .--2.556E+03 & 1.949E+03 & 4.41E-08 \\
3 & 7.41 & 7.01--8.00 & 1.00 & 3.579E+03 & . . .--5.176E+03 & 2.620E+03 & 7.48E-09 \\
4 & 8.45 & 8.01--9.00 & 1.00 & 6.650E+03 & . . .--8.576E+03 & 3.399E+03 & 2.71E-09 \\
5 & 9.45 & 9.01--10.00 & 1.00 & 1.045E+04 & . . .--1.286E+04 & 4.289E+03 & 1.53E-09 \\
6 & 10.46 & 10.01--11.00 & 1.00 & 1.523E+04 & . . .--1.816E+04 & 5.292E+03 & 9.54E-10 \\
7 & 11.47 & 11.01--12.00 & 1.00 & 2.109E+04 & . . .--2.457E+04 & 6.412E+03 & 6.75E-10 \\
8 & 12.49 & 12.01--13.00 & 1.00 & 2.825E+04 & . . .--3.222E+04 & 7.651E+03 & 4.65E-10 \\
\enddata
\end{deluxetable}

\end{document}